\documentclass[12pt]{article}
\usepackage{amsmath,amssymb,array}
\newcommand{\pa}{\partial}
\newcommand{\py}{-i\frac{\pa}{\pa y}}
\newcommand{\x}{\stackrel{2}{x}}
\newcommand{\p}{\stackrel{1}{-i\mu\frac{\pa}{\pa x}}}
\newcommand{\ph}{\stackrel{1}{-ih\frac{\pa}{\pa x}}}

\newcommand{\const}{\rm const}
\makeatletter
\renewcommand{\@biblabel}[1]{#1.}
\makeatother
\newcommand{\h}{\hspace*{1pt}}
\makeatletter

\@addtoreset{equation}{section}
\makeatother

\begin{document}
\sloppy

\title{OPERATOR SEPARATION OF VARIABLES FOR ADIABATIC PROBLEMS
IN QUANTUM AND WAVE MECHANICS}

\author{Belov V.V.\thanks{Institute for Electronics and Mathematics,
Trekhsvyatitelskiy per., 3/12, 109028, Moscow, Russia, E-mail: belov@miem.edu.ru} \and Dobrokhotov S.Yu.\thanks{Institute for Problems in
Mechanics  Russian Academy of Sciences,  prosp.Vernadskogo, 101-1, Moscow
119526 Russia; E-mail: dobr{@}ipmnet.ru} \and
Tudorovskiy T.Ya.\thanks{Institute for Problems in Mechanics  of
Russian Academy of Sciences,  prosp.Vernadskogo, 101-1, Moscow 119526 Russia;
E-mail: timtudor{@}mail.ru }}


\date{}
\maketitle
\renewcommand{\theequation}{\thesection .\arabic{equation}}

\bigskip


\bigskip

\begin{abstract}
We study linear problems of mathematical physics in which the adiabatic approximation is used in the wide sense.
Using the idea that all these problems can be treated as problems with operator-valued symbol, we propose a general
regular scheme of adiabatic approximation based on operator methods. This scheme is a generalization
of the Born--Oppenheimer and Maslov methods, the Peierls substitution, etc. The approach proposed in this paper
allows one to obtain ``effective" reduced equations for a wide class of states inside terms (i.e., inside modes,
subregions of dimensional quantization, etc.) with the possible degeneration taken into account. Next, applying
the asymptotic methods in particular, the semiclassical approximation method, to the reduced equation, one can classify
the states corresponding to a distinguished term (effective Hamiltonian). We show that the adiabatic effective
Hamiltonian and the semiclassical Hamiltonian can be different, which results in the appearance of  ``nonstandard
characteristics" while one passes to classical mechanics. This approach is used to construct solutions of several
problems in wave and quantum mechanics, in particular, problems in molecular physics, solid-state physics, nanophysics,
hydrodynamics.
\end{abstract}

\section{Introduction}
Many linear problems of mathematical and theoretical physics
contain different spatio-temporal scales.
Among them there are problems of molecular physics,
problems concerning electron waves in crystals,
wave propagation in media with rapidly varying
characteristics,
surface and internal waves in fluids,
electron--phonon interaction,
electromagnetic waves and quantum particles propagation in
waveguides, etc.
The main instruments
for investigating such problems are contained
in the adiabatic approximation, which is based on the idea
of separation of ``fast" and ``slow" modes by means of
``freezing" the slowly varying variables.
For instance, the ``slow" variables describe the nuclear motion in
molecules and the ``fast" variables concern the electron motion,
or the ``slow" variables describe the longitudinal motion
and the ``fast" variables describe the transverse motion
in thin waveguides. Needless to say that
there are many different versions of adiabatic approximation
and thousands papers and monographs related to this approach and
its applications in different fields of mechanics and physics.
Among them we mention \cite{Born,Peierls,BornHuang,LandLif,
LifPit,MaslovAsymptMethods,Berry1,Percival,Hagedorn,Mir}.
Nevertheless,
we take the liberty to present
a general regular scheme of the adiabatic approximation
suggested in \cite{DobrWaterWaves,DobrOperVal,BerlyandDobr} and
to combine different approaches including the Born--Oppenheimer method,
the Maslov operator method, the Peierls substitution, etc.
From the mathematical viewpoint,
the equations (or the system of equations) describing all these
phenomena have the same structure.
Namely, following~\cite{MaslovAsymptMethods},
these equations (systems) can be treated as
equations with ``operator-valued symbol.''
Our idea of the study of such equations is not new:
the asymptotic analysis of the original problem
can be divided into two parts:
(1) the ``operator'' reduction to simpler differential
or pseudodifferential equations with the principal symbol,
known in different fields of physics as an effective Hamiltonian,
or therm, or dispersion relation, or mode, etc.,
and with corrections to this symbol;
(2) asymptotic constructions of the solutions
of this simpler reduced equation based on different variants
of the semiclassical approaches,
like the WKB-method, Born-method, oscillatory approximations,
ray expansions,
the Maslov canonical operator, averaging, etc.
Here we present the first part of this concept
in the form of a regular rigorous algorithm (in \S 3),
based on operator methods \cite{MaslovOperMethods}.

The result of the first step is the reduced equation;
it has different names in different fields of physics,
we call it the {\it effective equation of adiabatic
motion.} We illustrate the ``operator" reduction or
the ``operator separation" of variables
by using the above-mentioned problems from different fields
of physics and, in \S4,
present the corresponding equations
for the wave functions of adiabatic motion. The examples
given in items 4.1-4.4 of the paragraph were studied
long ago, whereas the results of items 4.5-4.6 (as well as
\S 5) were obtained by authors recently.

To realize the second step,
it is necessary to take into account that usually
the original problem includes several parameters:
some of them, like the transverse and longitudinal
characteristic sizes of a waveguide
or the ratio between the masses of light and heavy particles,
allow one to use the adiabatic approximation
and do not crucially correlate
with the energy of adiabatic motion,
and the other ones, like magnitudes of the external
electromagnetic field, the momentum of the incoming wave in
the scattering problem, etc., determine the energy.
This fact implies different forms of (asymptotic) wave functions
of adiabatic motion and, as a consequence,
a redefinition of the principal symbol and the effective
Hamiltonians depending on the relations
between the above-mentioned parameters.
In turn, it gives different types of characteristics
(trajectories of Hamiltonian systems)
which V.~Maslov \cite{MaslovNonstandard} called ``nonstandard characteristics"
and which must be used in asymptotic constructions.
We discuss the possible classification of these characteristics
using, as the main example, the quantum wave propagation in thin
(or {\it nano}) tubes.
In spite of the fact that the given arguments
seem to be natural, and in some way
contain in physical literature,
we didn't find their systematic consideration.
The methods for constructing
asymptotic or exact solutions of the ``redefined" equation for
the wave functions of adiabatic motion are well known
and here the results must be connected with a concrete physical problem.
Therefore, we do not construct asymptotic solutions
for most of the derived reduced equations
and in \S 5  only briefly describe different solutions for the
equations of quantum particles in nanostructures.

The main results of the ``operator separation of variables" are
realized in formulas \eqref{PsiIn2}, \eqref{psieq0} and Eq. \eqref{chieq}.
Although they simply develop the approach of
Born-Oppenhiemer, Peierls and Maslov, nevertheless,
they allow us to consider
a wide range of adiabatic problems
uniformly and in a rather compact form.
We believe that this approach is very useful in different
situations, since it gives not only a general regular scheme
for deriving the reduced equations exactly but allows one to
obtain qualitative and quantitative estimates of the range of
applicability of any approximation.
Naturally,
the argument resulting in formulas~\eqref{PsiIn2}
and~\eqref{psieq0},
the classification of different approximations,
the relations between the adiabatic and semiclassical
asymptotics, etc. can be better illustrated
with a simple example. These considerations,
some of which are well known in physics
and some of which are well known in mathematics,
are given in \S 2 and \S\S 3.1-3.2.
In \S 2, we present
a minimal amount of the required information
from the operator calculus of noncommuting
operators~\cite{MaslovOperMethods} (see also~\cite{DobrZh}).
We point out that the facts
from~\cite{MaslovOperMethods} used here
are not simply arguments
of ``mathematical justification and verification" type,
but are completely constructive and developed algorithms
well adjusted to the problems studied here.

Finally we can formulate the main result of the paper in the following way. We suggest the regular asymptotic (adiabatic) procedure which allows one 1) to determine correctly the leading part of asymptotic solution corresponding to wide diapason of energies (or frequencies) 2) to construct and estimate if necessary the ``adiabatic" corrections. Needless to say that in this work we understand an asymptotic solution in the formal sense, i.e. in the sense of small ``right hand side" (discrepancy). The proof of the fact that the constructed asymptotic solution approximate some exact solution of original equation is out of scope of our consideration here and we touch this problem only very briefly.

\section{Differential and pseudodifferential operators with
a parameter and their symbols. Elementary formulas from calculus
of noncommuting operators.}

We want to study some asymptotic solutions of (systems of) partial
differential equations with small parameter $\mu$ in the
configuration space with coordinates
$x=(x_1,\ldots,x_N)$ which can be written in general form:
\begin{gather}\label{eq1}
i\Psi_t=\hat {\cal H}\Psi.
\end{gather}

Here $\Psi(x,t)$ can be a scalar or vector function,
${\cal H}$ is a partial differential scalar or matrix operator.
It is convenient for us to present the operator ${\cal H}$
as a function of noncommuting operators
$-i\pa/\pa x=(-i\pa/\pa x_1,\ldots,-i\pa/\pa x_N)$
and $x=(x_1,\ldots,x_N)$ and, generally speaking, of time $t$:
$\hat{\cal H}={\cal H}(-i\pa/\pa x_1,\ldots,-i\pa/\pa
x_N,x_1,\ldots,x_N,t)$,  where the function
${\cal H}(p_1,\ldots,p_n,x_1,\ldots,x_N,t)$ is usually called
the symbol of the operator $\hat{\cal H}$.
Actually, we shall consider
the situation in which the function ${\cal H}$ can depend on
the parameter $\mu$ and also on some other ones.
Quite often a small parameter $\mu$ appears
as a factor before the derivatives $\pa/\pa x_j$, say,
before $\pa/\pa x_1,\ldots,\pa/\pa x_n,\ n\leq N$.
It follows from the considerations given below
that there is always a parameter $\mu$ before
$\pa/\pa t$.
We denote the other variables by $y_1,\ldots,y_m$, $m=N-n$.
So finally Eq.~\eqref{eq1} takes the form
\begin{gather}\label{InEq}
i\mu\Psi_t=
{\cal H}\left(-i\mu\frac{\pa}{\pa
x_1},\ldots,-i\mu\frac{\pa}{\pa x_n},x_1,\ldots,x_n,
-i\frac{\pa}{\pa y_1},\ldots,-i\frac{\pa}{\pa
y_m},y_1,\ldots,y_m,t,\mu\right)\Psi
\end{gather}
As the operators $\pa/\pa x_j$ and $x_j$,
as well as operators $\pa/\pa y_k$ and $y_k$
do not commute,
one has to agree about the order of action of
$x_j$ and $\pa/\pa x_j$ and, analogously, of $y_k$ and $\pa/\pa y_k$.
The theory of functions of noncommuting operators is very well
developed \cite{MaslovOperMethods},
see also \cite{DobrZh,KarasevMaslov1,KarasevMaslov2}.
For the completeness of consideration,
let us present
a minimal amount of the required information
from the operator calculus and recall the terminology.

First, let
${R}(x,p)=\sum_{k=0}^l R_k(x)p^k$
be a polynomial in variables $p$ with coefficients smooth in~$x$.
This function generates the operator
$\hat R=\sum _{k=0}^lR_k(x)(-i\mu \frac{\pa }{\pa x})^k$.
The function $R(x,p)$ is called the symbol of the differential
operator $\hat R$ with a parameter $\mu$.
It is clear that the way of the construction of the operator
$\hat R$ by means of the symbol $R$ is not unique.
For example, one can build an operator
${\hat R}'=\sum _{k=0}^l (-i\mu \frac{\pa}{\pa x})^k R_k(x)$,
different from $\hat R$.
Using the Feynman notation, we can write
$\hat R=R(\stackrel{1}{\hat p},\stackrel{2}{x})$ and
$\hat R=R'(\stackrel{2}{\hat p},\stackrel{1}{x})$,
where the numbers above ${\hat p}$ and $x$ determine the order
of their action. (About other ways of ordering, for example, by Weyl,
see \cite{MaslovOperMethods}). In this work,
we will always use the first way of ordering.
Under this agreement, the definition
of the differential operator is equivalent to the definition of
its symbol. By letting the order of the polynomial $k$
tend to infinity,
one can obtain, at least, ``naive" operators
whose symbols are not polynomials.
Such operators are called pseudodifferential.
Their rigorous definition is given by means
of the ``$\mu$-Fourier transform"
\cite{MaslovAsymptMethods,MaslovOperMethods}:
\begin{equation}\label{PDO}
A(\stackrel{1}{\hat p},\stackrel{2}{x})\varphi(x)=
F_{p\to x}^\mu [A(p,x) [F_{x\to p}^\mu\varphi(x)](p)](x);
\end{equation}
where the direct and inverse ``$\mu$-Fourier transforms"
$F_{x\to p}^\mu$ and $F_{p\to x}^\mu$ are defined
by the equalities:
\begin{equation*}
[F_{x\to p}^\mu\varphi(x)](p)=\frac{1}{(2\pi i \mu)^{n/2}}
\int_{\mathbb{R}^n_x}
e^{-i\langle p,x\rangle/\mu}\varphi(x)d x, \,
[F_{p\to x}^\mu\tilde\varphi(p)](x)=\frac{1}{(-2\pi i \mu)^{n/2}}
\int_{\mathbb{R}^n_p}e^{i\langle p,x\rangle/\mu}\tilde\varphi(p)d p.
\end{equation*}
From now on, $\langle,\rangle$ is the inner product
in the Euclidean space of the corresponding dimension.

The replacement of operators by their symbols
turns out to be very useful in practical calculations.
As a result,
the calculations concerning operators
are replaced
by significantly simpler work
(which can be algorithmized)
with symbols, i.e., functions  (``with $c$-numbers").
Since, in asymptotic approaches,
defining an operator is practically equivalent
to defining its symbol,
in the process of obtaining asymptotic formulas
{\it one can manipulate only with symbols
and ``recall" the operators corresponding to these symbols
only in studying refined problems such as, for example,
justification of the asymptotic accuracy
of the solutions constructed.}
Of course,
the main difficulties in dealing
with functions of operators
arise due to the fact that
the operators~$\hat p$ and~$x$ do not commute.
On the other hand,
their commutator is~$i\mu$,
and it is small, which allows one to use
asymptotic expansions in the constructions.
In view of this fact, it is natural to consider
the symbols~$R$ depending on the parameter~$\mu$
and to assume that
$R(p,x,\mu)=R_0(p,x)+\mu R_1(p,x)+\cdots$.
Moreover, the right-hand side in this relation
is understood as an asymptotic expansion
in the parameter~$\mu$.
The function $R_0(p,x)$ is called
the leading symbol or, sometimes, a Hamiltonian,
and~$R_j$ are called $j$th-order corrections.

The next generalization consists in the assumption that
the symbol of the operator $\hat R$ may be {\it an operator}.
A simple example appears in the situation in which
$R(p,x,\mu)$ is a matrix
(or an operator acting in a finite-dimensional space).

{\bf Example 1. The Klein--Gordon equation.}
Consider, for instance, the Klein--Gordon equation
$\mu^2\varphi_{tt}-\mu^2\varphi_{xx}+v(x)\varphi=0$
written in vector form for the vector function
$\Psi=\begin{pmatrix} \Psi_1 \\
\Psi_2\end{pmatrix}=\begin{pmatrix}\varphi \\i\mu\varphi_t\end{pmatrix}$:
\begin{gather}i\mu\Psi_t={\cal H}(-i\mu\pa/\pa x,x)\Psi,
\quad \quad\Leftrightarrow\quad
\left\{\begin{array}{l}
i\mu\Psi_{1t} = \Psi_2,\\
i\mu\Psi_{2t} = v(x)\Psi_1+\mu^2\Psi_{2xx}.
\end{array}\right.
\end{gather}
The symbol of the operator ${\cal H}(p,x)$
is the $2\times 2$ matrix function
\begin{gather*}{\cal H}(p,x)=\begin{pmatrix}
0 && 1 \\
p^2+v(x) && 0
\end{pmatrix}.
\end{gather*}

From this viewpoint,  one can consider many fundamental
physical equations like the Dirac and Pauli equations,
the Lam\'e equation in linear elasticity theory,
the linearized hydrodynamics equations, etc.
(If they include a small parameter in an appropriate way.)
The appearance of a small parameter $\mu$
before the derivative $\pa/\pa x$ is very important in our
constructions. As we have mentioned,
there exist many problems with different scales in which
a small parameter appears only in front of some derivatives.
Problems of such types give the majority of nontrivial
$\mu$-differential operators with operator-valued symbols.

{\bf Example 2. Molecular physics.}
Consider, for instance,
the Schr\"odinger equation for two groups of particles:
heavy atomic nuclei with mass~$M$ and
light electrons with mass~$m$.
We denote the coordinates of nuclei and electrons
by~$x'$ and~$y'$, respectively.
Let us assume that $l_0$ is the linear size of a molecule and
$d_0$ is the amplitude of nuclear oscillations.
Thus the characteristic magnitude of the
electron energy is $\varepsilon_e\sim\hbar^2/(2ml_0^2)$.
By physical reasons (stated by Born and
Oppenheimer), the motion of a nucleus could be considered
in the oscillatory approximation and its energy is
$\varepsilon_n\sim\hbar^2/(2Md_0^2)\sim k d_0^2/2$
with the elasticity coefficient $k$.
To estimate $k$, one has to remember that,
in the adiabatic approximation,
the potential energy of a nucleus
is the total energy of electrons \cite{Shiff},
so $k\sim\pa^2\varepsilon_e/\pa x^2\sim \hbar^2/(ml_0^4)$.
Thus we have $\hbar^2/(2Md_0^2)\sim \hbar^2 d_0^2/(2ml_0^4)$.
From this, we obtain $d_0/l_0\sim (m/M)^{1/4}$.
Oscillatory energies of nuclei and electrons relate as
$\varepsilon_n/\varepsilon_e\sim (m/M)(l_0^2/d_0^2)\sim
(m/M)^{1/2}$.
Let us introduce the parameter $\mu=(m/M)^{1/2}$
and divide both sides of the Schr\"odinger equation by
$\hbar^2/(2ml_0^2)$. After passage to dimensionless variables
$x=x'/l_0,\ y=y'/l_0$, the stationary Schr\"odinger equation
takes the form
\begin{gather}
\label{BornOpp} \hat {\cal H}\Psi=E\Psi,\qquad\hat {\cal H}=\left(
-\frac{1}{2}\mu^2\Delta_x
-\frac{1}{2}
\Delta_y
+v(x,y)\right)\Psi=E\Psi.
\end{gather}
The symbol ${\cal H}$ of the $\mu$-differential\footnote{Born
and Oppenheimer in their famous paper \cite{Born} used the parameter
$\varkappa=\sqrt{\mu}$, which is the ratio
$d_0/l_0\sim\varkappa$ of the characteristic wavelength
to the wave function and the linear size of the molecule.
}
operator $\hat {\cal H}$ is again the operator
\begin{gather}
\label{Smb} {\cal H}(p,x)=\frac{1}{2}p^2-\frac{1}{2}\Delta_y
+v(x,y).
\end{gather}
Usually, {\it $x$ are called slow variables and $y$ are called
fast ones}.
Ideologically close approaches to the determination of electron
states ({\it terms}) in molecule one can find in \cite{MaslovAsymptMethods,Percival,Hagedorn,BelovMolecule}.

{\bf Example 3. Quantum 2-D waveguide.}
One can meet an equation with closed structure
considering  a ``narrow" straight quantum waveguide.
The word ``narrow" means that the characteristic width
of the waveguide $d_0$ is much smaller than its length $l_0$.
We introduce the small parameter $\mu=d_0/l_0$.
The dynamics of a spinless quantum (quasi)particle
in a plane waveguide is determined by the 2-D Shr\"odinger
equation with the potential $v=v(x,y)$ inside the waveguide.
Due to two different scales, there appear two different
characteristic energies: the characteristic energy of the lower
transverse levels (which is usually called the characteristic
energy of the ``transverse quantization") and the
characteristic longitudinal energy $\varepsilon_\parallel$.
One can estimate $\varepsilon_\perp$
from the uncertainty principle, which gives
$\varepsilon_\perp=\hbar^2/(2md_0^2)$.
Let us introduce dimensionless variables $x'=x/l_0$, $y'=y/d_0$,
$t=(\mu\omega_\perp)^{-1},\ \omega_\perp=\varepsilon_\perp/\hbar$
and dimensionless potentials
$v'=v/\varepsilon_\perp$.
Then the corresponding Schr\"odinger equation takes the form
(we omit the primes of the dimensionless variables):
\begin{gather}
\label{2DWaveGuide} i\mu\frac{\pa \Psi}{\pa t}
=\hat {\cal H}\Psi,\qquad\hat {\cal H}=\left(
-\frac{\mu^2}{2}\frac{\pa^2}{\pa x^2}
-\frac{1}{2}\frac{\pa^2}{\pa y^2}+v(x,y)\right)\Psi.
\end{gather}
The symbol of the $\mu$-differential operator is the operator
\eqref{Smb} with $\Delta_y=\pa^2/\pa y^2$.

Generalizations of the plane quantum waveguide
are quantum thin tubes (nanotubes) and thin films (nanofilms),
and their symbols are matrix-operators if one includes
spin into consideration. These more complicated examples, as
well as several examples from other fields, will be considered later.

Let us again stress that the {\it definition of the symbol of
a $\mu$-differential operator differs from the standard definition
of the symbol of an operator without parameter}.
Namely, we construct the symbol taking only
the slow variables into account.
This is why the symbols of the $\mu$-differential operator
$\hat{\cal H}$ in Examples~2 and~3 are
again differential operators
acting in some appropriate Hilbert space with coordinates~$y$.
The transition to $\mu$-differential symbols
is a formalization of the idea of
``freezing the slow variables."
We discuss the related problems later.
The introduction of a small parameter $\mu$ formally
ensures that the commutator  $[x,\hat{\cal H}]$ is small.
There is no universal interpretation of this fact;
this depends on each concrete physical problem.

Let us also note that one can consider the equations from
Examples 2 and~3 as infinite vector ones.
To show this, let us assume, for simplicity,
that, for each $x\in\mathbb{R}^n$,
the spectrum of the operator
$-(1/2)\pa^2/\pa y^2+v(x,y)$ is discrete and simple
and that the corresponding eigenfunctions
$\{w_n(x,y)\}$ and eigenvalues $\lambda_n(x)$
depend smoothly on~$x$.
Then one can expand any solution $\Psi(x,y,t)$ of
Eq.~\eqref{2DWaveGuide} in the Fourier series
\begin{gather}
\Psi=\sum_k w_k(x,y)\psi_k(x,t).
\label{PsiFast}
\end{gather}

Substituting solution \eqref{PsiFast} into
Eq.~\eqref{2DWaveGuide}, we obtain:
\begin{gather}
i\mu\frac{\pa\psi_k}{\pa t}=-\frac{\mu^2}{2}\frac{\pa^2\psi_k}
{\pa x^2}-\mu^2\sum_n\langle w_k,\frac{\pa w_n}{\pa x}
\rangle_y\frac{\pa\psi_n}{\pa x}
-\frac{\mu^2}{2}\sum_n\langle w_k,\frac{\pa^2 w_n}{\pa x^2}\rangle_y\psi_n
\label{Schred2}
\end{gather}

If we introduce the infinite vector
$\psi=(\psi_1,\psi_2,\ldots)^T$,
then we can represent Eq.~\eqref{Schred2} as the following
infinite vector equation with infinite-dimensional matrix
Hamiltonian $\hat{\cal H}$:
\begin{gather*}
i\mu\psi_t=\hat {\cal H}\psi, \quad \hat {\cal H}={\cal H} _0(p,x) +
\mu {\cal H}_1(p,x) + \mu^2 {\cal H}_2(p,x), \\
({\cal H}_0)_{kn}=\left(\frac{p^2}{2}+\lambda_n(x)\right)\delta_{kn}, \quad
({\cal H}_1)_{kn}=-i\langle w_k,\frac{\pa w_n}{\pa x}\rangle_y p, \quad
({\cal H}_2)_{kn}=-\frac{1}{2}\langle w_k,\frac{\pa^2 w_n}{\pa x^2}\rangle_y.
\end{gather*}

In all examples considered above,
the momentum operators corresponding to the slow variables $x_j$
are $-i\mu\pa/\pa x_j$.
Of course, one can consider a general situation
in which the Hamiltonian depends on the operators
$\hat x_j,\ \hat p_j$ generating the Heisenberg algebra
with commutators
$[\hat p_j,\hat x_k]=\mu\delta_{j,k},\quad\mu\ll 1$.
Such a situation appears in the electron-phonon interaction,
we shall discuss it in \S 4.
The other obvious generalizations of the equations
with operator-valued symbols
are vector equations containing
``slow" and ``fast" variables.
For instance, we can consider the Pauli equation
in a thin quantum waveguide. In this case (see \S4.6),
the symbol is a matrix operator differential
with respect to fast variables.

To conclude this section,
we present a useful formula which plays an important role
in the future consideration. Let $\hat A$ and $\hat B$
be pseudodifferential operators
$$\hat A=A(\stackrel{1}{-i\mu\pa/\pa x},\x,\mu),\qquad
\hat B=B(\stackrel{1}{-i\mu\pa/\pa x},\x,\mu),$$
then the symbol $smb (\hat A \hat B)$ of their product
$\hat A \hat B$ is equal to (see \cite{MaslovOperMethods})
\begin{equation}\label{smb}
smb (\hat A \hat B)={A(p\stackrel{1}{-i\mu\pa/\pa x},\x,\mu)
B(p,x,\mu)}. \end{equation}

\section{General scheme of the operator separation of variables
in adiabatic problems. }

\subsection{General statement of the problem with
operator-valued symbols and parameters.}

We are going to construct a certain asymptotic solution
${\Psi=(\Psi^1,\ldots,\Psi^s)^T}$, $s\geq2$, to vector equation
\eqref{InEq} with a small parameter $\mu\ll 1$
or its stationary variant
\begin{gather}\label{InEqSt}
\hat{\cal H}\Psi=E\Psi.
\end{gather}
In \eqref{InEq} and in \eqref{InEqSt}, the matrix operator
(quantum matrix Hamiltonian) $\hat{\cal H}$
is generated by its operator-valued symbol
\begin{gather*}
{\mathcal H}={\cal H}(p,x,-i\pa/\pa y,y,t,\mu)=
\begin{Vmatrix}
{\cal H}_{11} & \ldots & {\cal H}_{1s} \\
\hdotsfor{3} \\
{\cal H}_{s1} & \ldots & {\cal H}_{ss}
\end{Vmatrix},
\\
{\cal H}_{ij}={\cal H}_{ij}(p,x,\py,y,t,\mu),\qquad
1\leq i,j\leq s.
\end{gather*}
(In the stationary case, ${\cal H}_{ij}$ do not depend on time~$t$.)
We assume that the operator-valued symbol (the matrix-operator)
${\cal H}=\|{\cal H}_{ij}(p,x,\py,y,\mu)\|$
smoothly depends on $p,x,t$ and acts in an appropriate vector
Hilbert space $\mathbb{H}_y$ with coordinates $y$
from some domain $\mathbb{M}_y$ and with the inner product
$\langle\cdot,\cdot\rangle|_y$
(for instance, in
$L_2(\mathbb{M}_y)\times L_2(\mathbb{M}_y)\times\ldots \times
L_2(\mathbb{M}_y)$).
Another natural assumption is that
the symbol ${\cal H}(p,x,\py,y,\mu)$ can be expanded
into a regular series with respect to the parameter~$\mu$:
\begin{gather}
\mathcal{H}(p,x,\py,y,t,\mu)=\mathcal{H}_0(p,x,\py,y,t)+
\mu\mathcal{H}_1(p,x,\py,y,t)+\ldots.
\end{gather}
We also assume that the (pseudo)differential operator
$\hat{\cal H}$ acts in an appropriate expanded Hilbert space
$\mathbb{H}_{x,y}$ with coordinates
$(x,y)\in \mathbb{R}^n_x\times \mathbb{M}_y$
and all the future operations related to them are valid.
Of course, one has to verify the last assumption in each concrete
problem.
Usually (but not always), we shall consider situations
in which ${\cal H}$, as well as the operator $\hat{\mathcal H}$,
are essentially self-adjoint.

It is important to emphasize again that in \eqref{InEq} there
is a small ``adiabatic" parameter $\mu$ in front of the
derivatives with respect to ``slow" variables $x$,
but there is no small parameter in front of the derivatives
with respect to ``fast" variables $y$.

Of course, one has to add additional boundary and initial
conditions to Eq.~\eqref{InEq}. We shall do it later after
the discussion in \S 4, and now we only note that
we are going to consider only special problems
interesting from the physical viewpoint.
The statements of these problems follow the
adiabatic separation of the original Eq.~\eqref{InEq}
into a set of reduced equations corresponding to different
``terms" or ``modes" and determined
by  ``effective Hamiltonians" or ``dispersion relations."
We present our concept of this separation
(the ``operator separation of variables")
together with the corresponding formulas
in the two subsequent sections.

\subsection{Anzatz of the operator separation of variables.}

Let us illustrate the main ideas of the operator separation
of variables in adiabatic problems with an example of
a ``quantum waveguide" \eqref{2DWaveGuide}.
If the potential $v(x,y)$ is the sum $v_1(x)+v_2(y)$,
one can separate the variables and find
a special solution to Eq.~\eqref{2DWaveGuide} as a product
of two functions (modes) ${\chi}(y,\mu)\psi(x,t,\mu)$.
It is clear that this representation is not true if
$v(x,y)\neq v_1(x)+v_2(y)$,
nevertheless,
since there are different scales in the longitudinal and
transverse directions, we can separate the modes adiabatically.
According to the standard adiabatic approach based on the
fundamental papers by Born and Oppenheimer,
the leading term of the wave function in the adiabatic
approximation is sought in the form of the product
\begin{gather}
\Psi(x,y,t,\mu)\approx{\chi}(x,y,\mu)\psi(x,t,\mu).
\label{PsiIn0}
\end{gather}
But this representation can be used in a situation when
the function $\psi(x,t,\mu)$ is quite smooth and
works poorly for large enough energies of longitudinal motion.
If the function $\psi(x,t,\mu)$ exhibits fast oscillations,
for instance, if $\psi$ is the WKB-solution
$\psi(x,t,\mu)=\exp(iS(x,t,\mu)/\mu)\varphi(x,t,\mu,h)$, then
representation \eqref{PsiIn0} is not convenient
for the asymptotic expansion and,
instead of formula \eqref{PsiIn0}, one has to include
the classical momentum $\pa S/\pa x$ into the factor
${\chi}(y,x,\mu)$ and use the formula~\cite{MaslovAsymptMethods}
\begin{gather}
\Psi(x,y,t,\mu)\approx \chi(\pa S/\pa x,x,y,\mu)\psi(x,t,\mu).
\label{PsiIn1}
\end{gather}
Recall that the phase $S$ is the solution of the Hamilton--Jacobi
equation $\pa S/\pa t+H_{\rm eff}(\pa S/\pa x,x,t)=0$
with the so-called effective Hamiltonian $H_{\rm eff}(p,x,t)$.
For the case in which $H_{\rm eff}(p,x,t)$
is a function of $p$ only and $S=-\omega t+ p x$,
the Hamilton--Jacoby equation is the {\it dispersion relation}.
Formula \eqref{PsiIn1} is still not satisfactory,
because for the case in which there exist focalization effects,
i.e., there are turning points or caustics,
the WKB-representation is not true, and it is necessary to change the
form of $\psi$ and the form of $\chi$. We propose to ``correct"
\eqref{PsiIn1} in such a way that a new formula would also work
in the case of focal and turning points.
This correction is based on the observation that,
in the WKB-case modulo a small correction,
the right-hand side in \eqref{PsiIn1} remains the same
(see, e.g., \cite{MaslovAsymptMethods})
if one assumes that the first factor
is the (pseudodifferential) {\it operator}
${\chi}(\stackrel{1}{-i\mu\pa/\pa x},\x,y,t,\mu)$
written as a function (its symbol) of
the noncommuting operators $x$ and $\hat p=-i\mu\pa/\pa x $.
Finally,
we suggest to look for the solution $\Psi(x,y,t)$
in the adiabatic approach in the following form
\cite{DobrWaterWaves,BerlyandDobr,DobrZh,DobrDoktor}:
\begin{gather}
\Psi(x,y,t,\mu)= {\chi}(\p,\x,y,t,\mu)\psi(x,t,\mu),
\label{PsiIn2}
\end{gather}
where $\hat{\chi}$ is the ``pseudodifferentional"  operator
whose symbol has the (asymptotic) expansion with respect to
the parameter $\mu$
\begin{equation}
\chi(p,x,y,t,\mu)=\chi_0(p,x,y,t)+\mu\chi_1(p,x,y,t)+\ldots.
\label{expchi}
\end{equation}
From the physical viewpoint, representation \eqref{PsiIn2}
means that we ``freeze" not only slow variables $x$
as in formula \eqref{PsiIn0}, but also slow momenta,
which are differential operators $-i\mu\pa/\pa x$
in quantum mechanics.
Note that in many situations
the leading term $\chi_0(p,x,y,t)$ in expansion \eqref{expchi}
does not depend on $p$, but the corrections usually do.
This dependence plays an important role
in the estimation of the limits
of the adiabatic approximation in concrete problems.

We still do not fix the equation for the function $\psi$
describing the longitudinal motion. Following the idea of
the so-called {\it Peierls substitution} in solid state physics
(see, e.g., \cite{Peierls,LifPit,Panatti}),
we assume that the wave function
$\psi$ is a solution of the following equation
(describing the longitudinal dynamics):
\begin{equation}
\label{psieq0} i\mu\psi_{t}=\hat L \psi,\qquad \hat L=L(\p,\x,t,\mu),
\end{equation}
where $\hat L$ is a pseudodifferential (sometimes, differential)
operator with symbol $L(p,x,t,\mu)$ having the expansion
\begin{equation}
L(p,x,t,\mu)=L_0(p,x,t)+\mu L_1(p,x,t)+\ldots.
\label{expL}
\end{equation}
The operator $\hat L$ is called the (full) quantum effective
Hamiltonian with the principal part $\hat L_0$. Sometimes, the
symbol $L_0$ is also called an effective classical Hamiltonian
and is denoted by $L_0=H_{\rm eff}(p,x,t)$.
The operator $\hat \chi$ will be called
an {\it intertwining\/} operator (cf.~\cite{MaslovOperMethods,KarasevMaslov1,KarasevMaslov2}).
Equation~\eqref{psieq0} can also be understood
as the {\it quantization of the Hamilton--Jacobi equation
or the dispersion relation}.
The wave function $\psi$ has different names
in different fields. For instance,
$\psi$ is a nuclear function in molecular physics,
a longitudinal wave function in waveguides,
an electron function in crystals, etc.
We shall call it a {\it wave function of
adiabatic motion} and we shall call Eq.~\eqref{psieq0}
the effective equation of adiabatic motion.

Representation \eqref{PsiIn2} together with Eq.~\eqref{psieq0}
(a generalization of the Peierls substitution)
is a formalization of the operator separation of variables
in the adiabatic approximation. Of course, the corrections~$L_1$,
$L_2$, $\ldots$ appear in the problems in which the variables
cannot be separated exactly.

The reduced equation \eqref{psieq0} contains less independent
variables and hence should be simpler than the original one.
Thus, we see (and we mentioned this in Introduction)
that solving the original equation can be divided into two
parts:
{\it {\bf 1)} the ``operator {\rm(}adiabatic\/{\rm)}
separation of variables"
based on formula~{\eqref{PsiIn2}},
which reduces the original equation to Eq.~\eqref{psieq0},
and~{\bf 2)} the process of solving this simpler equation}.

The realization of the first step consists in finding the
symbols (functions) $\chi_j$ and $L_j$. We shall state the
general scheme of their construction and discuss different
related questions
(e.g., concerning the reasonable number of terms
in expansions \eqref{expchi}, \eqref{expL})
in the next sections.
Now we discuss a natural generalization of the operators
$\hat \chi$ and $\hat L$.

It is easy to see that, in the case of exact separation of
variables, $\chi=\chi_0$ is an eigenfunction of some additional
spectral problem.
The same fact holds for the functions $\chi_0(x,p,y,t)$;
later we shall numerate them by a multiindex $\nu$.
Thus formula \eqref{PsiIn2} describes only some special
solutions of the original equation corresponding to the term
with the index~$\nu$.
It is possible to construct more general ones summing
solutions  \eqref{PsiIn2} with different indices $\nu$ and
the corresponding $\hat\chi$, $\psi$. Another conclusion is that,
in the case of exact separation of variables,
the spectrum of the above-mentioned additional spectral problem
can be degenerate and
several eigenfunctions can correspond to the same eigenvalue.
Then, instead of the product ${\chi}(y,\mu)\psi(x,t,\mu)$,
one should write the sum
$\sum_{j=1}^k{\chi}_j(y,\mu)\psi_j(x,t,\mu)$,
where $k$ is the multiplicity of the corresponding eigenvalue.
The same generalization should be performed in formula
\eqref{PsiIn2}. Also if the original problem is a vector one
(i.e., if \eqref{InEq} is a system of PDE for $s$ unknown
functions), then $\hat\chi_j$ has $s$ components.
Finally, in formula \eqref{PsiIn2} and in Eq.~\eqref{psieq0}
we mean the following:

1) $\hat \chi$ is a matrix pseudodifferential operator
with $s$ rows and $r$ columns,

2) $\psi$ is an $k$-dimensional vector function
$\psi=(\psi_1,\ldots,\psi_k)^T$,

3) $L$ is an $k\times k$ matrix pseudodifferential operator
with the principal symbol $L_0= H_{\rm{eff}}(p,x) E_k$,
where the number $r$ determines the multiplicity of the
corresponding effective Hamiltonian $H_{\rm{eff}}$ and $E_k$
is the $k\times k$ identity matrix.
The corrections $L_j$ usually are not diagonal,
which means that interaction is present inside
the mode (or the term) determined by this effective Hamiltonian
$H_{\rm{eff}}(p,x)$.

The number of terms in the expansions
of the intertwining operator $\hat\chi$
and the operator~$\hat L$ (with fixed index $\nu$)
can be arbitrarily large. However,
it is, as a rule, a very complicated problem
to calculate the terms of these series explicitly,
even terms with small numbers.
Therefore, it is natural to consider
only the terms required to estimate correctly
the {\it leading} term of the asymptotics
of the wave function or of the energy value.

However, the notion of the ``leading" term of an asymptotics
can be determined not only by the adiabatic parameter~$\mu$,
but also by the other ones, for instance, by the so-called
``semiclassical parameter"~$h$,
which is related to the form of the effective potential
and the solution of the reduced equation~(\ref{psieq0}).
The appearance of this new parameter is very important
for future constructions of the asymptotics.
We shall discuss the corresponding questions in detail
later in \S 5.
Now we only say that, usually, for the construction of the
leading term of asymptotic solution,
it is sufficient to find $L_0$, $L_1$, and $L_2|_{p=0}$.
Another interesting fact is that
the effects of a semiclassical splitting
of the effective Hamiltonian
and a change in the classical characteristics
occur in the degenerate case (see \S 4.6).

\subsection{Scheme of the operator separation of variables}

To simplify the future consideration,
let us assume that, in Eqs.\eqref{InEq} and \eqref{InEqSt},
${\cal H}$ and the operator $\hat{\mathcal H}$
are essentially self-adjoint.
We shall seek the solution of Eq.~(\ref{InEq})
in the following form:
\begin{gather}
\Psi_i(x,y,t,\mu)=
\sum_{j=1}^{k}\chi_{ij}(\x,\p,y,t,\mu)\psi_j(x,t,\mu)= (\hat\chi\psi)_i,
\label{PsiIn}
\end{gather}
where $\psi=(\psi_1,\ldots,\psi_k)^T$ is the wave function
of some chosen {\it term} (or a chosen ``fast" mode)
with the degeneration multiplicity equal to~$k$
and
$\hat\chi$ is an {\it intertwining\/} matrix pseudodifferential
operator:
\begin{gather}
\hat\chi=
\begin{Vmatrix}
\hat\chi_{11} & \ldots & \hat\chi_{1k} \\
\hdotsfor{3} \\
\hat\chi_{s1} & \ldots & \hat\chi_{sk}
\end{Vmatrix}, \qquad
\chi(p,x,y,t,\mu)=\chi_0(p,x,y,t)+\mu\chi_1(p,x,y,t)+\ldots.
\label{operIN}
\end{gather}
We assume that the vector function $\psi$ satisfies
the ``effective equation of adiabatic motion"~\eqref{psieq0}
generated by the matrix operator $\hat L$
\begin{equation*}
\hat L=
\begin{Vmatrix}
\hat L_{11} & \ldots &  \hat L_{1k} \\
\hdotsfor{3} \\
\hat L_{k1} & \ldots &  \hat L_{kk}
\end{Vmatrix},
\qquad L(p,x,t,\mu)=L_0(p,x,t)+\mu L_1(p,x,t)+\ldots,
\end{equation*}
where the matrix $L_0(p,x,t)$ is proportional
to the unitary $k\times k$ matrix $E_k$:
$L_0(p,x,t)=H_{\rm eff}E_k$.
The coefficient of proportionality $H_{\rm eff}$
is an {\it effective Hamiltonian}.
Hence the problem is reduced to finding
the operators $\hat \chi$ and $\hat L$
or their {\it symbols} $\chi$ and~$L$.
After we find them, we can reduce the initial problem
to a more simple (reduced) equation (\ref{psieq0})
for the vector function $\psi$.
The original solution $\Psi$ can be reconstructed
in accordance with (\ref{PsiIn}).

Substituting the function $\Psi$ from (\ref{PsiIn}) into
Eq.~(\ref{InEq}), we obtain:
$$
i\mu\hat\chi\psi_t + i\mu\hat\chi_t\psi=\hat{\cal H}\hat\chi\psi.
$$
Using condition (\ref{psieq0}), rewrite this equation in the
following form:
$(\hat\chi\hat L + i\mu\hat\chi_t-\hat{\cal H}\hat\chi)\psi=0$.
A sufficient condition for the last equality to be valid
is the operator relation
$\hat\chi\hat L + i\mu\hat\chi_t-\hat{\cal H}\hat\chi=0$.
Let us pass from operators to symbols \cite{MaslovAsymptMethods}
in this relation using formula \eqref{smb}.
This leads to the equation
\begin{gather}
\chi(p\p,\x,y,t,\mu)L(x,p,t,\mu)+i\mu\chi_t(p,x,y,t,\mu)-\nonumber\\
-{\cal H}(p\p,\x,\py,y,t,\mu)\chi(p,x,y,t,\mu)=0.
\label{chieq}
\end{gather}

It can be solved using regular perturbation theory, i.e.,
expanding the items into series with respect to $\mu$.
Collecting terms of order $\mu^0=1$, we obtain a family of
spectral problems for the self-adjoint operator
${\cal H}_0(p,x,y,-i\pa/\pa y,t)$ depending on $x,p,t$:
\begin{equation}
\label{term}
{\cal H}_0(p,x,\py,y,t)\chi_0(p,x,y,t)=\chi_0(p,x,y,t)L_0(p,x,t).
\end{equation}
We shall assume that the asymptotics (\ref{PsiIn})
is completely determined by the eigenvalue (term)
$H_{\rm eff}(p,x,t)$
whose multiplicity $k$ does not depend on $p,x,t$.
Moreover, we shall assume that
the value $H_{\rm eff}$ is separated
from the other eigenvalues or a part of the spectrum of ${\cal H}_0$
(if the spectrum contains a continuous component)
uniformly with respect to $(p,x,t)$
in a certain fixed domain $(p,x,t)\in \cal M$.

So
\begin{equation}
\label{L0}
L_0(p,x,t)=H_{\rm eff}(p,x,t)E,
\end{equation}
where $E$ is a unitary $k\times k$ matrix.
The matrix $\chi_0(x,p,y,t)$
consisting of orthonormal vector columns,
i.e., eigenfunctions of the operator ${\cal H}_0$
corresponding to the eigenvalue $H_{\rm eff}(x,p,t)$,
is the {\it intertwining operator} on the proper subspace
induced by this eigenvalue.
It is natural to assume that $\chi_0(x,p,y,t)$
depends smoothly on all its arguments.

Collecting terms of order $\mu$, we obtain inhomogeneous
equations for $\chi_j$ and $L_j$:
\begin{equation}
({\cal H}_0-H_{\rm eff}E)\chi_j=F_j-{\cal H}_j\chi_0+\chi_0 L_j, \qquad
j=1,2,\ldots
\label{chi1}
\end{equation}
where $F_j$ depend on $\chi_{0},\ldots,\chi_{j-1}$
and $ L_0\ldots,L_{j-1}$, in particular,
\begin{gather}
\nonumber F_1=\hat{\cal D}\chi_0,\quad
F_2=\hat {\cal D}\chi^\nu_1-{\cal H}_1\chi^\nu_1+\chi^\nu_1 L_1+
i\sum\limits_j\left[\frac{\pa {\cal H}_1}{\pa p_j}
\frac{\pa \chi^\nu_0}{\pa x^j}-
\frac{\pa\chi_0^\nu}{\pa p_j}\frac{\pa L_1}{\pa x^j}\right]+\\
+\frac{1}{2}\sum\limits_{i,j}\left[\frac{\pa^2 {\cal H}_0}{\pa p_i\pa p_j}
\frac{\pa^2\chi_0^\nu}{\pa x^i\pa x^j}-
\frac{\pa^2 H_{\rm eff}}{\pa x^i\pa x^j}
\frac{\pa^2\chi_0^\nu}{\pa p_i\pa p_j}\right].\label{F2}
\end{gather}
Here
\begin{gather*}
\hat {\cal D}=i\frac{\pa}{\pa t}
+i\sum_j\left[\frac{\pa {\cal H}_0}{\pa p_j}\frac{\pa}{\pa x^j}
-\frac{\pa H_{\rm eff}}{\pa x^j}\frac{\pa}{\pa p_j}\right]=
i\frac{d}{dt}+i\sum_j\left[\frac{\pa {\cal H}_0}{\pa p_j}-
\frac{\pa H_{\rm eff}}{\pa p_j}\right]\frac{\pa}{\pa x^j},
\\
\frac{d}{dt}=\frac{\pa}{\pa t}-
\sum\limits_j\frac{\pa H_{\rm eff}}{\pa x^j}\frac{\pa}{\pa p_j}+
\sum\limits_j\frac{\pa H_{\rm eff}}{\pa p_j}\frac{\pa}{\pa x^j}.
\end{gather*}

Due to the self-adjointness of the operator
$({\cal H}_0-H_{\rm eff}E)$
and the Fredholm alternative,
the solvability condition for this equation
is equivalent to the condition that its right-hand part
is orthogonal to the vector-columns of the matrix $\chi_0$.
It follows that
$L_j=\langle\chi_0^T,{\cal H}_j\chi_0\rangle|_y
-\langle\chi_0^T,F_j\rangle|_y$.
In particular, one can obtain:
\begin{gather}
L_1=\left\langle\chi_0^T,{\cal H}_1\chi_0\right>_y
-i\left<\chi_0^T,\frac{d\chi_{0}}{d t}\right>_y
-i\left\langle\chi_0^T,\sum_{j=1}^n\left[\frac{\pa {\cal H}_0}{\pa p_j}-
\frac{\pa H_{\rm eff}}{\pa p_j}\right]\frac{\pa \chi_0}{\pa x^j}\right>_y.
\label{L1}
\end{gather}
Assuming that $L_1$ has form (\ref{L1}),
one can find the correction, i.e.,
the matrix $\chi_1=({\cal H}_0-H_{\rm eff}E)^{-1}
(F_1-{\cal H}_1\chi_0+\chi_0 L_1)$,
fixing it for determinacy by means of the condition of
orthogonality of vector-columns of the matrices $\chi_0$ and
$\chi_1$.
The reiteration of this procedure leads to calculation of
$L_j,\chi_j$. Formulas (\ref{L0}), (\ref{L1}), etc. give
the coefficients of expansion of the symbol of the reduced
equation (\ref{psieq0}).
Note that the construction of the correction
$L_1$ includes only functions of zero approximation
(like in the standard perturbation theory).
In general, the symbol $L_2$ includes $\chi_1$,
so, to find it, one has to invert
the operator $({\cal H}_0-H_{\rm eff}E)$.

{\bf Remark 1.} The methods from \cite{MaslovOperMethods} allow one
to consider more general situations in which
the quantum Hamiltonian
$\hat{\cal H}(\hat p,\hat x,\hat p_y,\hat y,\mu)$
is a function of vector {\it operators}
($\hat p,\hat x,\hat p_y,\hat y)$
with commutator relations
$[\hat x_j,\hat p_{j}]=i\mu$,  $[\hat y_j,\hat p_{yj}]=i$,
$\mu\ll 1$, or even more complicated ones
(see Example in Section~4.2).
However in this paper we basically consider the situation
in which $\hat x=x, \hat p_x=-i\mu\pa/\pa x$.

{\bf Remark 2.} It is not difficult to modify
the presented formal scheme
for the non-self-adjoint original operator
$\cal H$.
In particular, one has to use the eigenfunctions of the adjoint
operator in the orthogonality conditions. But, of course, it is
necessary to add some additional conditions like the existence
of real-valued effective Hamiltonians, etc.
(see Example in section 4.3).

{\bf Remark 3.\label{Rem3} Operator separation of variables and adiabatic
approximation in classical mechanics.}
There exist a certain classical analog of adiabatic
approach based on the ``operator separation of variables" (\cite{Anosov,ArnKozlNeish,Meunier,Neish-ManyFreq,Neisht}).
The main idea can be illustrated by means of the Hamiltonian
${\cal H}(\mu p_x,x,p_y,y,\mu)$ with a small parameter $\mu$.
If we change the variables $x$, $p_x$ by $\xi=x/\mu$ and
$p_\xi=\mu p_x$,
then we obtain a Hamiltonian of the form
${\cal H}(p,\mu\xi,p_y,y,\mu)$.
It is convenient to write
the Hamiltonian in noncanonical variables $x$ and $p$,
$dp\wedge dx=\mu dp_x\wedge dx=\mu dp\wedge d\xi$:
${\cal H}(p,x,p_y,y,\mu)$.
The Hamiltonian equations for the variables $p,\,x,\,p_y,\,y$
have the form
\begin{gather}
\dot x=\mu \frac{\pa {\cal H}}{\pa p_\xi}\ll 1, \qquad
\dot p=-\mu\frac{\pa {\cal H}}{\pa x}\ll 1,\qquad
\dot y=\frac{\pa {\cal H}}{\pa p_y}, \qquad
\dot p_y=-\frac{\pa {\cal H}}{\pa y}.
\label{xpypy}
\end{gather}

Since we have $\dot p,\,\dot x\sim \mu$
for the derivatives, while $\dot p_y,\,\dot y\sim 1$,
it is natural to say that the variables $p,\,x$ are ``slow variables"
and the variables $p_y,\,y$ are ``fast variables."
Taking into account that there are variables of two types,
it is natural to ``freeze" slow variables and obtain a family of
Hamiltonians with $k$ degrees of freedom
depending on the parameters $(p,\,x)$.
We do not consider the resonanace problems
and restrict our consideration to the case in which
$k=1$ and $(p_y,y)\in\mathbb{R}^2$.
Let us assume that, in some domain $(p,x)\in\Omega$,
the trajectories of ${\cal H}(\{p,x\},p_y,y,0)$ are closed.
The braces $\{\cdot\}$ mean that the included variables are
considered as parameters (i.e., are ``frozen").
Then it is possible to introduce ``action-angle" variables
$(J,\,\varphi)$ corresponding to these closed trajectories.
The passage to these variables is determined
by the change of variables
$y=Y_0(J,\varphi,p,x),\ p_y=P_y^0(J,\varphi,p,x)$.
Unfortunately, this change of variables is not canonical and,
to make it canonical, one has to add corrections and write
$y=Y_0(J,\varphi,P,X)+\mu Y_1(J,\varphi,P,X)+\ldots$,
$p_y=P_y^0(J,\varphi,P,X)+\mu P_y^1(J,\varphi,P,X)+\ldots$,
$p=P+\mu P_1(J,\varphi,P,X)+\ldots$, and
$x=X+\mu X_1(J,\varphi,P,X)+\ldots$.
Then the original Hamiltonian can be written in the form
\begin{gather}
{\cal H}(J,\varphi,P,X)={\cal H}_0(J,P,X)+
\mu\Biggl[
\frac{\pa {\cal H}_0}{\pa y}(P,X,P_y^0,Y_0)Y_1+
\frac{\pa {\cal H}_0}{\pa p_y}(P,X,P_y^0,Y_0)P_y^1+\nonumber\\
+\frac{\pa {\cal H}_0}{\pa x}(P,X,P_y^0,Y_0)X_1+
\frac{\pa {\cal H}_0}{\pa p}(P,X,P_y^0,Y_0)P^1
+{\cal H}_1(P,X,P_y^0,Y_0)\Biggr] + O(\mu^2),\nonumber\\
P_y^1=P_y^1(J,\varphi,P,X), \quad Y_1=Y_1(J,\varphi,P,X),
\quad P^1=P^1(J,\varphi,P,X),\quad X_1=X_1(J,\varphi,P,X)\nonumber\\
P_y^0=P_y^0(J,\varphi,P,X),\qquad Y_0=Y_0(J,\varphi,P,X),
\end{gather}
where ${\cal H}(p,x,p_y,y,\mu)={\cal H}_0(p,x,p_y,y)+\mu
{\cal H}_1(p,x,p_y,y)+O(\mu^2)$.
Now we have a typical problem from averaging theory.
After averaging,
we obtain two terms of the expansion
of the effective Hamiltonian
$L(J,P,X,\mu)=L_0(J,P,X)+\mu L_1(J,P,X)+O(\mu^2)$:
\begin{gather}
L_0(J,P,X)={\cal H}_0(J,P,X),\nonumber\\
L_1(J,P,X)=\int_0^{2\pi} d\varphi\Biggl[
\frac{\pa {\cal H}_0}{\pa y}(P,X,P_y^0,Y_0)Y_1+
\frac{\pa {\cal H}_0}{\pa p_y}(P,X,P_y^0,Y_0)P_y^1+\nonumber\\
+\frac{\pa {\cal H}_0}{\pa x}(P,X,P_y^0,Y_0)X_1+
\frac{\pa {\cal H}_0}{\pa p}(P,X,P_y^0,Y_0)P^1
+{\cal H}_1(P,X,P_y^0,Y_0)\Biggr],
\end{gather}
so $J={\rm const}$ and the integration of the original system
is reduced to solving a system with $n$ degrees of freedom.
The action $J$ corresponds to the ``quantum" number $\nu$
of the term $\chi^\nu$,
the term  $\int_0^{2\pi}d\varphi\,
{\cal H}_1\Bigl(P,X,P_y^0(J,\varphi,P,X),Y_0(J,\varphi,P,X)\Bigr)$
corresponds to $\langle\chi_0,{\cal H}_1\chi_0\rangle_y$, and
the corrections related to the canonical change of variables
in the classical problem correspond to other terms in $L_1$
in quantum problem. Of course, this is simply an analogy
(cf. \cite{Berry1,Taylor,Berry2}, etc.).

{\bf Remark 4.} The classical analogue of the reduction at the first stage is well known \cite{Arnold}: excluding the fast variables, we obtain a system in the zeroth approximation with holonomic constraints; this system is equivalent to the $n$-dimensional Lagrangian system. Thus the adiabatic reduction to Eq.\eqref{psieq0} could be interpreted as the ``excluding of quantum constraints" (see \cite{deWitt,JensenKoppe,daCosta1,daCosta2,EM1,EM2,SchusterJaffe}). But the classical system corresponding to the reduced quantum system generally does not coincide with the result of the classical reduction (in the sense \cite{Arnold}).
Moreover, the classical systems arising in the adiabatic reduction turn out to be different for different relations between $\mu$ and ``semiclassical" parameter $h$, which will be introduced in \S5; using the terminology in \cite{MaslovNonstandard,MaslovCharactsPseododiff}, we can say that they correspond to different nonstandard characteristics of the quantum problem. For example, the classical equation of motion in nanotubes can sometimes include terms arising because spin exists. This is explainable physically because longitudinal motion is already determined by rather small energies that are quite comparable to the spin energy.

{\bf Remark 5.} In some problems one can apply the semiclassical approximation to solve Eq.\eqref{term} (see e.g. \cite{BerlyandDobr,MaslovOpticsMIEM}).

\section{Examples of problems with operator-valued symbols
and parameters.}

Let us illustrate the general scheme with several examples.

\subsection{Equation with rapidly oscillating coefficients
and electron waves in crystals}

One can meet equations with rapidly oscillating coefficients
in many problems of solid physics and continuum mechanics.
For instance, these types of equations describe
propagation of electron waves in crystals, of elastic waves in
composite materials, etc.
For constructing asymptotic solutions to equations
with rapidly oscillating coefficients,
there exist different approachs adapted to their behavior and
their properties.
Among these approaches, one can at least mention
averaging methods, homogenization, and adiabatic
approximation. There exists a very extensive literature
concerning this topic and a review of all these approaches
is very far from our aims. We only want to show that one can
look at the equations with rapidly oscillating coefficients
from the viewpoint of equations with operator-valued symbols
\cite{DobrWaterWaves,DobrOperVal} and,
for the construction of their asymptotics,
use the method described in \S 4.
So here we mention only the general monographs
\cite{Born,BornHuang,BahvPanas,LandLif,Vaks,Tref} and
ideologically close papers \cite{Bus1,Bus2,KlFed}.

As an example, we consider the Schr\"odinger equation
with fast oscillating potential
\begin{gather}\label{period}
i\mu\psi_t=-\frac{\mu^2}{2}\Delta\psi+v(\frac{\Phi(x)}{\mu},x)\psi,
\qquad x\in \mathbb{R}^n, \quad \Phi\in\mathbb{R}^m,
\end{gather}
where $v(y,x)$ is a smooth function $2\pi$-periodic with
respect to each ``fast" variable $y_j$, $j=1,\dots,m$.
The given phases $\Phi_j(x)$ are smooth functions.
Generally speaking, their number $k$ can be arbitrary.
In some problems, the phases are
linear functions $\Phi_j=\langle k_j,x_j\rangle$;
the case of nonlinear phases $\Phi_j$ describes
the case of a nonuniform potential $v$.
The additional dependence of the potential
on the variable $x$ implies its slow deformation.
Eq.~\eqref{period} with such a potential
simulates the propagation of electron waves
in a lattice or, for instance, if $m=1$, in stratified media.
The simplest example of a rapidly oscillating potential is
given by the formula $(n=1, m=1)$:
$v=v_0(x)+a(x)\cos\frac{\Phi(x)}{\mu}$,
where $v_0(x),a(x)$ are smooth functions.

Let us find the unknown function $\psi(x,t,\mu)$
in the form
\begin{equation}\label{period1}
\psi(x,t,\mu)=\Psi
\left(\frac{\Phi(x)}{\mu},x,t,\mu\right),
\end{equation}
where the new unknown function $\Psi(y,x,t,\mu)$
is $2\pi$-periodic with respect to each variable $y_j$.
Substituting \eqref{period1} into Eq.~\eqref{period},
we see that the function $\psi(x,t,\mu)$ \eqref{period1}
satisfies Eq.~\eqref{period}
if the function $\Psi (y,x,t,\mu)$ is a solution
of Eq.~\eqref{InEq} with
\begin{equation}\label{period3}
\hat{\cal H}=\left(-i\mu\frac{\pa}{\pa x}-i\frac{\pa \Phi}{\pa
x}\frac{\pa}{\pa y}\right)^2+v(y,x).
\end{equation}
We again see that the small parameter $\mu$ is in front
of the derivative $\pa/\pa x$, but there is no small parameter
in front of the derivative $\pa/\pa y$.
Thus the equation with Hamiltonian \eqref{period3}
is an equation with operator-valued symbol, namely,
\begin{gather}
{\cal H}={\cal H}_0-\mu \sum_{j=1}^m\Delta\Phi_j\frac{\pa}{\pa y_j},\quad
{\cal H}_0=\left(p-i\sum_{j=1}^m\frac{\pa\Phi_j}{\pa x}
\frac{\pa}{\pa y_j}\right)^2 +v(y,x).
\end{gather}
For each fixed $(p,x)$,
the operator ${\cal H}$ acts in the $L_2$-space
on a $k$-dimensional torus.

If the vectors $\frac{\pa \Phi_j}{\pa x}$ are linearly
independent for each $x$, then the spectrum of the operator
$\cal H$ is discrete, but, to obtain the reduced equation,
one must also use the assumption that the multiplicity of
eigenvalues is independent of $(p,x)$.
In particular,
if $m=1$ and $\pa \Phi/\pa x\neq 0$,
then the spectral problem \eqref{term} for determining
the effective Hamiltonians is
a periodic problem and can be reduced to
the 1-D Schr\"odinger equation on a circle
for Bloch solutions (see \cite{DobrDoktor,dobrRapOsc}).
To realize this reduction,
we change the variables in the equation
${\cal H}_0 \chi_0=H_{\rm{eff}}\chi_0$
for $\chi_0$ and $H_{\rm{eff}}$ as follows:
$$
y=U\xi , \quad \chi_0=u e^{-i P\xi},\qquad
U=\left|\frac{\pa \Phi}{\pa x}\right|,
$$
and put
\begin{equation}\label{P}
P=P(p,x)\equiv\left.\left<\frac{\pa \Phi_j}{\pa x},p\right>\right/
\left|\frac{\pa\Phi}{\pa x}\right|^2.
\end{equation}
Then this equation takes the form
\begin{equation*}
-u_{\xi \xi}+v(U\xi,x)u={\cal E}u,\quad {\cal E}=H_{\rm{eff}}-p^2+P^2,
\end{equation*}
and the periodicity condition becomes the Bloch condition:
$$
u( \xi+2 \pi/U,x)= e^{2\pi i P}u( \xi,x).
$$
The variable $x$ is contained in the reduced problem as a
parameter.
The variable (number) $P$ is called the ${\it quasimomentum}$
of the corresponding Bloch solution.
It is a well-known fact that the spectrum of the operator
$-\pa^2/\pa \xi^2 +v(U\xi,x)$ on a circle consists
of bands and gaps.
Let us enumerate the bands by the number $\nu$
and denote the ends of the $\nu$th band by $E_-^\nu$ and
$E_+^\nu$.
The spectral parameter ${\cal E}$ and the quasimomentum $P$
in each $\nu$th band are connected by the {\it dispersion relation}
$$
{\cal E}={\cal E}^\nu(P,x).
$$
The assumption on the potential $v(y,x)$
that the $\nu$th effective Hamiltonian (eigenvalue)
of the operator $\cal {H}^\nu$ is simple
(or does not intersect with other effective Hamiltonians
at some points $(p,x)$)
is equivalent to the assumption that, for each $x$,
the {\it $\nu$th band does not stick together with
the $\nu-1$st and $\nu+1$st bands}.
Under this assumption,
one can find the reduced equation describing solutions
corresponding to the $\nu$th term (effective Hamiltonian)
$$
H_{\rm{eff}}^\nu={\cal E}^\nu(P(p,x),x)+p^2-(P(p,x))^{2}.
$$
The corresponding function $\chi_0^\nu(y,p,x)$ is expressed via
the Bloch function $u^\nu$, $u^\nu(\xi,P,x)$, by the formula
$$
\chi_0= u^\nu(\frac{y}{U},P(p,x),x)\exp(-i\frac{P(p,x) y}{U}).
$$
If $n=1$ and $\Phi=x$, then $P=p$,
which leads to the well-known fact in solid state physics:
the quasimomentum becomes the momentum for the equation for
electron waves in crystals.
In contrast to Examples~1, 2, and~3,
the effective Hamiltonian here is not a polynomial in $p$ and
the function $\chi_0$ depends on the momentum~$p$.

Let us write the first correction $L_1$ to the effective
Hamiltonian. Using formula \eqref{L1} we obtain
$$
L_1=\langle\chi_0,\frac{d\chi_0}{dt}\rangle+
\langle\chi_0,(2p-\frac{\pa H}{\pa p})\nabla \chi_0\rangle+
2i{\rm Re}\langle\nabla\chi_0\frac{\pa\chi_0}{\pa y},\nabla \chi_0\rangle.
$$

Note that it is possible to meet a situation in which
the number of phases in the potential is greater than
the dimension of the configuration space~$n$.
For instance,  consider the case $m=2$, $n=1$.
In this situation, the operator $\cal H$ is degenerate
and its spectrum has a rather complicated structure.
In particular, the spectrum can be an everywhere dense set on
the spectral axis (the so-called devil's stair)
and the multiplicity of its eigenvalues can depend on~$x$.
These problems are related to different types
of complicated resonances and to problems with intersecting
characteristics.
Some results for the solutions of the Cauchy problem
in this situation are obtained in \cite{DobrOperVal,DobrDoktor,dobrRapOsc}.

{\bf Remark. Bloch electrons in a weak magnetic field and the
Peierls substitution.}
The original Peierls substitution was first proposed for the
problem with Bloch electrons in a weak magnetic field (see,
e.g.,~\cite{LifPit}). Using the terminology of this paper, this
problem can be stated as a spectral problem for the magnetic
Schr\"odinger operator with the periodic electric potential:
$$
\hat{\cal H}=\frac{1}{2}\left(-i\mu\frac{\pa}{\pa x}-A(x)\right)^2+v(\frac{x}{\mu}), \qquad
\langle \nabla,A\rangle=0.
$$
For simplicity, we restrict ourselves by simple cubic lattice,
i.e., assume that $v=v(y_1,y_2,y_3)$ is $2\pi$-periodic with
respect to each variable $y_j=x_j/\mu,\ j=1,2,3$.
The fact that the magnetic field is weak means that $A(x)$ does
not contain any irregular dependence on the parameter $\mu$ in
contrast to the crystalline potential $v(\frac{x}{\mu})$. After
the space regularization \eqref{period1} with phase vector
$\Phi(x)\equiv x$ similarly to \eqref{period3}, we obtain the
problem with operator-valued symbol
$
{\cal H}=(1/2)(p-i{\pa}/{\pa y}-A)^2+v(y).
$
The procedure based on the formulas \eqref{P} leads to the
problem for Bloch solutions for the operator
$
{\cal H}=-(1/2)\Delta_y+v(y).
$
Let ${\cal E}={\cal E}^\nu(P)$, $P=(P_1,P_2,P_3)$ determine
the dispersion relation for the Bloch solutions with
quasimomentum $P$. Then, according to this subsection,
$L_0^\nu\equiv H_{\rm{eff}}^\nu={\cal E}^\nu(p-A),\
L_1^\nu=[{\rm tr}(\frac{\pa^2{\cal E}^\nu}{\pa p^2}\frac{\pa
A}{\pa x})](p-A)$, and in \eqref{psieq0} $\hat L=\hat
L_0^\nu+\mu \hat L_1^\nu+O(\mu^2)={\cal
E}^\nu(-i\mu\frac{\pa}{\pa x}-A)+O(\mu^2)$ which
is exactly the {\it Peierls substitution}.
The higher order corrections look much more complicated than the
leading term \cite{Blount}. It seems to us that our approach
allows one to calculate these corrections easier
than in \cite{Blount}. Semiclassical analysis of the  reduced equation
\eqref{psieq0} reveals a very complicated and striking topology of
surfaces invariant to the corresponding phase flow. The recent
results and bibliography can be found in \cite{Novikov}.

\subsection{Electron-phonon interaction}
As was noted in Remark 1 in Section~3.3, \S 3,
one can consider adiabatic problems as problems
containing ``slightly noncommuting" operators.
In the zeroth-order approximation,
these operators can be substituted by ``$c$-numbers," which
allows one to determine a term.
For ``slightly noncommuting" operators,
there are physical quantities slowly varying in time.
The slightly commuting operators can generate
a certain Lie algebra. In the simplest case,
this Lie algebra is the Heisenberg algebra,
and we can directly use the scheme and formulas proposed in~\S 3.
Sometimes, it is possible to consider the same problem
from a different viewpoint, which depends on the choice of
the operators. For instance, to simplify the form of the
original Hamiltonian, one can use a (noncanonical) change of
variables, which, in turn, leads to the replacement
of the Heisenberg commutation relations
by different ones.
We consider such an example which can be analyzed
from these different viewpoints,
but since the study of problems based on non-Heisenberg
commutation relations requires
new nontrivial algebraic and geometric constructions,
we restrict ourselves to the approach described in~\S 3,
although, from some viewpoint, the approach based on
non-Heisenberg commutation relations can be more readily
realized in some concrete problems.

The electron-phonon interaction is the interaction between
light fermions (electrons) and heavy bosons (phonons).
Here the lattice modes (bosons) are slow and the electrons
(fermions) are fast. The Hamiltonian of electron-phonon
interaction is (see, e.g., \cite{Schrieffer})
\begin{gather}
\hat{\cal H}=\sum_n \hat c_n(\hat\psi_{n+1}^{+}\hat\psi_n +
\hat\psi_n^{+}\hat\psi_{n+1})
+\sum_n K(\hat x_{n+1}-\hat x_n)^2 + \hat p_n^2/2,\label {Ham-e-ph}\\
\label{psicommut}
[\hat x_n,\hat p_{n'}]=i\mu\delta_{nn'},
\quad \mu=\hbar/\alpha\sqrt{m_0 c_0}\\
\nonumber
[\hat\psi_j^+,\hat\psi_k]_+\equiv\hat\psi_j^+\hat\psi_k+
\hat\psi_k^+\hat\psi_j=i\delta_{jk}.
\end{gather}
Here $m_0,c_0,K,\alpha$ are physical constants;
moreover, $\mu\ll 1$;
the linear operators $\hat \Psi^+_1,\ldots,\hat\Psi^+_M$ and
$\hat\Psi_1,\ldots,\hat\Psi_M$ act on the Hilbert space
$\mathbb{H}_1$,
the linear operators $\hat p_1,\ldots,\hat p_M$ and
$\hat x_1,\ldots,\hat x_M$ act on the Hilbert space
$\mathbb{H}_2$.
The full quantum Hamiltonian $\hat{\cal H}$ acts on the Hilbert
space $\mathbb{H}=\mathbb{H}_1\otimes\mathbb{H}_2$.
The typical situation is given by the operators
\begin{equation}\label{c}
\hat c_n=f(\hat x_{n+1}-\hat x_n),
\end{equation}
where $f(z)$ is a smooth function; in particular,
$\hat c_n=1-\alpha(\hat x_{n+1}-\hat x_n)$.
Let us set $\hat X=(\hat x_1,\ldots,\hat x_N)$,
$\hat P=(\hat p_1,\ldots,\hat p_N)$,
$\hat\Psi=(\hat \psi_1,\ldots,\hat \psi_M)$,
$\hat\Psi^+=(\hat \psi_1^+,\ldots,\hat \psi_M^+)$.
In a more general case, the Hamiltonian  $\hat{\cal H}$ of
electron-phonon interaction can be written as
\begin{gather}
\hat{\cal H}=\langle \hat \Psi^+,{\cal L}(\hat X,\hat P)\hat
\Psi\rangle+\Phi(\hat X)+\hat P^2/2,
\end{gather}
where  ${\cal L}(P,X)$ is a Hermitian $M\times M$ matrix with
coefficients depending on $X$ and $P$,
and the operator ${\cal L}(\hat X,\hat P)$
is understood in the sense of Weyl calculus
(see \cite{MaslovOperMethods,KarasevMaslov1,KarasevMaslov2}).
To simplify the consideration,
we restrict ourselves to the case in which
${\cal L}={\cal L}_1(X)+{\cal L}_2(P)$;
then the question about the ordering of the operators
$\hat X$ and $\hat P$ does not appear.
As we have just said, it is possible to develop
the ``operator separation of variables" based on the algebra
of operators $\hat c_j,\hat p_j$ with the commutation relations
$$
[\hat p_j,\hat c_k]=i\mu(\delta_{jk}-\delta_{jk+1})\hat c_k,
$$
but here  we use the standard representation  $\hat x_j=x_j$ and
$\hat p_j=-i \mu \pa/\pa x_j$ and close the lattice by the
Born--Karman periodicity condition identifying the operators
with the numbers $j$ and $j+M$.

Denote the electron-phonon wave function by ${\Upsilon}$
and consider the stationary problem
\begin{gather}\label{Sp}
\hat {\cal H}{\Upsilon}={\cal E} {\Upsilon},\quad {\Upsilon}\in
\mathbb{H}.
\end{gather}
We have an equation with operator-valued symbol
which, obviously, is the operator
\begin{gather}\label{el-ph}
{\cal H}(X,P)=\langle\Psi^+,{\cal L}(X,P)\Psi\rangle+(\Phi(X)+P^2/2)\hat I_1
\end{gather}
acing on the Hilbert space $\mathbb{H}_1$.
We also denote the identity operator acting on~$\mathbb{H}_j$
by $\hat I_j$.

To realize the scheme of the operator separation of variables,
it is necessary to find the spectrum of the operator-valued symbol
${\cal H}(X,P)$. A nice fact is that this spectrum can be
expressed via the eigenvalues of the matrix ${\cal L}$.
Namely, suppose that  $\psi(E_j)=(\psi_1(E_j),\ldots,\psi_M(E_j))$
are the eigenvectors of the matrix
${\cal L}(X,P)$ corresponding to its eigenvalues
$E_j(X,P)$ $E_1\leq \ldots \leq E_M$ and satisfying the
normalization conditions
$\langle\psi(E_j),\psi(E_k) \rangle=\delta_{jk}$.
Using the basis $\{\psi(E_j)\}$, one can expand the operators
$\hat\Psi^+$ and $\hat\Psi$
\begin{gather}
\hat\psi^+=\sum_{j=1}^M \psi^*(E_j)\hat a_j^+, \qquad
\hat\psi=\sum_{j=1}^M \psi(E_j)\hat a_j.
\end{gather}
The coefficients (operators) of this expansion,
$\hat a_j^+$ and $\hat a_j$, are called the creation and
annihilation operators \cite{Ziman}.
They determine the operators of the number of particles
$\hat{\cal N}_j=\hat a^+_j \hat a_j$.
Finally,
the operator-valued symbol can be written as
\begin{gather*}
{\cal H}(X,P)=\sum_{j=1}^M E_j(X,P)\hat {\cal N}_j+\Phi(X)+P^2/2,
\qquad
\hat {\cal N}_j=\hat a^+_j \hat a_j.
\end{gather*}
From this, we find the $\nu$th effective Hamiltonian
\begin{gather}\label{ElPhH}
H^\nu_{\rm eff}(X,P)=\sum_{j=1}^\nu E_j(X,P)+\Phi(X)+P^2/2,
\end{gather}
and the symbols of the corresponding intertwining operators
$\chi_0^\nu$:
\begin{gather}
\chi_0^\nu=\hat{\cal N}_1\ldots \hat{\cal N}_\nu
(I-\hat{\cal N}_{\nu+1})\ldots(I-\hat{\cal N}_M)\cdot 1.
\end{gather}
Let us analyze the formula for $L_1$.
If the effective Hamiltonian $H^\nu_{\rm eff}$ is degenerate
or ${\cal L}_1(X), {\cal L}_2(P)$ are complex matrices,
then one must use the general formula \eqref{L1}.
For the case in which ${\cal L}_2(P)=0$ and
${\cal L}_1(X)$ is a real-valued matrix and its spectrum
$(E_1(x),\ldots,E_M(x))$ is nondegenerate,
it follows from formulas \eqref{ElPhH}
that the effective Hamiltonians are nondegenerate
and $H^\nu_{\rm eff}$ and $\chi_0^\nu$ are real.
Thus, taking into account the relations
${\cal H}_1=0$,
$\left<\chi_0^T,\frac{d\chi_{0}}{dt}\right>_y
=\frac{1}{2}\frac{d}{dt}\left<\chi_0^T,\chi_{0}\right>_y=0$,
$\frac{\pa {\cal H}_0}{\pa p_l}-
\frac{\pa H_{\rm eff}}{\pa p_l}I=0$,
we obtain
$$
L_1=0.
$$
This equality holds for the operators $\hat c_n$ given by formula
\eqref{c}, in particular, if $c_n= e^{x_n-x_{n-1}}$.

It was pointed out that in this case it is possible to relate
problem \eqref{Sp} to the integrable Toda lattice model
\cite{Toda1,Krichever}. Then this observation was used to construct
its semiclassical asymptotics \cite{Toda2},
and at the same time,
commutation relations of different type were chosen.

\subsection{Internal waves in ocean in a pycnocline}

The next example concerns a situation in which the operator
$\cal{H}$ is not self-adjoint, moreover,
the system of equations under study differs a little from
\eqref{InEq}. Thus the scheme of \S 3 requires a slight
modification.

We consider a system of hydrodynamic equations
for an ideal incompressible liquid
linearized on the flow with velocity $U$ and density~$\rho_0$.
Let $x=(x_1,x_2)$ be the horizontal coordinates,
$z$ be the vertical coordinate,
$u=(u_1,u_2,u_3)$, and $\rho$ be perturbations
of the velocity and density,
$\Pi$ be the pressure,
$g=(0,0,|g|)$ be the gravity acceleration.
We introduce dimensionless variables and parameters
by the formulas: $U=U'\omega_1\lambda$,
$u=u'\omega_1\lambda$, $t=t'\omega_2$, $x=x'L$, $z=z'L$,
$g=g'\omega_1^2\lambda$, $\Pi=\Pi'\omega_1^2\lambda^2$,
and
$\rho_0=\rho_0'\bar\rho$, $\rho=\rho'\bar\rho$,
where
$\omega_1$ is the characteristic frequency of the internal wave,
$\lambda$ is the characteristic wavelength,
$L$ is the characteristic distance in the horizontal direction
within which the characteristics of the liquid vary,
$\omega_2$ is the average value of the
V\"ais\"al\"a--Brunt frequency,
$\bar{\rho}$ and $\omega_1^2\lambda^2$ are the
characteristic values of density and pressure,
and
$\mu=\lambda/L=\omega_1/\omega_2\ll 1$ is a small parameter.

In the dimensionless variables,
the linearized system for waves in liquids
has the form \cite{Mir,Badulin}:
\begin{gather}\label{HD}
\left\{
\begin{array}{l}
\mu\rho_0\frac{\pa u}{\pa t}+\mu\rho_0
\langle U,\nabla\rangle u+\mu\rho_0\langle u,\nabla\rangle U+
\mu\nabla\Pi+\rho g=0,\\
\mu\frac{\pa\rho}{\pa t}+\mu\langle U,\nabla\rangle\rho+
\mu\langle u,\nabla\rangle\rho_0=0,\\
\mu\langle\nabla,u\rangle=0,
\end{array}\right.
\end{gather}
where $\langle\cdot,\cdot\rangle$ is the inner product
in $\mathbb{R}^3$.
The difference between Eqs.~\eqref{HD} and Eq.~\eqref{InEq}
is that the last equation of this system
contains the time-derivative.

We shall consider that the width of the picnocline $\Delta z\sim \mu$,
and that the depth of its location varies at distances $\sim 1$,
i.e.,
$\rho_0=\rho_0(z/\mu-f(x),x)$
(the equation for the surface of a picnocline is $z=\mu f$).
The functions $U(x)=(U_1,U_2,0)$,
$\rho_0(y,x)$, $f(x)$ are assumed to be smooth,
$0<\delta_0<\rho_0<c_0$ ($\delta_0,\ c_0$ are constants),
the square of the V\"ais\"al\"a--Brunt frequency
$\omega_0^2=-|g|\frac{\pa\rho_0}{\pa y}/\rho_0$ is positive
and vanishes rather fast as $|y|\to\infty$.
We assume that the functions $u_j$ and $\rho $ decay quite fast
as $|z|\to \infty$.
The other boundary and initial conditions for Eqs.\eqref{HD}
are chosen in a special way and should be formulated for the
corresponding reduced effective equation of adiabatic motion
\eqref{PsiIn0}.

Let us introduce a new independent variable $y=z/\mu-f(x)$
and a vector with five components
$\Psi(x,y,t,\mu)=(u,\rho,\Pi)$.
Into the equations for the vector $\Psi$,
we must substitute the differential operators
as $\frac{\pa}{\pa x_i}\to\frac{\pa}{\pa x_i}-\frac{\pa f}{\pa x_i}\frac{\pa}{\pa y}$ and
$\mu\frac{\pa}{\pa z}\to\frac{\pa}{\pa y}$.
Then, for $\Psi$, we obtain a system of equations
containing ``fast" variable $y$ and slow variables $x,\ t$.
For convenience,
we multiply this system by $i=\sqrt{-1}$:
\begin{gather}\label{HD1}
i\mu B\Psi_t={\cal H}(\x,\p,y,-i\frac{\pa}{\pa y},\mu)\Psi(x,y,t,\mu),
\qquad
B=
\begin{pmatrix}
\rho_0 & 0 & 0 & 0 & 0 \\
0 & \rho_0 & 0 & 0 & 0 \\
0 & 0 & \rho_0 & 0 & 0 \\
0 & 0 & 0 & 1 & 0 \\
0 & 0 & 0 & 0 & 0
\end{pmatrix}.
\end{gather}

The matrix symbols ${\cal H}_0$ and ${\cal H}_1$
have the form
\begin{gather}\nonumber
{\cal H}_0(x,p,y,-i\frac{\pa}{\pa y},t)=
\begin{pmatrix}
\rho_0\langle U,p\rangle & 0 & 0 & 0 & p_1 \\
0 & \rho_0\langle U,p\rangle & 0 & 0 & p_2 \\
0 & 0 & \rho_0\langle U,p\rangle & -i|g| & -i\frac{\pa}{\pa y} \\
0 & 0 & -i\frac{\pa\rho_0}{\pa y} & \langle U,p\rangle & 0 \\
p_1 & p_2 & -i\frac{\pa}{\pa y} & 0 & 0
\end{pmatrix},\\
{\cal H}_1=-i\rho_0
\begin{pmatrix}
\frac{\pa U_1}{\pa x_1} & \frac{\pa U_1}{\pa x_2} &
0 & 0 & 0 \\
\frac{\pa U_2}{\pa x_1} &
\frac{\pa U_2}{\pa x_2} & 0 & 0 & 0\\
0 & 0 & 0 & 0 & 0 \\
0 & 0 & 0 & 0 & 0 \\
0 & 0 & 0 & 0 & 0
\end{pmatrix}+
\begin{pmatrix}
0 & 0 & 0 & 0 & \frac{\pa f}{\pa x_1}(i\frac{\pa}{\pa y}) \\
0 & 0 & 0 & 0 & \frac{\pa f}{\pa x_2}(i\frac{\pa}{\pa y}) \\
0 & 0 & 0 & 0 & 0 \\
-i\frac{\pa\rho_0}{\pa x_1}+i\frac{\pa f}{\pa x_1}\frac{\pa\rho_0}{\pa y} &
-i\frac{\pa\rho_0}{\pa x_2}+i\frac{\pa f}{\pa x_2}\frac{\pa\rho_0}{\pa y} &
0 & 0 & 0 \\
\frac{\pa f}{\pa x_1}(i\frac{\pa}{\pa y}) &
\frac{\pa f}{\pa x_2}(i\frac{\pa}{\pa y}) & 0 & 0 & 0
\end{pmatrix}+\nonumber\\
+\langle U,\nabla f\rangle B \left(i\frac{\pa}{\pa y}\right),\nonumber
\end{gather}
where $p=(p_1,p_2,0)$.
Let us show  that the general scheme of the operator separation can
be easily modified for this situation,
although one of the equations in the system
does not contain the time-derivative.

We seek the solution in the form \eqref{PsiIn} and assume
that $\psi(x,t,\mu)$ satisfies an effective equation of the form
\eqref{PsiIn0}. Then, instead of \eqref{chieq},
we obtain the following relation for symbols:
\begin{gather*}
B\chi(\x,p\p,y,t,\mu)L(x,p,t,\mu)+i\mu B\chi_t(x,p,t,y,\mu)
-\\{\cal H}(\x,p\p,y,\py,t,\mu)\chi(x,p,y,t,\mu)=0.
\end{gather*}
Expanding this relation into a series with respect to a small
parameter $\mu$,
instead of the eigenvalue problem,
we obtain the problem
for the spectral parameter $H_{\rm eff}$
of the operator sheaf
$({\cal H}_0-BH_{\rm eff})$,
\begin{gather*}
\left({\cal H}_0(x,p,y,-i\frac{\pa}{\pa y},t)-
B H_{\rm eff}(x,p,t)\right)\chi_0(x,p,y,t)=0,
\end{gather*}
and the following equation for the corrections $\chi_1$ and $L_1$:
\begin{gather}
\left({\cal H}_0(x,p,y,-i\frac{\pa}{\pa y},t)-
B H_{\rm eff}(x,p,t)\right)\chi_1=F_1-{\cal H}_1\chi_0+B\chi_0 L_1,
\qquad F_1=\hat {\cal D}\chi_0\label{HDL1}\\
\hat {\cal D}=iB\frac{\pa}{\pa t}
+i\sum_j\left[\frac{\pa {\cal H}_0}{\pa p_j}\frac{\pa}{\pa x^j}
-B\frac{\pa H_{\rm eff}}{\pa x^j}\frac{\pa}{\pa p_j}\right]=
iB\frac{d}{dt}+i\sum_j
\frac{\pa ({\cal H}_0-BH_{\rm eff})}{\pa p_j}\frac{\pa}{\pa x^j}.
\nonumber
\end{gather}

We assume that the chosen spectral parameter
$H_{\rm eff}(x,p,t)$ is nondegenerate,
so $\chi_0$ is a vector with five components
$\{\chi_{0j}\}$, $j=1,\ldots,5$. Then the solution
of the problem for the operator sheaf has the form
\begin{gather}
\left\{\begin{array}{l}
\chi_{01}=i\frac{p_1}{p^2}\frac{\pa w\left(y,x,\alpha\right)}{\pa y}, \\
\chi_{02}=i\frac{p_2}{p^2}\frac{\pa w\left(y,x,\alpha\right)}{\pa y}, \\
\chi_{03}=w\left(y,x,\alpha\right),\\
\chi_{04}=i\frac{\pa\rho_0}{\pa y}
\frac{w\left(y,x,\alpha\right)}{\Lambda}, \\
\chi_{05}=-i\rho_0\frac{\Lambda}{p^2}
\frac{\pa w\left(y,x,\alpha\right)}{\pa y}.
\end{array}\right.
\end{gather}
Here $\Lambda=\langle U,p\rangle-H_{\rm eff}(x,p)$,
$\alpha=\Lambda^2/p^2$,  $p^2=p_1^2+p_2^2$,
$w(y,x,\alpha)$ is an eigenfunction of the problem
\begin{gather*}
\frac{1}{\rho_0(y,x)}\frac{\pa}{\pa y}\rho_0(y,x)
\frac{\pa}{\pa y}w(y,x,\alpha)+\frac{\omega_0^2(y,x)}{\alpha}
w(y,x,\alpha)=\varkappa(x,\alpha) w(y,x,\alpha),
\end{gather*}
and $H_{\rm eff}(x,p,t)$ is a solution of the algebraic equation
\begin{gather*}
\varkappa\left(x,\frac{(\langle U,p\rangle
-H_{\rm eff}(x,p))^2}{p^2}\right)=p^2.
\end{gather*}
We choose an eigenvalue $\varkappa$ and consider the
corresponding function $H_{\rm eff}(x,p)$.
In general, this function is multi-valued;
we fix one of its branches and assume that
this branch is a smooth function of $x,\, p$.

The first correction $L_1$ is found from the
solvability condition for Eq.~\eqref{HDL1}.
Its right-hand part must be orthogonal
to the kernel of the adjoint operator
$({\cal H}-BH_{\rm eff})^*$.
We denote a function from its kernel by $\chi_0^*$.
Then the first correction to the effective Hamiltonian is
\begin{gather*}
L_1(x,p,t)=
\frac{1}{\langle\chi_0^*,B\chi_0\rangle}
\left\langle\chi_0^*,\left[{\cal H}_1-iB\frac{d}{dt}
-i\sum_j\frac{\pa ({\cal H}_0-BH_{\rm eff})}{\pa p_j}
\frac{\pa}{\pa x^j}\right]\chi_0
\right\rangle_y.
\end{gather*}

To determine $\chi^*_0$, we note that the construction
of the operator adjoint to $({\cal H}_0-BH_{\rm eff})$
is equivalent to the replacement
$|g|\leftrightarrow-{\pa\rho_0}/{\pa y}$.
This gives
\begin{gather*}
\chi_{0k}^*=\chi_{0k}, \quad k=1,2,3,5, \quad \chi_{04}^*=
-i|g|\frac{w(y,x,\alpha)}{\Lambda}.
\end{gather*}
Using these relations, we obtain
\begin{gather}
\langle\chi_0^*,B\chi_0\rangle_y=
\frac{1}{p^2}\int\rho_0\left|\frac{\pa w}{\pa y}\right|^2 dy+
\int \rho_0\left|w\right|^2 dy +
\int\frac{\rho_0\omega_0^2}{\Lambda^2}|w|^2dy, \label{picnoB}\\
\langle\chi_0^*,{\cal H}_1\chi_0\rangle_y=
-\frac{i}{p^4}\sum_{j,k=1}^2 \frac{\pa U_j}{\pa x_k}p_j p_k
\int \rho_0\left|\frac{\pa w}{\pa y}\right|^2 dy - \label{picnoH1}\\
- i\frac{\Lambda}{p^4}\langle p,\nabla f\rangle\left(
\int\frac{\pa \bar w}{\pa y}\frac{\pa}{\pa y}
\left(\rho_0\frac{\pa w}{\pa y}\right)dy+c.c.\right)
-i\frac{1}{p^2\Lambda}\int\left< p,
\nabla\rho_0-\nabla f\frac{\pa\rho_0}{\pa y}\right>\bar w\frac{\pa\rho_0}{\pa y}
\frac{\pa w}{\pa y}dy+ \nonumber\\
+ \frac{i}{p^2}\langle U,\nabla f\rangle
\int\rho_0\frac{\pa\bar w}{\pa y}\frac{\pa^2 w}{\pa y^2} dy
+ i\langle U,\nabla f\rangle
\int\rho_0\bar w\frac{\pa w}{\pa y} dy
- i\langle U,\nabla f\rangle\frac{|g|}{\Lambda^2}\int\bar w
\frac{\pa}{\pa y}\left(\frac{\pa \rho_0}{\pa y} w\right)dy. \nonumber
\end{gather}
Using formulas \eqref{picnoB}-\eqref{picnoH1} we can calculate the first correction $L_1$. We don't give here the explicit formula for $L_1$ in the general case because of its bulk.

\subsection{Electromagnetic waveguides, integral optics,
surface gravity water waves and shells.}

The electromagnetic wave propagation in waveguides
is described by wave equation containing the second
time-derivative. In the two-dimensional case,
we have a situation similar to that considered
in Example~2, \S 2.
One can easily generalize the scheme of \S 3
to this situation. The change consists in the following:
instead of $i\mu\psi_{t}$, one must write the second
time-derivative $\mu^2\psi_{tt}$ in the left-hand side
of \eqref{psieq0}. The same change allows one to consider
the three-dimensional waveguide problems. But now it is
possible to consider waves in thin films
(integral optics) or in thin tubes.
The stationary variant of such equations is the Helmholtz equation
\begin{gather}
(\Delta+k^2{n}(x))\Psi=0
\end{gather}
with the refractive index ${n}(x)$ and, e.g.,
the Dirichlet conditions $\Psi=0$ on the boundary of the film or
the tube.
The parameter $\mu$ characterizes the ratio between the
transverse and longitudinal dimensions of the waveguide;
one can apply the adiabatic approximation if the boundary of the
waveguide changes slowly.
V.~P.~Maslov considered problems of this type in 1958
in \cite{MaslovWaveGuide}, where he constructed asymptotic
solutions and predicted the possibility of the construction of
one-mode resonators by means of the waveguide geometry.
Later on, problems of this type were considered
in more general situations in optics and quantum mechanics
(see, e.g., \cite{JensenKoppe,daCosta1,daCosta2,EM1,EM2,SchusterJaffe,Maslov-Vorob,Exner,LinJaffe,DellAnt}).

The consideration of the Helmholtz equation is very similar
to the consideration of the stationary Schr\"odinger equation.
In what follows, we consider the Schr\"odinger equation
in a quantum waveguide in a more complicated situation.

More complicated examples similar to planar waveguides
(thin films) lead
to problems about the wave propagations in shells.
Instead of the wave equations,
one must consider the Lam\'e equations in elastic theory.
The operator separation of variables can also be used
in such problems, but the study of these problems is far beyond
the aim of this paper.

A more exotic example of the operator separation of variables
is given by the theory of surface gravity water waves
over an uneven bottom (see, e.g., \cite{Voronovich}). Actually, this problem is
the linearization of the problem with free boundary
and the anzatz \eqref{PsiIn2} was first used
in this situation \cite{DobrWaterWaves}.
The operator approach is discussed in detail
in \cite{DobrWaterWaves,KarasevMaslov1,KarasevMaslov2,DobrKor},
so we do not consider this problem here.

\subsection{Nanophysics: wave propagation in nanofilms.}

The two subsequent examples
(quantum waves in nanofilms and nanotubes)
are probably the simplest ones in adiabatic problems.
An interest in these problems appeared recently
because of great progress in nanotechnologies.
It seems that most of the results described below,
as well as many recent mathematical results
(e.g., \cite{JensenKoppe,daCosta1,daCosta2,SchusterJaffe,Exner,LinJaffe,DellAnt}),
did not appear many years ago,
because there was no deep physical interest
in the corresponding problems.
Now the question is to study concrete applied problems.
Needless to say that the representation
of the solution in the form
appropriate for practical analysis
is an additional and sometimes nontrivial problem
(even in the case of nanofilms and nanotubes with simple
structure, e.g., without branching).
So below we discuss some specific properties concerning
quantum waveguides, present the effective equation of adiabatic
(longitudinal) motion in thin films and tubes,
and briefly touch upon only a few possible applications
of the general theory to problems of quantum waves in nanotubes
with spin taken into account. The results of this subsection represent the
particular case of the general results concerning quantum wave propagation in thin films taking spin into account. These more general results are obtained together with J. Br\"uning \cite{BrunDobrTud}.

Thin crystalline films of width $\sim 10$nm
(a few monoatomic layers), synthesized recently,
give a more complicated example of quantum waveguide.
Such a film is a waveguide for a quasiparticle with charge $e$
propagating along the film,
and we can affect this particle by means of an external
electromagnetic field. In reality,
a quasi-particle has spin,
but we shall not consider spin effects for nanofilms.

The nanofilm width $d_0\sim1$\,nm (10\,\r{A})
is comparable with the de Broglie wavelength
$\lambda=2\pi/k_F\sim1$\,nm of an electron with energy
of the order of the Fermi energy
$\varepsilon_F\sim1$\,eV.
This circumstance leads to the following effect
of ``dimensional quantization" of low-dimensional systems:
the domain of the wave function localization
in the normal direction to the film
has dimensions $\sim\lambda$,
and the energy corresponding to the motion
in this direction is quantized.
Therefore,
the total three-dimensional problem
of describing the quantum states
can be divided into several reduced problems
(on ``subbands of dimensional quantization")
already with two-dimensional quantum effective Hamiltonians
(along the film surface),
which, in the end, allows one to obtain
a sufficiently explicit description of these states
by using asymptotic formulas.

The film boundaries play an important role in the future constructions.
A natural idea is to simulate the boundaries of the film by means
of the Dirichlet conditions or ``rigid walls" for the wave
function. But it is more convenient to simulate by using
the so-called ``soft walls." Boundaries of this type are related
to the physical mechanism of confinement of electrons
near the physical film. The confinement appears as
a result of the electrostatic interaction between the film and
the quasi-particle. One usually simulates this interaction by
introducing a confinement potential $v_{\rm int}$ in the
direction normal to the film.
The potential $v_{\rm int}$ increases very fast
near the imaginary boundaries of the film (the ``walls").
Thus the wave function decays very fast outside the film
and the confinement potential $v_{\rm int}$ replaces the
``rigid" walls simulated by the Dirichlet conditions.
From the ``mechanical" viewpoint,
the confinement potential represents the interaction
with imaginary walls. The same idea is used in simulation of
nanotubes. We shall present the corresponding formulas somewhat
later.

The effective dynamics of quantum states in the approximation
of the strong coupling method is determined by the Schr\"odinger
equation:
\begin{gather}
\label{H} i\hbar\Psi_t=\widehat{\cal H}\Psi,\qquad \widehat{\cal
H}= \dfrac{\widehat{\bf P}^2}{2m}+v_{\rm int}({\bf r})
+v_{\rm ext}({\bf r},t),
\end{gather}
where $\widehat{\bf P}=-i\hbar\nabla-(e/c){\bf A}({\bf r},t)$,
$e=-e_0$ is the charge of electron, $m$ is the effective
mass of quasi-particle, $c$ is the velocity of light,
and $(v_{\rm ext}({\bf r},t),{\bf A})$ are the potentials
of external electromagnetic field.
We shall consider a space-uniform time-dependent magnetic field
${\bf H}={\bf H}(t)$.

The characteristic value of the transverse
energy~$\varepsilon_\perp$ in the tube can be found
from the uncertainty relation:
since the ``transverse" momentum is $\sim\hbar/d_0$,
we have $\varepsilon_\perp\sim \hbar^2/(md_0^2)$.
Let us introduce the characteristic ``longitudinal" length
$l_0$. Depending on the problem considered,
$l_0$ can be, for instance, the radius of curvature
of the film or the radius of the solution localization area,
etc.
We assume that $l_0\gg d_0$.
We introduce the magnetic length
$l_M=\sqrt{\hbar c/\bigl(e|\mathbf{H}|\bigr)}$,
the magnetic field quantum $\Phi_0=2\pi\hbar c/e$,
and the dimensionless magnetic field
as the number of magnetic flux quanta passing
through the characteristic area $l_0 d_0$:
${\bf H}'=l_0 d_0/{l_M}^2\cdot\mathbf{H}/|\mathbf{H}|
=2\pi l_0 d_0\cdot\mathbf{H}/\Phi_0$.
We introduce the new variables ${\bf r}'={\bf r}/l_0$,
the dimensionless time $t'=t/T$, $T=md_0 l_0/\hbar$,
the dimensionless potentials
$v_{\rm int}'=v_{\rm int}/\varepsilon_\perp$,
$v_{\rm ext}'=v_{\rm ext}/\varepsilon_\perp$,
${\bf A}'=ed_0(\hbar c)^{-1}{\bf A}$,
and
the dimensionless constant $\alpha'=\hbar\alpha/d_0^2$
and divide both sides of Eq.~(\ref{H}) by the energy of
transverse motion $\varepsilon_0$.
Below, we shall omit the primes.
Then the equation describing the motion of a quantum particle
(or a quasi-particle) in a quasi-two-dimensional crystal
takes the form
\begin{gather}
\label{Pauli1}
i\mu \Psi_t=\widehat{\cal H}\Psi, \qquad
\widehat{\cal H}=1/2(-i\mu\nabla-{\bf A})^2
+v_{\rm int}({\bf r})+v_{\rm ext}({\bf r},t).
\end{gather}
The fact that we consider the last equation in a film
is determined by the boundary conditions.
We shall assume that the film is determined
by some smooth surface $\Gamma$. This means that
Eq.~\eqref{Pauli1} holds and the boundary conditions are
formulated in some neighborhood of $\Gamma$.
It is convenient to use the special curvilinear coordinates
for the description of these conditions,
as well as for all future investigations.

{\bf Curvilinear coordinates.}
By $x=(x^1,x^2)$ we denote the (dimensionless) local coordinates
on the surface $\Gamma$, then each point ${\bf r}$
in a neighborhood of $\Gamma$ can be determined
by three values $(x^1,x^2,y)$, where $y$ is the distance between
the point ${\bf r}$ and its projection $R(x)\in \Gamma$.
Then we have
$$
{\bf r}={\bf R}(x)+y {\bf n}(x),
$$
where, as above, ${\bf n}(x)$ is a unit normal vector on~$\Gamma$.
Note that, in general,
the coordinates $x^1,x^2$ are not orthogonal, but always
$\langle {\bf n},{\bf n}\rangle=1$,
$\langle {\bf n},\pa_i {\bf R}\rangle=0$, $i=1,2$.
Thus the metric tensor is
\begin{gather}
\label{Metric}
G_{ab}=
\begin{Vmatrix}
\gamma_{ij} && 0 \\
0 && 1
\end{Vmatrix},  \quad  G=\det{G_{ab}}, \quad a,b=1,2,3,
\end{gather}
where
$\gamma_{ij}=\langle \pa_i {\bf r},\pa_j {\bf r} \rangle$,
$\gamma=\det{\gamma_{ij}}$, $i,j=1,2$, and $G=\gamma$.
Now let us present the components of the vector potential
${\bf A}$ in the coordinates $(x^1,x^2,y)$.
We choose the symmetric
form of the vector potential ${\bf A}=1/2[{\bf H},{\bf r}]$.
Hence the vector potential satisfies the Lorentz gauge:
$\pa A^a/\pa r^a=0$.
From now on, it is convenient to use the Einstein notation
and the summation rule.

{\bf Soft and rigid walls.}
Using the curvilinear coordinates,
one can consider an ``empty" film with ``rigid" walls:
$v_{\rm int}=0$, $\Psi|_{\pa\Omega}=0$
(the Dirichlet condition)
or a film with ``soft" walls:
$v_{\rm int}\neq 0$, $\Psi(x,y)\in L_2(y)$ at each $x$.
However, the last definition requires
$v_{\rm int}(x,y)$ to be identically defined
in the entire $\mathbb{R}^3$.
The last condition is too strong,
since $\Psi(x,y)$ is exponentially small for $y\gg\mu$
and any conditions on the function $\Psi$ in this region
affect its behavior negligibly.
To be definite, in what follows, we assume that
$\Psi(x,y)|_{\pa \Omega}=0$.
The ``empty" film with ``rigid" walls
can be considered as the limit of soft walls
described by the potential rapidly increasing
near the boundary.
As an example, let us consider the potential
$v_{\rm int}(x,y')=(y'/D(x))^{2m},\,y'=y/\mu,\,m>0$.
As $m\to\infty$,
we have $v_{\rm int}(x,y<d(x))\to 0$ and
$v_{\rm int}(x,y>d(x))\to \infty$.

{\bf Operator-valued symbol in a nanofilm.}
It is well known that the use of the function $\Psi'=\Psi G^{1/4}$
instead of the function $\Psi$ can significantly simplify
the corresponding calculations.
Substituting the function $\Psi=G^{-1/4}\Psi'$ into
Eq.~(\ref{Pauli1}), we obtain the following equation
for the function $\Psi'$:
\begin{gather}
i\mu \Psi'_t=\widehat{\cal H}'\Psi', \qquad
\widehat{\cal H}'=G^{1/4}\widehat{\cal H}G^{-1/4}.
\label{Pauli3}
\end{gather}
Using the formula
$G^{1/4}\Delta G^{-1/4}=G^{ab}\pa_a\pa_b+G^{ab}_{,a}\pa_b+
G^{-1/4}\pa_a\bigl(G^{1/2}G^{ab}\pa_b(G^{-1/4})\bigr)$,
we obtain
\begin{gather*}
\widehat{\cal H}'
=\frac{1}{2}G^{ab}\hat p_a\hat p_b - \frac{i\mu}{2} G^{ab}_{,a}\hat p_b
-\frac{\mu^2}{2}\frac{1}{G^{1/4}}\pa_a\left[
G^{1/2}G^{ab}\pa_b\left(\frac{1}{G^{1/4}}\right)\right]-\nonumber\\
-G^{ab}A_a \hat p_b-\frac{i\mu}{4} G^{ab}A_a \pa_b(\ln{G})
+ \frac{1}{2}G^{ab}A_a A_b+v_{\rm ext}(x,y,t)+v_{\rm int}(x,y/\mu).
\end{gather*}

We want to study solutions to Eq.~(\ref{Pauli3})
that have only a few oscillations in the transverse direction.
From the physical viewpoint, it is clear that, in general,
the nontrivial behavior of such solutions should be determined
by two-dimensional effective equations of adiabatic motion
\eqref{psieq0} on the surface $\Gamma$
in a neighborhood of the physical film.
Recall that our goal is to find these reduced equations
corresponding to solutions with different numbers of transverse
oscillations.
As we have different scales in the transverse and longitudinal
direction, it is natural to use the variable $y'=y/\mu $
instead of~$y$. To simplify the notation, we omit the prime.
Then the operator $\widehat{\cal H}'$ in
Eq.~(\ref{Pauli3}) is
\begin{gather}
\widehat{\cal H}'
=\frac{\gamma^{ij}}{2}(\hat p_i\hat p_j - 2 A_i \hat p_j + A_i A_j)
+\frac{1}{2}(\hat p_y^2- 2 A_y \hat p_y + A_y^2)+v_{\rm ext}({\bf r},t)+
v_{\rm int}(x,y)-\nonumber\\
-\frac{i\mu}{2}\gamma^{ij}_{,i} \hat p_j -
\frac{i\mu}{4}\gamma^{ij} A_i \pa_j (\ln \gamma)-
\frac{i}{4}A_y \pa_y (\ln \gamma)- \nonumber \\
-\frac{\mu^2}{2}\frac{1}{\gamma^{1/4}}\pa_i\left[
\gamma^{1/2} \gamma^{ij}\pa_j\left(\frac{1}{\gamma^{1/4}}\right)\right]
-\frac{1}{2}\frac{1}{\gamma^{1/4}}\pa_y\left[
\gamma^{1/2}\pa_y\left(\frac{1}{\gamma^{1/4}}\right)\right],
\label{Pauli4}
\end{gather}
where $A_i=\langle\pa_i{\bf r},{\bf A}\rangle$ and
$A_y=\langle{\bf n},{\bf A}\rangle$.
Eq.~(\ref{Pauli3}), with the Hamiltonian determined
by formula (\ref{Pauli4}),
is the object of our future study.

Using these formulas, we find the first and second terms
of the expansion of the symbol of operator
$\widehat {\cal H}'$:
\begin{gather}
{\cal H}_0'\left(x,p,y,-i\frac{\pa}{\pa y},t\right)=
\frac{1}{2} g^{ij}{\cal P}_i{\cal P}_j+
\frac{1}{2}\hat{\cal P}_y^2+
v_{\rm ext}\bigl({\bf R}(x),t\bigr)+v_{\rm int}(x,y) \\
{\cal H}_1'\left(x,p,y,-i\frac{\pa}{\pa y},t\right)=
\frac{1}{2}y\gamma^{ij}_1{\cal P}_i{\cal P}_j - yg^{ij}{\cal P}_i A^1_j +
\langle\nabla v_{\rm ext}({\bf R}(x),t),y{\bf n}\rangle-\nonumber\\
-i\left(\frac{1}{2}g^{ij}_{,i} p_j+
\frac{1}{4}g^{ij}A_i^0\pa_j (\ln g)+
\frac{1}{4}A^0_y[\pa_y(\ln\gamma)]_{y=0}\right),
\end{gather}
where $p=(p_1,p_2)$,
${\bf A}={\bf A}_0+\mu y {\bf A}_1$, ${\bf A}_0=1/2[{\bf H},{\bf R}]$,
${\bf A}_1=1/2[{\bf H},{\bf n}]$,
$A_i=A^0_i+\mu y A^1_i+O(\mu^2)$,
$A^0_{i}=\langle\pa_i{\bf R},{\bf A}_0\rangle$,
$A^1_i=\langle\pa_i {\bf R},{\bf A}_1\rangle
+\langle \pa_i {\bf n},{\bf A}_0\rangle$,
$A_y=A^0_y+\mu y A^1_y+O(\mu^2)$,
$A^0_{y}=\langle {\bf n},{\bf A}_0\rangle$,
$A^1_y=\langle{\bf n},{\bf A}_1\rangle=0$,
${\cal P}_i=p_i-A^0_{i}$, and
$\hat{\cal P}_y=p_y-A^0_{y}$.

To describe the so-called slow modes,
we need to compute ${\cal H}_2'$
under the assumption that
${\cal P}_i=0$, $v_{\rm ext}=0$, and $\pa{\bf H}/\pa t=0$.
We obtain
\begin{gather}
\label{geompot}{\cal G}(x)=
-\frac{(\varkappa_1-\varkappa_2)^2}{8}-
\frac{1}{2g^{1/4}}\pa_i\left[
g^{1/2} g^{ij}\pa_j\left(\frac{1}{g^{1/4}}\right)\right]
\end{gather}
This term is independent of $y$ and $\hat p_y$
and contains only geometric characteristics of the embedding
(the first summand)
and the limiting manifold (the second summand).
We call it a {\it geometric potential}.

{\bf Effective Hamiltonians of longitudinal motion.}
Now we present $\chi_0^\nu$ and the  effective ``adiabatic"
Hamiltonians $H^\nu_{\rm eff}$.
The index $\nu$ enumerates the Hamiltonians
$H^\nu_{\rm eff}$ which, in our problem,
is called the effective adiabatic Hamiltonian
on the $\nu$-th subband of the size quantization.
Substituting the function
$\chi_0^\nu=\exp(iy\langle{\bf n},{\bf A}_0\rangle)w^\nu$
into Eq.~\eqref{term}
with ${\cal H}_0={\cal H}_0'$, we obtain
\begin{gather}
\label{Hmu}
H^\nu_{\rm eff}(p,x,t)=\frac{1}{2} g^{ij}{\cal P}_i{\cal P}_j+
v_{\rm ext}\bigl({\bf R}(x),t\bigr)+\varepsilon_\perp^\nu(x),\qquad
\chi_0=\exp(iy\langle{\bf n},{\bf A}_0\rangle)w^\nu.
\end{gather}
where $w^\nu(x,y)$ and $\varepsilon_\perp^\nu(x)$
are the respective eigenfunction and eigenvalue of the following
problem:
\begin{gather}
\label{term0-film}
\left(-\frac{1}{2}\frac{\pa^2}{\pa y^2} + v_{\rm int}(x,y)\right)w^\nu(x,y)
=\varepsilon_\perp^\nu(x) w^\nu(x,y),\quad w^\nu(x,Y_1(x))=w^\nu(x,Y_2(x))=0.
\end{gather}
It is well known that the spectrum of this problem is
nondegenerate, thus the symbols $L$ and $\chi^\nu$ are scalar
functions.
For the model potential $v_{\rm int}(x,y)=(y/D(x))^{2m},\,m>0$
considered in item~2,
we obtain
$\varepsilon_\perp^\nu(x)=(d(0)/d(x))^2\varepsilon_\perp^\nu(0)$,
where $d(x)=D(x)^\frac{m}{m+1}d(0)$ is the dispersion of the
state with energy $\varepsilon_\perp^\nu(x)$.
Assuming that the width of the film is proportional to~$d(x)$,
we conclude that $D(x)^\frac{m}{m+1}$
is the coefficient of homothety.
As $m\to\infty$, this coefficient tends to $D(x)$.
So we obtain the natural result stating
that, in the model of empty film with rigid walls,
the width of the film is equal to the distance between the walls.

Let us present the first correction $\mu L_1$ in an expansion
of the symbol of the effective Hamiltonian of longitudinal motion.
It is given by formula \eqref{L1}.

Using the formula for $\chi_0^\nu$ and the expansion of the gauge
condition
$\pa_i(\gamma^{ij}A_j)+1/2 \gamma^{ij}A_i\pa_j(\ln{\gamma})
+ \pa_y A_y + 1/2 A_y\pa_y(\ln{\gamma})=0$
with respect to $y=\mu y'$,
it is easy to obtain the relations
\begin{gather}
\left< w^\nu,\pa_j w^\nu\right>_y=0, \quad
\left<\chi_0^\nu,\frac{\pa \chi_0^\nu}{\pa t}\right>_y=
iY\left< {\bf n},\frac{\pa {\bf A}_0}{\pa t}\right>, \quad
\left<\chi_0^\nu,\frac{\pa H_{\rm eff}}{\pa p_j}
\frac{\pa \chi_0^\nu}{\pa x^j}\right>_y
= i Y g^{ij} {\cal P}_i\pa_j\left<{\bf n},{\bf A}_0\right>,
\nonumber
\end{gather}
where $Y=\langle \chi_0^\nu,y\chi_0^\nu\rangle_y$,  and
\begin{gather}
Yg^{ij}{\cal P}_i \langle \pa_j {\bf R}, {\bf A}_1 \rangle=
-Yg^{ij} {\cal P}_i\left<{\bf n},\pa_j {\bf A}_0\right>=
1/2\langle{\bf H},\boldsymbol{\Lambda}\rangle,
\nonumber\\
\pa_i(g^{ij}A^0_j)+\frac{1}{2}g^{ij}A^0_i\pa_j(\ln g)
+A^0_y[\pa_y(\ln{\gamma})]_{y=0}=0,\quad
g^{ij}_{,i} p_j-\pa_i (g^{ij}A^0_j)=\pa_i (g^{ij}{\cal P}_j).
\end{gather}
From this, we find
\begin{gather}
\label{L1film}
L_1=
-Y\alpha^i_j g^{jk}{\cal P}_i{\cal P}_k
-\left<{\bf E}\bigl({\bf R}(x),t\bigr),Y{\bf n}\right>
-\langle {\bf H},\boldsymbol{\Lambda}\rangle
-\frac{i}{2}\pa_i(g^{ij}{\cal P}_j),
\end{gather}
where $\boldsymbol{\Lambda}=[Y{\bf n},\boldsymbol{\cal P}]$,
$\boldsymbol{\cal P}=g^{ij}{\cal P}_i \pa_j {\bf R}$,
and
${\bf E}=-\nabla v_{\rm ext}-T({\pa {\bf A}_0}/{\pa t})$.
We shall see below that the correction $L_2$
is important in the construction of the leading term
of the asymptotic solution
to the effective equation of adiabatic (longitudinal) motion
only under the assumptions
${\cal P}_i=0$, $v_{\rm ext}=0$, and $\pa{\bf H}/\pa t=0$.
In this case, it coincides with the ``geometric" potential
${\cal G}(x)$~\eqref{geompot}.

\subsection{Nanophysics: wave dynamics in nanotubes}

{\bf The Pauli operator.}
Lengthy molecules consisting of a great many atoms
situated on cylinder-type spatial surfaces are called
{\it nanotubes} \cite{Iijima,DresselNanotubes,Avouris,Chernozatonskii,Stankevich,Eletskii1,Eletskii2,Prinz}.
The surface of such tubes can have some additional internal
torsion. The nanotube diameter $d_0\sim1$\,nm (10\,\r{A})
is comparable
with the de Broglie wavelength $\lambda=2\pi/k_F\sim1$\,nm
of an electron with energy of the order of the Fermi energy
$\varepsilon_F\sim1$\,eV,
and the nanotube characteristic length~$l_0$
is significantly larger than~$d_0$.

In the approximation of the strong coupling method,
the wave functions in nanotubes are determined by
the nonrelativistic one-particle Hamiltonian,
i.e., by the Pauli operator with the spin-orbit interaction
taken into account:
\begin{gather}
\label{Htube}
\widehat{\cal H}=
\dfrac{\widehat{\bf P}^2}{2m}
+ v_{\rm int}({\bf r})+v_{\rm ext}({\bf r},t)
- \dfrac{e\hbar}{2mc}\langle\boldsymbol{\sigma},{\bf H}\rangle
+\widehat{\cal H}_{\rm SO},
\quad
{\widehat{\bf P}}=-i\hbar\nabla-\dfrac{e}{c}{\bf A}({\bf r},t).
\end{gather}
Here ${\bf r}\in\mathbb{R}^3$ is the radius vector of a point in
a neighborhood of the tube,
$\widehat{\cal H}_{\rm SO}$ is the operator of interaction
of the electron spin
with the electric field of the crystal \cite{Gantmakher}:
$\widehat{\cal H}_{\rm SO}=\alpha
\left<\boldsymbol{\sigma},\left[\nabla v_{\rm int},
\widehat{\bf P}\right]\right>$,
and $\alpha$ is the constant of spin-orbit interaction.
This Hamiltonian differs from that in a nanofilm \eqref{H} only
by the presence of terms describing the spin effects.
Thus, all the notation is the same.

In this section we consider some of results published in
\cite{BDS,BDST-DAN,BDT-RJMP}.

{\bf Curvilinear coordinates in tubes.}
As in the case of thin films, it is convenient to perform
all arguments by using a special system of curvilinear
coordinates.  We assume that the tube axis (the curve)~$\gamma$
is given by the equation ${\bf r}=l_0 {\bf R}(x),\
{\bf r}\in\mathbb{R}^3$, where ${\bf R}(x)$ is a smooth vector
function and $x\in\mathbb{R}$ is a natural parameter on~$\gamma$
(the tube length is counted off from a certain point $x^*$),
$|\pa_x{\bf R}(x)|=1$, $\pa_x=\pa/\pa x$.
If $|\pa^2_x {\bf R}|\neq 0$, it is determined the Frenet
trihedron. The curvature $k(x)=|\pa^2_x{\bf R}|$ and the torsion
$\varkappa(x)$ of the curve~$\gamma$ are connected by the Frenet
trihedron
$\bigl\{\pa_x{\bf R}, {\bf n}=\pa^2_x{\bf R}/|\pa^2_x{\bf R}|,{\bf b}
=[\pa_x{\bf R},{\bf n}]\bigr\}$ at each point~$x$
by formulas $\pa_x {\bf n}=-\varkappa {\bf b}-k \pa_x {\bf R}$
and $\pa_x {\bf b}=\varkappa {\bf n}$.

By $\Pi(x)$ we denote the plane intersecting
the tube axis at the point
${\bf R}(x)$ orthogonally to the axis; the section of the tube by
this plane (the area in $\Pi(x)$) we denote by $\Omega(x)$,
the boundary of $\Omega(x)$ we denote by $\pa\Omega(x)$.
Then the tube is the union of areas $\Omega(x)$,
and its boundary is the union of $\pa\Omega(x)$.
The ``physical meaning" of the boundary
$\pa\Omega(x)$ and of the boundary conditions
will be discussed later.
We introduce dimensionless coordinates $(x,y_1,y_2)$
determined by the relations ${\bf r}=l_0{\bf R}(x)+{\bf y},\
{\bf y}=d_0 y_1 {\bf n}_1(x)+d_0 y_2 {\bf n}_2(x)$, where
$\{{\bf n}_1(x),{\bf n}_2(x)\}$ is the basis
in the plane $\Pi(x)$.

If we put ${\bf n}_1={\bf n},\,{\bf n}_2={\bf b}$,
then the coordinates thus introduced will be nonorthogonal.
It is convenient to introduce orthogonal coordinates
(see \cite{Babich,Rash}).
First, let $\{{\bf n}_1(x),{\bf n}_2(x)\}$
be a certain orthonormal basis in the plane $\Pi(x)$
smoothly depending on $x$
(in general, this basis does not coincide with ${\bf n},{\bf b}$);
and let $\theta(x)$ be the angle between the vectors
${\bf n}$ and ${\bf n}_1$.
Then, along with the torsion $\varkappa$,
we can introduce an ``effective torsion"
$\varkappa_{\rm eff}=-\langle \pa_x{\bf n}_{1},{\bf n}_2\rangle
=\varkappa-\pa_x\theta$.
Choosing the angle $\theta(x)$ (along with
$\{{\bf n}_1(x),{\bf n}_2(x)\}$)
so that $\pa_x\theta=\varkappa$,
we let $\varkappa_{\rm eff}$ be zero.
The coordinates thus constructed are orthogonal
(around the tube axis, where they are specified).
The components of the metric tensor
$g_{ij},\,i,j=\{x,y_1,y_2\}$ in these coordinates
are determined as follows:
$g_{00}=G=(1-k\langle {\bf y},{\bf n}\rangle)^2$,
$g_{11}=g_{22}=1$, and $g_{ij}=0,i\neq j$.
Everywhere below we shall use these coordinates.
All formulas obtained below are valid
in the case of a straight axis if we set $k(x)=0$
and $\varkappa(x)=0$.
If $k(x)\neq 0$, then $y_1$ and $y_2$ are the coordinates
only in the area where $1-k\langle {\bf y},{\bf n}\rangle>0$.
It follows from the considerations about the tube curvature
given below that these coordinates are determined
in the area of the tube axis under study.

{\bf Boundary conditions and geometry of nanotubes.}
As in the case of a nanofilm,  the ``surface" of a nanotube can
be simulated by ``rigid" and ``soft" walls.
The rigid walls are determined by the imaginary surface of the
tube and the Dirichlet conditions on this surface.
The soft walls are simulated by an appropriate choice of the
crystal potential $v_{\rm int}(y,x)$ rapidly increasing
while approaching the imaginary surface of the tube
and creating a potential well where the electron wave function
is localized.
Outside this well, the wave function is exponentially small.
We shall consider tubes whose cross-section
by the plane~$\boldsymbol{\Pi}(x)$ rotates with respect
to the basis $\bigl\{{\bf n}_1(x),{\bf n}_2(x)\bigr\}$
in which the metric tensor is diagonal,
and simultaneously expands in the plane $x=\const$
with respect to the point~${\bf R}(x)$.
We define the tube chirality as follows:
we fix a cross-section~$\Omega_{x^*}$
for some $x^*\in\gamma$ and assume that,
at a point $x\neq x^*$,
the cross-section~$\Omega_x$ is obtained
from~$\Omega_{x^*}$ by a turn through an angle~$\Phi(x)$
(i.e., through an ``angle of internal torsion"
with respect to the basis
$\bigl\{{\bf n}_1(x),{\bf n}_2(x)\bigr\}$)
and by expanding by a factor~$D(x)$.

The domain $\Omega(x)$ can be introduced
as a multiply connected domain,
for example,
in the form of a circular or an elliptic annulus.
The adequacy of this representation depends
on the form of the crystal potential
in a concrete nanotube.
If the domain~$\Omega(x)$ is simply connected,
then, in the physical literature,
such a nanotube is called a ``quantum
wire"~\cite{EM1,EM2}.

{\bf Operator-valued symbol.}
The way of introducing dimensionless variables in a nanotube
is the same as in the previous Example 4.5.
It is convenient to pick out the factor $G^{-1/4}$
from the wave function $\Psi$,
where $G$ is the determinant of the metric tensor in the
variables $(x,y_1,y_2)$, i.e., to substitute
$\Psi=G^{-1/4}\Psi'$ into the original equation.
Then the function $\Psi'$ satisfies the equation
$i\mu\Psi'_t=\hat {\cal H}'\Psi'$,
$\hat {\cal H}'=G^{1/4}{\cal H}G^{-1/4}$.
In what follows, we shall use the wave function $\Psi'$
and the Hamiltonian ${\cal H}'$. After some transformations,
the quantum Hamiltonian $\hat {\cal H}'$ takes the standard form
\eqref{InEq} with the operator-valued symbol
\begin{equation*}
{\cal H}'={\cal H}'_0+\mu{\cal H}'_1+\mu^2({\cal G}(x)
+\widetilde{\cal H}'_2)+O(\mu^3),
\end{equation*}
where
\begin{equation}
\begin{array}{lr}
\mathcal{H}'_0=\frac{{\cal P}_0^2}{2}+v_{\rm ext}({\bf R}(x),t)+
\sum_{j=1}^2\frac{\widehat{\mathcal{P}}_j^2}{2}+v_{\rm int}(x,y),
\\
\mathcal{H}'_1=-1/2\langle \pa_x {\bf R},{\bf H}\rangle \hat{l}+
i k/2\langle {\bf n},{\bf A}_0 \rangle +
\Bigl(k\langle {\bf y},{\bf n}\rangle{\cal P}_0-1/2
\langle {\bf y}_\perp,{\bf H}\rangle\Bigr)p +
\\
+ \Bigl\langle \nabla v_{\rm ext}\bigl({\bf R}(x),t\bigr)
+ 1/2[{\bf A}_0,{\bf H}],{\bf y}\Bigr\rangle
- 1/2\langle\boldsymbol{\sigma},{\bf H}\rangle
+ \mu^{-1}\alpha\langle\boldsymbol{\sigma},\hat{\bf M}\rangle,
\end{array} \label{HExpansion}
\end{equation}
and we introduced the notation
\begin{equation}
\begin{array}{lr}{\cal P}_0=p-\langle \pa_x{\bf R},{\bf
A}_0\rangle,\quad \widehat{\mathcal{P}}_j=-i\pa/\pa y_j-\langle
{\bf n}_j,{\bf A}_0\rangle,\quad j=1,2,
\\
\hat l=i(y_2\pa/\pa y_1-y_1\pa/\pa y_2),\quad
{\bf y}_\perp=[{\bf y},\pa_x{\bf R}]=y_1 {\bf n}_2-y_2 {\bf n}_1,
{\bf A}_0=1/2\bigl[{\bf H}(t),{\bf R}(x)\bigr],
\\
\hat{\bf M}=\pa_x{\bf R}
\left(\frac{\pa v_{\rm int}}{\pa y_1}\widehat {\cal P}_2-
\frac{\pa v_{\rm int}}{\pa y_2}\widehat {\cal P}_1\right)+
{\bf n}_1\frac{\pa v_{\rm int}}{\pa y_2}{\cal P}_0 -
{\bf n}_2\frac{\pa v_{\rm int}}{\pa y_1}{\cal P}_0.
\end{array} \label{Note}
\end{equation}

For the symbol $\mathcal{H}'_2$,
we present only its ``geometric" part ${\cal G}(x)=-k^2/8$,
the ``remainder"
$\widetilde{\mathcal{H}}'_2=\mathcal{H}'_2-{\cal G}$
is a polynomial with respect to the momentum~${\cal P}_0$
and the components of the magnetic field ${\bf H}(t)$
with zero constant term and with coefficients
smoothly depending on $(x, y)$.
In what follows, we shall see that the explicit form
of this ``remainder" is not necessary for the construction of
the leading term of asymptotic solutions to the effective
equation of adiabatic (longitudinal) motion.

The boundary conditions (rigid and soft walls)
defining the nanotube are similar to those
in the case of a nanofilm.
The corresponding change will be discussed somewhat later.

{\bf Reduction to equations on the tube axis
and the adiabatic Hamiltonian.}
Now we want to use the scheme of \S 3 and to find the symbol~$L$
of the effective equation of adiabatic motion (along the tube
axis). This case is characterized by the fact
that the reduced equation contains a single spatial variable.
Thus, it is natural immediately to separate
the factor
$\exp(i\int_{x^*}^x\langle \pa_x{\bf R},{\bf A}_0\rangle
dx/\mu)$ in the wave function.
This separation takes
the extended momentum operator $\hat{\cal P}$
into the ``short" operator $\hat p=-i\mu \pa/\pa x$
but,
in the case of a magnetic field depending on time~$t$,
gives the correction
$\int_{x^*}^x\left< \pa_x{\bf R},\pa {\bf A}_0/\pa t\right>dx$
to the effective potential.
Next,
because the function~$\Psi$ is a spinor and ${\cal H}_0$
is a scalar operator, the true multiplicity of degeneration
of the term determining the reduced equation
is equal to~$2r$ (the definition of~$r$ is given later).

Taking these remarks into account,
we present the solution~$\Psi$ of Eq.~(\ref{InEq})
in the form
\begin{equation}
\label{newrepr}
\Psi(x,y,t,\mu)=\hat\chi^\nu\left[\exp\left(i\int_{x^*}^x
\langle \pa_x{\bf R},{\bf A}_0\rangle dx/\mu\right)\psi^\nu\right],
\quad
\hat\chi^\nu=\chi^\nu(\p,\x,y,t,\mu),
\end{equation}
where the symbol $\chi^\nu(x,p,y,t,\mu)=\chi_0^\nu(x,p,y,t,\mu)+\mu
\chi_1^\nu(x,p,y,t,\mu) + \ldots$
of the (pseudo)differential operator
$\hat \chi^\nu(x,p,y,t,\mu)$ is a matrix function consisting of
$2r$ columns and $2$ rows and~$\psi$ is a vector function
with~$2r$ (interacting) components $\psi^{\nu j}$ satisfying
Eq.~\eqref{psieq0}.
As was mentioned above,
to construct the leading term of the asymptotic solution
to this equation, we need only to have
its {\it essential part\/}
$L^\nu_0(p,x)+\mu L^\nu_1(p,x) + \mu^2 {\cal G}(x)$.

Equation \eqref{term} can be reduced to the equation
\begin{equation}
\label{ModelEq}
\left(-\frac{\Delta_y}{2}+
v_{\rm int}(x,y)\right)w^\nu=\varepsilon_\perp^\nu(x)w^\nu
\end{equation}
by the substitutions $\chi_0^\nu
=\exp(i\langle{\bf y},{\bf A}_0\rangle)w^\nu$ and
\begin{gather}\label{Heff}
H_{\rm eff}^\nu=\frac{p^2}{2}+v_{\rm ext}({\bf R}(x),t)
+\varepsilon_\perp^\nu(x)+
\int_{0}^x\left< \pa_x{\bf R}(x'),
\frac{\pa {\bf A}_0}{\pa t}(x',t)\right>dx'.
\end{gather}
Here $\nu$ is just the number of the (classical) effective
Hamiltonian (or the adiabatic term)
which is also called the number
of a {\it subband of dimensional quantization}.
The eigenvalue $\varepsilon^\nu_{\perp}(x)$
is the energy of the $\nu$-th transverse mode at a point~$x$.
In contrast to the case of nanofilms, the eigenvalues
$\varepsilon_\perp^\nu$ (and hence the effective Hamiltonians)
can be degenerate.
The number $r$, which appeared above, is precisely
their multiplicity.
Generally speaking, $r$ can depend on $x$,
and in this case the effect called
``the intersection of terms or effective Hamiltonians"
can occur \cite{Kucherenko,Avron1,Avron2,ColinDeVerd}.
Here {\it we assume that $r$ is independent of~$x$}.
Finally, we have
\begin{gather}\label{chi0}
\chi_0^\nu=\exp(i\langle{\bf y},{\bf A}_0\rangle)
\|w_1^\nu,\ldots,w_r^\nu\|\otimes E_s,\\
\|w_1^\nu,\ldots,w_r^\nu\|\otimes E_s=
\begin{Vmatrix}
w_1^\nu (x,y)& 0 & \cdots & w_r^\nu (x,y)& 0 \\
0 & w_1^\nu(x,y) & \cdots & 0 & w_r^\nu(x,y)
\end{Vmatrix},\nonumber
\end{gather}
where $\otimes$ is the tensor product of matrices
and~$E_s$ is the unit $2\times 2$ matrix.
The matrix function $\chi^\nu_1$
can be found from Eq.~(\ref{chi1}).

Now we discuss the choice of the model potential.
We will consider a tube with soft walls
and with the same elliptic cross-section \cite{van-der-Waals},
which can be modeled by using the potential
\begin{equation}
v_{\rm int}(x,y)=v_{\rm int}
\left(x^*,\frac{\boldsymbol{\Phi}(x)^{-1}{\bf y}}{D(x)}\right),
\qquad
v_{\rm int}(x^*,y)=
\left[
\left(\frac{y_1}{a}\right)^2+\left(\frac{y_2}{b}\right)^2
\right]^{m}, \quad m>0.
\label{vint}
\end{equation}

Passing from the variables $y=(y_1,y_2)$
to the new variables $y'=(y_1',y_2')$
determined by the relation $y=D^\gamma y'$,
$\gamma=m/(m+1)$,
we obtain
$\varepsilon_\perp^\nu(x)
=D(x)^{-2\gamma}\varepsilon^\nu_\perp(0)$.
It is easy to see that the dispersion $d(x)$
with respect to the coordinates~$y$
in the state~$w_n^\nu$
depends on~$x$ according to the relation
$d(x)=D(x)^{-\gamma} d(0)$.
It is natural to assume that~$d(x)$ is {\it proportional\/}
to the linear dimensions of the tube section.
Then
$D^{\gamma}$ is the ``soft" coefficient of extension
of the section,
and~$\gamma$ is the stiffness coefficient
of the walls.
The dependence of the energy on~$x$
can be represented as
\begin{gather}
\varepsilon^\nu_\perp(x)
=\varepsilon^\nu_\perp(0)\frac{d(0)^2}{d(x)^2}.
\label{edep}
\end{gather}
As $m\to\infty$,
the potential~(\ref{vint}) disappears
in the interior of the domain
and tends to $\infty$ outside this domain;
the coefficient $\gamma\to 1$ and
$d(x)\to D(x)^{-1}d(0)$.
In the limit, we obtain the ``empty cylinder" model:
$v_{\rm int}(x^*,y)=0$
for $({y_1}/{a})^2+({y_2}/{b})^2\leq 1$
and $v_{\rm int}(x^*,y)=\infty$
for $({y_1}/{a})^2+({y_2}/{b})^2> 1$,
where $D(x)$ is the coefficient of extension
(of homothety).
As in the case of nanofilms,
we introduce additional ``rigid" walls in the area
where the wave function is exponentially small.

Taking into account the form of potential \eqref{vint},
we obtain the relation
\begin{gather}
w^\nu_j(x,y)=\frac{1}{D(x)}
w^\nu_j\left(x^*,\frac{\boldsymbol{\Phi}(x)^{-1}{\bf y}}{D(x)}\right),
\qquad j=1,\dots,r.
\end{gather}

{\bf Remark.} Calculating $\chi_0^\nu$,
we do not fix any special form of the functions $w_1^\nu,\ldots,w_r^\nu$.
We assume that they {\it form an orthonormal basis
in the eigenspace of problem (\ref{ModelEq})
corresponding to the eigenvalue (term)
$\varepsilon_\perp^\nu(x)$ with the number~$\nu$}
and depend smoothly on all its variables.
Of course,
such a basis is not unique,
and it is convenient to make its final choice
in the subsequent construction
of asymptotic solutions.
For example, sometimes,
$w_j^\nu$ can be taken to be the eigenfunctions of the momentum
operator~$\hat l$, i.e., in this case,
it is necessary to distinguish the states
inside the term according to the projections of the orbital
momentum in these states on the tube axis.
Then the momentum matrix~$\Lambda$ is diagonal.
Of course,
it is also possible to change the basis in the space of spinors;
this is convenient for the case in which
the spin affects the classical dynamics
(see below  the ``medium-wave regime").
Obviously, the choice of a new basis
is equivalent to the inclusion
of some unitary $2r\times 2r$ matrix depending on~$x$
into formula~\eqref{newrepr}
after the operator $\hat\chi^\nu$.

{\bf Effective Hamiltonians of longitudinal motion.}
Using formulas from \S 3,
we obtain  $L^\nu_0$ and $L^\nu_1$.
In general,
the objects  $\Lambda$, $L_y$, etc.
introduced below also depend on the number~$\nu$
(as well as $\chi_j$, $L^\nu_j$, and~$\psi^\nu$).
Sometimes, we omit this dependence to simplify the notation.

The symbols $L^\nu_0$ and $L^\nu_1$ are determined as follows:
\begin{equation}
\begin{array}{lr}
L^\nu_0(p,x)=H_{\rm eff} \ E_r\otimes E_s,\\
L^\nu_1(p,x)=ik/2\langle{\bf n},{\bf A}_0\rangle
E_r\otimes E_s + L_y\otimes E_s+E_r\otimes L_s+L_{sy},
\quad L_s=-\frac{1}{2}\langle\boldsymbol{\sigma},{\bf H}\rangle,
\\
L_y(p,x)=\Bigl((\pa_x\Phi) p
-1/2\langle \pa_x{\bf R},{\bf H}\rangle\Bigr)\Lambda-
\langle {\bf Y}_\perp,{\bf H}\rangle p +
\left\langle {\bf Y}, \nabla v_{\rm ext} +
\frac{\pa {\bf A}_0}{\pa t}+kp^2{\bf n} \right\rangle,
\\
L_{sy}(p,x)={\mu}^{-1}{\alpha}\Bigl(
M^0\otimes\langle\boldsymbol{\sigma},\pa_x{\bf R}\rangle
+M^1\otimes\langle\boldsymbol{\sigma},{\bf n}_1\rangle
+M^2\otimes\langle\boldsymbol{\sigma},{\bf n}_2\rangle \Bigr).
\end{array} \label{HamAdiabatic}
\end{equation}
By $E_r$ we denote the unit $r\times r$ matrix,
by $\Lambda(x)$ we denote the $r\times r$ momentum matrix
with elements $\Lambda_{jj'}
=\left< w^\nu_j,\hat l w^\nu_{j'}\right>_ y$,
by $M^j(x)$ we denote the $r\times r$ matrix of the from
$(M^0)_{jj'}=-i\left<w^\nu_j,\bigl((\pa_1 v_{\rm int})\pa_2
-(\pa_2 v_{\rm int})\pa_1\bigr)w^\nu_{j'}\right>_y,\,
(M^1)_{jj'}=\left<w^\nu_j,(\pa_2 v_{\rm int})w^\nu_{j'}\right>_y p,\,
(M^2)_{jj'}=-\left<w^\nu_j,(\pa_1 v_{\rm int})w^\nu_{j'}\right>_y p$,
where $\pa_i=\pa/\pa y_i$,
and
by ${\bf Y}(x)=Y_1 {\bf n}_1 + Y_2 {\bf n}_2,\,
{\bf Y}_\perp(x)=Y_2 {\bf n}_1-Y_1 {\bf n}_2$
we denote the three-dimensional ``vectors"
whose components are the $2\times 2$ ``dipole" matrices
$(Y_i)_{jj'}(x)
=\Bigl\langle w^\nu_j,y_i w^\nu_{j'}\Bigr\rangle_y$,\
$i=1,2$.
As above,
$\left<\cdot,\cdot\right>_y$ denotes the integration
over the variables~$y$.
The symbol $L^\nu_2(p,x)$ is significantly more complicated,
but we need only a part of it, i.e.,
the so-called ``geometric potential"
${\cal G}(x)=-(k^2(x)/8) E_r\otimes E_s$.
In the long-wave approximation,
it is necessary to take this term into account.
Precisely this term generates bound states
in an empty waveguide~\cite{MaslovWaveGuide}.

{\bf Additional boundary and initial conditions.}
Formulas \eqref{HamAdiabatic} allow one to construct
the leading term of different asymptotic solutions to the
effective equation of adiabatic (longitudinal) motion of $\nu$th
subband of dimensional quantization.
To perform more complete constructions,
one has to sum solutions with different numbers $\nu$.
But from the physical viewpoint,
one is interested in the {\it reduced equations
with only small numbers}~$\nu$,
and it usually suffices to consider
the case of several~$\nu$.
(In nanotubes, $\nu$, as a rule, does not
exceed~$7$~\cite{DresselNanotubes}).
This fact turns out to be very important later
in the study of equations in curved waveguides and tubes
and allows one to ignore the applicability problem
for the asymptotic formulas obtained for large~$\nu$,
and the convergence problem for the corresponding series
with respect to the number~$\nu$.
For this reason,
{\it it suffices to pose
the additional boundary and initial conditions
already not for the original equation,
but for finite (here one-dimensional)\/}
({\it simplified\/})
{\it equations of the form} \eqref{psieq0}.
Using the physical terminology,
we can say that it is of interest to study
the longitudinal dynamics of a small set of subregions
of transverse quantization.
The corresponding additional conditions
for the case of spatial waveguides
will be posed accurately below.

\section{Asymptotic solutions to the effective equations of
adiabatic motion}
Now we want to discuss the question about solutions
to the reduced effective equations of adiabatic motion.
The existence of the {\it adiabatic} parameter $\mu$ allows one
to separate the {\it fast motion} from the {\it adiabatic
motion} (the electron motion from the nuclei motion in a molecule,
the transverse motion from the longitudinal motion in waveguide,
electron waves from the lattice oscillations in crystals, etc.).
A very important fact is that the parameter $\mu$ slightly
depends on the energy of adiabatic motion
in a certain range where it varies.
The fact that the adiabatic approximation holds
for the entire region of energies of adiabatic motion
is well known in physical literature (see, e.g., \cite{Davydov}).
The adiabatic motion can be essentially
different for different energies from this region.
This fact is important if one is interested in the construction
of asymptotic or exact solutions to the reduced equation
of adiabatic motion describing different physical processes
(and corresponding to different energies).
This means that asymptotic and sometimes exact solutions
are of different type and thus  the process of determining
the leading (or essential) part of the symbol $L$ should be
revised.
For instance, some parts of the correction $L_1$
must be moved to the leading part of $L$.
As we also mentioned, this fact can, in turn, change the
definition of the characteristics of the reduced equation
and, in particular, lead to the ``semiclassical splitting" of
terms in the degenerate case (when $r\neq 1$).
In the case of nanotubes,
this effect shows how the spin affects the determination of
classical characteristics. We think that the best way to explain
these phenomena is to consider a simple nontrivial example
which, in our opinion, is the problem of quantum waves
in nanotubes. Thus, we restrict ourselves to this example
bearing in mind its importance.
Moreover, it seems advisable to explain the main ideas and
considerations with the example of the Schr\"odinger equation
with the Hamiltonian
\begin{gather}
\hat{\cal H}=
-\frac{\mu^2}{2}\Delta_x-\frac{1}{2}\Delta_y+v(x,y),
\label{simpH}
\end{gather}
where $x\in\mathbb{R}^n$ and $y\in\mathbb{R}^k$.
This is done in \S 5.1.
If one chooses an appropriate potential $v(x,y)$,
this equation describes problems of molecular physics,
as well as of quantum waveguides.
The ideas from \S 5.1 are applied to the problem of quantum
waves in nanotubes in \S 5.2.

As we mentioned above, the semiclassical analysis of the
reduced effective equation for adiabatic motion is well
developed and the solutions for the above-listed problems
are given in the simplest form by using the Maslov canonical
operator \cite{MaslovAsymptMethods,Vainberg}.
To obtain explicit formulas is a problem
which must be solved in concrete situations.
There are many publications devoted to the Maslov canonical
operator. Here we only note that this is actually a certain
algorithm whose realization,
as well as the process of obtaining an answer
appropriate from the viewpoint of the use in applied problems
(e.g., the plots of solutions, the calculation
of the scattering data, the frequency of beating, etc.),
even in the one-dimensional situation,
requires additional efforts and the use of computers.
A detailed description of the solutions based on this algorithm
and concrete physical results are not the goal of this work.
These results present the contents of other publications
(see, e.g., \cite{TMF-BDT,TMF-BDT2}).
Here, in \S 5.3, we only very briefly describe
the asymptotic solutions and the simplest physical results.
This remark also concerns all the examples considered
in this paper.

The majority of the ideas stated below in \S 5.1 can be
generalized to other examples discussed above.
Nevertheless,
it is necessary to emphasize that a ``simple" Hamiltonian
of the form \eqref{simpH} has a very special form,
and hence some effects related to this Hamiltonian
do not occur in the examples with other Hamiltonians.
On the other hand, if the Hamiltonian has a different form,
then other interesting effects can appear.

\subsection{General considerations.}

{\bf Internal and external parameters.}
In the examples under study,
we implicitly assume that the dimensionless coefficients
(e.g., the potential~$v_{\rm{ext}}$ in nanofilms and nanotubes)
is independent of~$\mu$.
Indeed, in real situations,
it is sometimes natural to assume that the coefficients
can depend on both~$\mu$ and other parameters.
These parameters characterize
the kinetic energy of adiabatic motion,
the strength of external fields,
the strength of interactions, etc.
In this case,
the functions $\chi_0^\nu$, $\chi_1^\nu$,
$\ldots$, $L_0^\nu$, $L_1^\nu$, $\ldots$ in formula \eqref{PsiIn}
also depend on these parameters.
Nevertheless,
under appropriate constraints,
the formula of separation of variables (\ref{PsiIn})
remains valid.
For purposes of mathematical rigor,
some constraints must be imposed
on these parameters of the problem
so as to connect them, for example,
with the parameter~$\mu$.
However, in this case,
one must bear in mind that, in concrete situations,
all these parameters are numbers,
and such constraint formulas
are of a very conventional character.
For this reason, to avoid cumbersome notation,
we present the explicit dependence on such parameters
only if it is necessary.

{\bf Semiclassical parameter $h$.}
The fact that the solutions of the equation of adiabatic motion
can be essentially different originates from the existence
of an additional parameter, which characterizes
the excitations of the adiabatic subsystem in the allowable range.
To introduce this parameter,
we consider the first well-known asymptotics
of Eq.~\eqref{BornOpp}
corresponding to different energies in the case of the Hamiltonian
$\hat{\cal H}_0=
\frac{1}{2}(-i\mu\frac{\pa}{\pa x})^2+\frac{1}{2}(\py)^2+v(x,y)$.
Denote, for a moment, by $\langle\cdot,\cdot\rangle$
the inner product in the original configuration space.
The kinetic energies of fast and adiabatic motions in this situation
are $\mathbf{K_f}=\langle\Psi,\frac{1}{2}(\py)^2\Psi\rangle$
and $\mathbf{K_a}
=\langle\Psi,\frac{1}{2}(-i\mu\frac{\pa}{\pa x})^2\Psi\rangle$,
respectively.

As we mentioned above, according to \cite{MaslovAsymptMethods},
the semiclassical solutions to Eq.\eqref{psieq0} have the WKB-form
$\Psi\approx\chi(\pa S/\pa x,x,y)\psi(x,t,\mu)$,
$\psi\approx\exp(iS(x,t)/\mu)\varphi(x,t,\mu)$,
where the function $\varphi(x,t,\mu)$ depends regularly on~$\mu$.
In this case, the kinetic energies of fast and adiabatic motions
have the {\it same order}:
$\mathbf{K_f}\sim \mathbf{K_a}\sim 1$.
This solution corresponds to the {\it excited state} of
adiabatic motion. On the other side, Born and Oppenheimer
\cite{Born,BornHuang} constructed the
harmonic oscillator type solution
$\Psi\approx\chi(x,y,\mu)\psi(x,\mu)$,
$\psi\approx\exp(-x^2/\mu)$.
This solution corresponds to the kinetic energy
$\mathbf{K}_a \sim\mu$, hence the energy of adiabatic (nuclei)
motion in this case is much smaller than the energy of fast
motion. Further, in the theory of waveguides,
we sometimes have solutions of the form
$\Psi\approx\chi(y,x,\mu)\psi(x,t,\mu)$,
where $\chi(y,x,\mu),\ \psi(x,t,\mu)$
depend regularly on $\mu$.
For these solutions, we obtain $\mathbf{K_a}\sim\mu^2$.
The above-listed different asymptotics can be classified by the
parameter $h=\mu\sqrt{\mathbf{K_f}/\mathbf{K_a}}
\Leftrightarrow
{\mathbf{K_a}}/{\mathbf{K_f}}\sim\mu^2/h^2$.
We call $h$ the {\it semiclassical} parameter. Let us emphasize
that the adiabatic parameter is always assumed to be small
and, conversely,
the parameter $h$ can be small but can also be $\sim 1$.

This parameter can be explained in another way.
For clarity, we consider the plain quantum straight waveguide
and, for a while, return back to dimensional variables.
We have the diameter $d_0$ and the length $l_0$ of the waveguide.
Recall that our goal is to construct asymptotic solutions
of the reduced equation
which describe the motion along the tube axis
in a sufficiently wide range of longitudinal energies
and the transverse wavelength $\lambda_\perp\sim d_0$.
To the longitudinal energy, there corresponds
the characteristic de Broglie wavelength
$\lambda_\parallel=\hbar/p_\parallel$,
where $p_\parallel$ is the dimensional momentum
of longitudinal motion.
Now the {\it ``semiclassical"} parameter is
$h=\lambda_\parallel/l_0$.
In other words, the parameter $h$ determines the ``smoothness"
of the function $\psi$ ($h^{-1}$ is the number of oscillations
at the distance $\sim l_0$) and agrees with the estimation of
its derivatives:
$\langle\psi,\frac{\pa\psi}{\pa x}\rangle\sim h^{-1}$.

We again consider the example of an empty waveguide.
Then the energy of the longitudinal motion on the
$\nu$-th subband of transversal quantization has the form:
$p_\parallel^2/2m+v_{\rm eff}(x)$,
where $v_{\rm eff}(x)=v_{\rm{ext}}(x)
+\nu^2\pi^2\varepsilon_\perp$,
$\varepsilon_\perp= \hbar^2/(2md_0^2)$.
Denote by $p_\perp$ the transverse momentum.
Taking into account the relation
between the de Broglie wavelength
and the corresponding momentum,
we obtain
$\frac{p_\parallel}{p_\perp}
\sim\frac{\hbar/\lambda_\parallel}{\hbar/\lambda_\perp}
\sim\frac{d/l_0}{\lambda_\parallel/l_0}\sim\frac{\mu}{h}$.
Thus,
the kinetic energies of longitudinal (adiabatic) and transverse (fast)
motions satisfy the relations
${\mathbf{K_a}}/{\mathbf{K_f}}\sim\mu^2/h^2$ and
${\mathbf{K_f}}\sim\varepsilon_\perp$.
Now we return to dimensionless variables.
Then the dimensionless longitudinal kinetic energy is
${\mathbf{K_a}}=p_{\parallel}^2/2\sim\mu^2/h^2$.
It is clear that if a particle moves along the waveguide,
then the kinetic energy can vary under the action
of the force $f=-\pa v_{\rm eff}/\pa x$.
For this force~$f$ not to accelerate the particle
so that its kinetic energy be of order
different from $\mu^2/h^2$,
it is necessary that its work does not exceed,
in the order of magnitude,
the parameter of the characteristic kinetic energy
corresponding to the initial momentum.
In dimensionless variables,
to the distance $\sim l_0$
there corresponds an interval $\sim1$.
Hence the work of the force $f$
is of the order of the derivative
$\pa v_{\rm eff}/\pa x$.
This implies that
the effective potential must have the form
$v_{\rm eff}=v^0_{\rm eff}+\frac{\mu^2}{h^2}v^1_{\rm eff}(x)$,
where $v^0_{\rm eff}=\const$ and $v^1_{\rm eff}(x)$ can,
in general, regularly depend on the parameters~$\mu$ and~$h$.
Moreover,
the work can be even equal to zero,
since the characteristic longitudinal momentum
is determined not only by the variable part
$v^1_{\rm eff}(x)$ of the effective potential,
but also by the ``input" momentum of the wave packet
under study
(i.e., by the gradient of the phase of the wave function
at the initial time instant in the Cauchy problem
or by the momentum of the incident wave
in the scattering problem).
In the last case,
the asymptotics of the wave function can be obtained
by using the well-known Born approximation.

{\bf Remark.} We point out that $v_{\rm eff}$
is determined by both the external field and the field of the
crystal.
Therefore in the case of a quantum waveguide,
the above constraints lead, in particular,
to the assumption that the geometric parameters of the waveguide,
i.e., the curvature (and torsion) of its axis, the width, etc.,
vary sufficiently ``slowly."

{\bf Characteristic time scale
and the reduced equation consistent with this scale.}
The question concerning the time scaling is nontrivial
and, generally speaking, can be solved separately
in each concrete problem.
It is natural to understand what characterisitc time
is required for a quantum particle to walk
through a certain characteristic distance.
For problems in nanotubes, the characteristic distance
is the total tube length
(e.g., in the scattering problem or in the problem of
the wavetrain propagation) or the size of the ``localization
area" of the wave function in the problem of bound states.
For a while, we assumee the characteristic distance
to be of the order of $l_0$ in dimensional variables
or to be $\sim 1$ in dimensionless variables.
One has to replace the time scaling by the energy scaling
in the case of stationary problems (for instance,
in problems of electron-phonon interaction or in molecular
physics).

{\bf Remark.}
To introduce the characteristic time scale in the general case,
we can use the following ideas.
It is clear that the characteristic time scale is
$t\sim a/\langle v\rangle$, where $a$ is the characteristic
distance for adiabatic motion
and $\langle v\rangle$ is the mean velocity.
Generally speaking, $a$ as well as $\langle v\rangle$
depend on the ``longitudinal" kinetic energy.
In quantum mechanics, we have
$\langle v\rangle=d\langle x\rangle/dt$. Using Eq.\eqref{InEq}, we obtain
$d\langle x\rangle/dt\sim i\mu^{-1}\langle[\hat p^2/2,x]\rangle
=\langle\hat p\rangle=\mu/h$ (cf.~Remark 3 in the \S \ref{Rem3}).
Thus we have $t\sim(h/\mu)a$. For the scattering problem,
the wavetrain propagation problem and some other problems,
we can set $a\sim 1$. For lower bound states and trapped modes
we have to set $a\sim h\sim\sqrt{\langle(\Delta x)^2\rangle}$.

The dimensionless time $t$ used in the general scheme for
Eq.\eqref{InEq} was actually chosen for the case
$p_\perp\sim p_\parallel$, i.e., for the case $\mu/h\sim 1$.
If this relation does not hold,
then the time of passage of a particle
through the waveguide,
which is naturally understood as the characteristic time of the
problem, must  be multiplied by the factor $(h/\mu)$.
Therefore, instead of~$t$,
it is convenient to introduce a new dimensionless time~$t'$
by the relation $t=(h/\mu)t'$.

In the case of nanotubes this redefining of the time scale
becomes consistent with the preceding physical argument
because of the following transformations
in Eq.~(\ref{psieq0}).
The term $v^0_{\rm eff}$ results only in a displacement
(renormalization) of the energy in the stationary problem
generated by the reduced equation;
or the factor
$\exp(-iv^0_{\rm eff}t'/\mu)$ appears
in the wave function~$\psi^{\nu}$
of the nonstationary Schr\"odinger equation~(\ref{psieq0}).
Taking this into account,
we represent the solution of this equation
in the form
$\psi^\nu=\exp(-iv^0_{\rm eff}t'/\mu){\psi'}^{\nu}$,
where ${\psi'}^{\nu}(x,t)$ is a new unknown function.
Since we assume that $p\sim \mu/h$,
it is natural to divide the equation
by the parameter~$\mu^2/h^2$.
In the left-hand side this gives
the derivatives
$i(h^2/\mu)\frac{\pa{\psi'}^{\nu}}{\pa t'}$,
which, after the above change of time,
take the form $ih\frac{\pa{\psi'}^\nu}{\pa t'}$.
It is important to point out that
this transformation concerns only the time variable:
the variables~$x$ and $y$ are not transformed.
As a result, instead of Eq.~(\ref{psieq0}) and
Eq.~\eqref{HamAdiabatic}
taking the corrections $L^\nu_1$, $L^\nu_2$, etc., into account,
we obtain the equation (the primes are omitted):
\begin{eqnarray}
\label{RedSchredEx3} ih\frac{\pa\psi^\nu}{\pa t}=
\Biggl\{\frac{1}{2}\left(-ih\frac{\pa}{\pa x}\right)^2+
v^1_{\rm eff}+
\frac{h^2}{\mu}\Bigl[L_1^\nu\Bigl(\frac{\mu}{h}(\ph),\x\Bigr)+
\mu
L_2^\nu\Bigl(\frac{\mu}{h}(\ph),\x\Bigr)+\ldots\Bigr]\Biggr\}\psi^\nu
\end{eqnarray}


{\bf Accuracy of asymptotic expansions.}
The number of terms in the expansion
of the intertwining operator $\hat\chi$
and the operator~$\hat L$ can be arbitrarily large.
However, as we already mentioned,
to calculate terms of these series explicitly,
even terms with small numbers,
is, as a rule, a very complicated problem.
Therefore, it is natural to consider
only the terms for which one can correctly estimate
the {\it leading term} of the asymptotics
of the wave function or of the energy value.
It is reasonable that the notion of the ``leading" term
of an asymptotics can be determined not only by the adiabatic
parameter~$\mu$,
but also by the ``semiclassical parameter"~$h$,
which is related to the form of the coefficients
and the solution of the effective equation of adiabatic motion.
We shall turn back to the question about numbers of terms
in the intertwining  operator $\hat\chi$ and
the operator~$\hat L$ later.
Now we recall well known estimates which allow us to
estimate these numbers.

Taking in account this fact, let us discuss the problem of
choosing the number of terms in the expansion of the symbols
of the operator~$\hat L$ and the intertwining
operator~$\hat\chi$. Again we restrict our consideration to the
case of nanotubes, although the main ideas can be generalized
for the majority of the adiabatic problems listed above,
including the non-self-adjoint problems
like water waves in a picnocline
(problems of such type usually appear in hydrodynamics).

Since the problem contains two parameters~$\mu$ and~$h$,
we shall calculate as many terms as we need
to construct the leading term of the asymptotics
with respect to $\max(h,\mu)$ if $h\ll 1$
and with respect to~$\mu$ if $h=1$.
(Recall that the parameter~$\mu$ is always assumed to be small,
and the parameter~$h$ can be either small
or of order $O(1)$.)
To find the minimal reasonable number of terms
in asymptotic expansions,
it is natural to use the well-known estimate
for the solution of the Cauchy problem
for a nonhomogeneous Schr\"odinger type equation:
$i\varepsilon\frac{\pa\phi}{\pa t}=\hat A\phi + f$,  \
$\phi|_{t=0}=0$.
Here $\hat A(t)$ is an essentially self-adjoint (for each~$t$)
operator in $L_2$, and $\varepsilon>0$.
Let~$\phi$ be a solution of this problem;
then the following inequality holds
for any~$t$ from the fixed interval $[0,T]$:
$\|\phi\|_{L_2}\leq
\frac{T}{\varepsilon}\max_{t\in[0,T]}\|f\|_{L_2}$.

We assume that $\psi_{\rm ex}$ is an exact solution
of the original equation (\ref{psieq0})
and $\Psi_{\rm as}$ is its asymptotic solution
of the form (\ref{PsiIn})
and these solutions coincide at zero instant;
moreover,
$\psi_{\rm as}$ satisfies the original equation
with discrepancy $f_{\rm as}$.
For the operator $\hat A$,
we choose the original quantum Hamiltonian
and set $\varepsilon=\mu$ and $T=\frac{h}{\mu}T_0$,
where $T_0$ is independent of~$h$ and~$\mu$.
Then we obtain the estimate
\begin{equation}
\label{Est2}\|\psi_{\rm as}-\psi_{\rm ex}\|\leq
\frac{h}{\mu^2}T_0 \max\|f_{\rm as}\|_{L_2}
\end{equation}
for the difference
$\phi=\psi_{\rm as}-\psi_{\rm ex}$.
This implies that the minimal reasonable number
of terms in the expansion of the symbols
of the operators~$\hat \chi$ and~$\hat L$
in formulas~(\ref{expchi}) and~(\ref{expL})
must at least satisfy the condition
$\frac{h}{\mu^2}\|f_{\rm as}\|_{L_2}\ll 1$ for $\mu\ll 1$.
Of course,
it should be remembered
that the norm of the discrepancy~$f_{\rm as}$
depends on~$\mu$ and~$h$.
As heuristic arguments, it is also useful
to apply the estimate (\ref{Est2})
to the reduced equation~(\ref{psieq0}).

{\bf Classification of quantum states
for longitudinal motion}
We return to the passage
from Eq.~\eqref{psieq0} to Eq.~\eqref{RedSchredEx3}.
For $h \ll 1$, to construct a wave function,
it is natural to use the semiclassical approximation.
Outside a neighborhood of the focal points
(the turning points),
the typical asymptotics of a wave function
with characteristic wavelength $\lambda_{\parallel}\sim h$
is given by the WKB-solution
\begin{equation}\label{wkb}
\psi(x,t)=A(x,t,h)\exp\left(\frac{iS(x,t)}{h}\right),\quad
 A(x,t,h)=A(x,t,0)+O(h),
\end{equation}
where $S(x,t)$ is the phase and $A(x,t,h)$ is, in general, the
vector amplitude.
As it is known~\cite{MaslovAsymptMethods},
in the first approximation,
after the substitution of this function
into the original equation,
the operator $-ih\frac{\pa}{\pa x}$ is,
in the leading part, replaced by $\frac{\pa S}{\pa x}$,
and thus the order of the terms in the operators
$h^2\mu^{j-2}L^\nu_j(\x, \frac{\mu}{h}(\ph))$
in~(\ref{RedSchredEx3})
is determined by the order of the functions
$L^\nu_j(x,\frac{\mu}{h}\frac{\pa S}{\pa x})$.
This fact leads to the well-known conclusion
that the phase $S(x,t)$ is determined
by the classical Hamiltonian system
whose Hamiltonian is the leading part
of the symbol expansion with respect to the parameter~$h$.
Bearing this in mind,
it is natural to define
the operator $\hat{p^h}=-i h \pa/\pa x$,
rather than the operator $\hat{p}=-i \mu \pa/\pa x$,
to be the momentum operator.
Clearly, for $\mu=h$,
the classical Hamiltonian is
the effective Hamiltonian~\eqref{L0},
but if the adiabatic and semiclassical parameters~$\mu$ and~$h$
are of different orders and $h\ll 1$,
then to construct semiclassical asymptotic
it is required to write the expansion
with respect to parameter $h$, assigning $\mu=\mu(h)$.
%
%
As we shall see later, in some cases,
additional terms from~$L^\nu_1$ will be included
in the classical Hamiltonian
(subject to the expansion with respect to~$h$).

Now let us discuss how many terms
in the expansion of the operator
in the right-hand side of~\eqref{RedSchredEx3}
we must have to find the leading part
of its asymptotic solution.
By setting $\varepsilon=h$
and applying the estimate (\ref{Est2})
to Eq.~(\ref{RedSchredEx3}),
we see that, at least intuitively,
it suffices to calculate
the effective Hamiltonian $(p^h)^2/2 + v^1_{\rm{eff}}$
and the first correction $L^\nu_1(x,\frac{\mu}{h} p^h) $.
This conclusion is consistent with the well-known fact
from the theory of semiclassical asymptotics:
terms of the order of~$h^2$
(and even of~$h^{1+\delta}$, $\delta>0$)
do not affect the phase $S(x,t)$
and the leading part of the amplitude $A(x,t,0)$.
This conclusion has a general character
and holds always if it is assumed that
$\mu\leq  h \ll 1$,
i.e., for the case in which
the semiclassical approximation can be used.
At the same time, as we just noted,
concrete formulas can essentially differ
in the following situations:
{\bf a)} if~$\mu$ and~$h$ have the same order
($\mu \sim h$) and
{\bf b)} if $\mu \ll h$.
If the parameter $h\sim 1$,
then the semiclassical approximation
cannot be used,
but Eq.~(\ref{RedSchredEx3}) remains to hold
and can even be simplified, although in this case
a part of~$L^\nu_2$ must be taken into account.
The existence of these differences results
in the following classification of asymptotic solutions
depending on the relation between the parameters~$\mu$ and~$h$
(or, which is equivalent, depending on the relation
between the longitudinal $\mathbf{K}_a$ and
transverse $\mathbf{K}_f$ kinetic energies in the waveguide).

{\bf a)} For $h=\mu$,
we have
the standard ``semiclassical" situation~\cite{MaslovAsymptMethods}
or the {\it ``short-wave" regime\/}
in which the ``longitudinal" energy is
of the same order as the energy of transverse motion and
$d\sim\lambda_\parallel\ll{l_0}$
in dimensional variables.
Then the effective adiabatic and semiclassical
Hamiltonians, as well as the corrections to them,
coincide, and to find the leading term of the asymptotics
of the wave function,
the complete description of the effective Hamiltonian
and the first correction is required.

{\bf b)} In this case,
which is naturally called the {\it ``medium-wave" regime},
$\mu\ll h\ll 1$,
the ``longitudinal" energy of the mode
is significantly less than that of  the ``transverse" mode
and
$d<\lambda_\parallel\leq\sqrt{l_0d}$
in dimensional variables.
Then, expanding the correction $L^\nu_1$
with respect to the parameter~$\mu/h$,
for the symbol of the operator,
we obtain
$$L^\nu = \left(\frac{(p^h)^2}{2}+v^1_{\rm eff}
+\frac {h^2}{\mu}L_1^\nu(x,0)\right)
+h\frac{\pa L_1^\nu}{\pa p}(x,0)p^h
+h\cdot{O}(\frac{\mu}{h}).$$
This implies that the nondifferential part
$\frac{h^2}{\mu}L^\nu_1(x,0)$
of the first adiabatic correction
in the expansion of the operator $\hat L^\nu_1$
can be transferred
into the semiclassical effective Hamiltonian.
This is clearly seen in the case $h =\sqrt{\mu}$.
Then the semiclassical effective Hamiltonian becomes
equal to $\frac{(p^h)^2}{2}+v^1_{\rm eff}+L^\nu_1(x,0)$.
Moreover, for $h^2\gg \mu$,
the term $\frac{h^2}{\mu}L^\nu_1(x,0)$
can play the determining role.
Then an argument similar to that in item~2.6
shows that this term ``accelerates"
the particle in the longitudinal direction
so that the characteristic longitudinal momentum
in dimensionless variables takes the value~$\sqrt{\mu}$.
In other words, in this case,
for the parameter~$h$ we must take
the parameter $\sqrt{\mu}$,
and we return to the situation considered above,
but with $v^1_{\rm eff}$ multiplied by a small value.
Clearly, if $L^\nu_1(x,0)=0$,
then the above argument is meaningless.
But, as we shall see later,
such a term appears in nanotubes
both due to their geometry
and due to the external electromagnetic field.
In this case,
there arise some additional parameters,
e.g., field amplitudes,
and these parameters can effectively decrease
the value of $L^\nu_1(x,0)$
and thus compensate the increase caused by
the parameter $h^2/\mu$.
We also note that the terms in the operator $\hat L^\nu_1$
containing the second- and higher-order derivatives
(corresponding to higher powers in the expansion of $L^\nu_1$
with respect to the variable~$p^h$)
can be omitted in the calculations of the leading term
of the semiclassical asymptotic solution,
although, of course,
these terms do not decrease the accuracy.

{\bf Remark.}
If the main part of the Hamiltonian is quadratic with respect to
the momenta, the Hamiltonian preserves its form.
In the other examples considered above
(e.g., for equations with rapidly oscillating coefficients,
for waves in picnocline),
the leading part of the adiabatic Hamiltonian $L_0$
depends on the momentum $p$. Therefore,
this expansion changes the structure of the leading part.
For example,  in the problem with rapidly oscillating
coefficients (electron waves in crystals),
$L_0$ is replaced by its expansion with respect to $p$
in a more complex way (non quadratically).
Moreover,
this expansion, as a rule, begins with terms quadratic with
respect to~$p$.
The coefficient before $p^2$ is inversely proportional
to the {\it effective mass} (see \cite{LandLif}).

{\bf c)} If the parameter $h\sim 1$,
then the semiclassical approximation cannot be used,
and the wave functions oscillate, if at all,
rather slowly and in the dimensional variables
$\lambda_\parallel\sim l_0$.
According to the above,
this situation is possible only if
$L^\nu_1(x,0)\equiv 0$.
But the adiabatic approximation works,
and from (\ref{RedSchredEx3})
one can easily derive the equation for
the leading term of the asymptotics
of the (smooth) wave function $\psi(x,t)$
on the waveguide axis.
For this, it suffices to set $h=1$
in~(\ref{RedSchredEx3})
and then to let $\mu\to0$.
As the result of this passage to the limit,
which is naturally called the {\it long-wave approximation\/},
we obtain the equation
$$
i\frac{\pa \psi}{\pa t}=
\left[\frac{1}{2}\frac{\pa^2}{\pa x^2}+v^1_{\rm eff}
-i\frac{\pa L^\nu_1}{\pa p }(x,0)\frac{\pa}{\pa x}
+ L^\nu_2(x,0)\right]\psi.
$$

{\bf Remark.} As already noted for the Helmholtz operator
in plane one-mode waveguides,
such an equation was first obtained in~\cite{MaslovWaveGuide},
where, in particular,
it was proved that one can organize
a single bound state in the waveguide
by choosing an appropriate curvature of the waveguide.
An equation similar to~\eqref{lw}
and several consequences of it
were obtained in~\cite{JensenKoppe,daCosta1,daCosta2,EM1,EM2,SchusterJaffe,Exner,LinJaffe}.
We also note that the equations of this type are close to the equations obtained as the result of averaging or homogenization on the theory of wave process in the media with rapidly varying characteristics
\cite{BahvPanas,Bensoussan,MarchKhrus,Jikov}.

{\bf d)} Finally, we can consider the case in which
$\mu\gg h$ or, in dimensional variables,
$\lambda_\parallel \ll \lambda_\perp \ll l_0$.
This case is naturally called the ``ultrashort-wave" regime.
For Eq.~(\ref{RedSchredEx3}) to be meaningful,
it is necessary to impose additional constraints
on the behavior of the functions $L^\nu_j$
in the variable~$p$.
We consider only the case for which
$L^\nu_j$ are polynomials of a degree not exceeding~$2$
with respect to~$p$.
Then it is easy to show that
the semiclassical approximation for any $\mu\gg h$
can be applied to Eq.~(\ref{RedSchredEx3}).
However,
this is not sufficient for reconstructing
the asymptotic solution of the original Schr\"odinger equation
in the waveguide
from the function $\psi(x,t)$
by formula~(\ref{PsiIn}).
For example,
if the first correction~$\chi^\nu_1$
in the expansion of the symbol of the intertwining operator
$\hat \chi^\nu$ in formula~(\ref{expchi})
depends linearly on~$p$,
$\mu\geq h^2$,
and a rapidly oscillating function of the form~(\ref{wkb})
is taken to be~$\psi$,
then
the function $\mu\hat \chi_1 \psi$ turns out to be not small.
In this sense,
the expansion of the operator $\hat\chi$
in powers of~$\mu$ is not an asymptotic expansion.
For this reason, as we shall see below,
the ultrashort-wave approximation for curved nanotubes
can be used for the case in which $h^2\ll\mu\ll h$.
We can also note that,
in the ultrashort-wave case,
the actual effective potential $v^1_{\rm {eff}}$
is small so that it can be transferred
into $L^\nu_1$ or even omitted at all.
Then the semiclassical effective Hamiltonian
coincides with the Hamiltonian of a free particle,
and hence, in this case, the semiclassics
is simply the Born approximation.

{\bf Remark.}
It should be remembered that,
in concrete calculations,
the above classification
(in the parameters~$\mu$ and~$h$)
must be made more precise,
which is related to the values of both
the external fields and the crystal field,
as well as the relations between them.
Of course, in this case,
the corrections can also be included
into the leading part of the symbol
(in the classical Hamiltonian),
which, however, can unnecessarily complicate
the procedure for constructing the asymptotic solutions.
On the other hand, as already noted,
in a real situation,
each parameter is a concrete small number,
and hence the further detailing
of how the effective Hamiltonian and corrections to it
depend on the relation between the parameters~$\mu$ and~$h$
has an academic, rather than practical, character.
Taking this consideration into account,
it is convenient,
from the mathematical viewpoint,
to fix the relations between the parameters~$\mu$ and~$h$,
assuming for the respective regimes
that
${\bf a)}\,h=\mu,\quad {\bf b)}\,h=\sqrt{\mu},
\quad{\bf c)}\, h=1,\quad{\bf d)}\,h=\mu^{3/2}$
and including the ``remaining" parts
of the relations between~$\mu$ and~$h$
into the coefficients of the equation
(such as the strengths of electric and magnetic fields,
curvature, etc.).

The suggested classification can be used in the general situation. But in some cases some regimes (like ultrashort wave region or long wave region) does not exist. Anyway the question of existence of solutions to a certain fixed $h$ should be solved individually.

{\bf The number of terms in the intertwining operator.}
We assume that we use the dimensionless time~$t$
consistent with the semiclassical parameter~$h$
(see subsection~{\bf Characteristic time scale
and the reduced equation consistent with this scale}).
Suppose that we have preserved~$N$ terms
in the expansion of the symbols
of the operators~$\hat \chi^\nu$ and $\hat L^\nu$
determined by the coefficients~$\chi^\nu_j$ and~$L^\nu_j$.
Suppose also that we have constructed a function $\psi^\nu$
satisfying the reduced equation (\ref{RedSchredEx3})
with accuracy up to a discrepancy~$f$.
It follows from the formulas of \S 3 that,
for appropriately defined~$\chi_j^\nu$ and~$L^\nu_j$,
the substitution of the function
$\Psi=\hat \chi^\nu \psi^\nu$
into the original equation gives
$$ih\frac{\pa \Psi}{\pa t}
-\frac{h^2} {\mu^{2}}\hat {\cal H}\Psi
={h^2} \mu^{N-1}\hat F\psi^\nu+\hat \chi^\nu f.$$
Here $\hat F$ is, in general, a pseudodifferential operator
that does not change
(as well as the operator~$\hat \chi^\nu$)
the order of the functions~$\psi^\nu$
with respect to the parameter~$h$
if the functions oscillate
with the characteristic wavelength not less than~$h$.
Applying the estimate~(\ref{Est2}) to this equation,
we readily obtain the following conclusion:
the function $\Psi$ differs from the exact solution
by a value of the order $O(h^\delta)$, $\delta>0$,
for $h\ll 1$
or of the order $O(\mu)$ for $h=1$
if $N\geq 1$ and the discrepancy~$f$
is equal to $O(h^{\delta+1})$ for $h\ll 1$
or to $O(\mu)$ for $h=1$.
Thus, the minimal reasonable number of terms
in the expansion of the operator~$\hat\chi^\nu$
in constructing the semiclassical asymptotics
is equal to~$2$
(i.e., we must consider the zeroth- and first-order terms).
But if we are interested in the long-wave approximation
(i.e., in the case $h=1$), then, obviously,
in the expansion of $\hat \chi^\nu$
we must consider three terms of the expansion
(i.e., $N=2$).
In this case, it suffices
to solve the reduced equation (\ref{RedSchredEx3})
up to~$O(\mu)$.

The problem of calculating the symbols
of the operators~$\hat \chi^\nu$ and $\hat L^\nu$
is close to the problems in perturbation theory
for operators with discrete spectrum
(in particular, matrices),
and the function $L^\nu$ is similar to an eigenvalue,
while~$\chi^\nu$ is similar to an eigenfunction.
The terms of the expansion
of the symbols~$\chi^\nu$ and~$L^\nu$
are calculated successively,
but the explicit calculation of~$L_j^\nu$
precedes the calculation of~$\chi_j^\nu$
and is based on the fact that~$\chi_j^\nu$ exists.
On the other hand,
the leading term of the asymptotics
is already determined by~$\chi_0^\nu$
(and, naturally, by the function $\psi^\nu$).
Thus,
in the construction of the semiclassical asymptotics
in the minimal reasonable approximation,
explicit formulas are required only
for $L_0^\nu,\,L_1^\nu$ and $\chi_0^\nu$,
while in the construction of the long-wave asymptotics,
explicit formulas are required
for $L_0^\nu,\,L_1^\nu,L_2^\nu$ and
$\chi^\nu_0,\chi^\nu_1$.
Moreover,
as was already discussed in item~2.9,
to obtain the medium-wave and long-wave approximations
($h\gg \mu$),
for~$L^\nu_1$,
it suffices to calculate this function
and its first-order derivative for $p=0$,
while for~$L^\nu_2$,
it suffices to calculate this function
for $p=0$.
This fact is very important
for deriving explicit formulas.
It should be noted that,
although the terms
$\mu \hat \chi^\nu_1 \psi^\nu$ for $h\ll1$
and $(\mu\hat \chi^\nu_1+\mu^2 \hat \chi^\nu_2)\psi^\nu$
for $h\sim 1$
are only corrections to the leading term~$\hat\chi_0\psi^\nu$
of the asymptotic expansion,
it can be described correctly
only if the existence of such corrections is guaranteed,
while the discrepancy obtained
by the direct substitution of the function
$\hat \chi^\nu_0 \psi^\nu$ into the original equation
is not sufficient to prove
that the difference between $\hat \chi^\nu_0 \psi^\nu$
and the exact solution is small.
Here we have the following distinction
from the standard semiclassical approximation
for the scalar Schr\"odinger equation:
in the latter case,
the substitution of the leading term of the asymptotics
into the original equation readily gives
the desired small discrepancy~$O(h^2)$.

{\bf Essential part of the quantum Hamiltonian
for longitudinal motion.}
Thus, it follows from the analysis performed
in the preceding items that,
in all cases {\bf a)}--{\bf d)},
the leading term of the asymptotic expansion of~$\psi^\nu$
(as well as of~$\Psi $)
is completely determined
(sometimes, with excessive accuracy)
by the quantum Hamiltonian
$$\hat {\cal L}^\nu=\frac{1}{2}\left(-ih\frac{\pa}{\pa x}\right)^2+
v^1_{\rm eff}+
\frac{h^2}{\mu}\left[L^\nu_1\Bigl(\x,\frac{\mu}{h}(\p)\Bigr)+
\mu L^\nu_2(x,0)\right].$$
This Hamiltonian is naturally called
the {\it essential part of the Hamiltonian of longitudinal motion
on the $\nu$-th subband of size quantization}.

{\bf Semiclassical splitting of adiabatic terms.}
For $h\gg\mu$,
the correction $L_1(x,0)$ can change the effective Hamiltonians.
This change can be essentially important
for the case in which the term is degenerate
or (and) the original problem is a vector problem
(i.e., the original quantum Hamiltonian is a matrix
Hamiltonian).
Let us consider this situation in more detail.
Because the original problem is self-adjoint,
the matrix $L_1(x,0)$ is Hermitian.
We assume that all its eigenvalues
$\lambda_1(x), \ldots, \lambda_r(x)$ are distinct
and, along with the eigenvectors,
smoothly depend on~$x$.
Moreover, to avoid any renormalization of energy,
for simplicity, we assume
that $\lambda_j(x)$ varies along the waveguide
and $\lambda_j=0$ at the beginning of the waveguide.
Since any adiabatic term is degenerate,
there is a great ambiguity in the choice of
vector functions $\chi^\nu_j, j=1,\ldots,r$,
and any variation in their choice naturally
leads to a change of the matrix~$L^\nu_1$.
We choose the vector functions $\chi^\nu_j, j=1,\ldots,r$,
so that the matrix $L^\nu_1(x,0)$ be diagonal.
If $1\gg h\gg\mu$
(we have the ``medium-wave" approximation),
then the expressions $\frac{h^2}{\mu}\lambda_j(x)$
must be added to the semiclassical Hamiltonian.
Following an argument similar to that in item~2.6,
we conclude that the parameter~$h$ must satisfy
the inequality $\sqrt{\mu}\geq h\gg\mu$.
We see that the semiclassical Hamiltonians (terms)
$\frac{1}{2}(p^h)^2+v^1_{\rm{eff}}+\frac{h^2}{\mu}\lambda_{j}$
become different for different~$j$
and thus the ``semiclassical" separation of ``adiabatic"
terms occurs.
The value of this splitting depends
on the number $\frac{h^2}{\mu}$
and attains its maximum at $h\sim \sqrt{\mu}$.
In the first approximation
with respect to the parameter~$\frac{h^2}{\mu}$,
the states corresponding to different semiclassical Hamiltonians
do not interact.
Thus, inside an adiabatic term,
we can asymptotically diagonalize
the system for the vector function~$\psi^\nu$.
Although this asymptotic fact turns out to be valid
for any values of~$h$ from the above interval,
it is clear that,
for values of~$h$ close to~$\mu$,
the subsequent terms of the expansion can be large
so that the above representation
of such an asymptotic ``diagonalization"
becomes meaningless.
Therefore,
the situation with~$h$ close to~$\mu$
must be considered as a situation with a degenerate term,
and it is more natural to leave the term
$\frac{h^2}{\mu}\lambda_{j}$
in the original correction $\frac{h^2}{\mu}L^\nu_1$,
which allows one more adequately
to take into account
the interaction between the states
inside an adiabatic term.
This reasoning concerns the case of
the matrix correction~$L_1^\nu$.
In the scalar situation,
the term $\frac{h^2}{\mu}L^\nu_1(x,0)$
can always be included
into the effective Hamiltonian,
but, in general,
this does not hold in the vector case.
We have no opportunity to discuss
the problems touched upon in this section
in more detail.
We only point out that all phenomena
mentioned in this and in the preceding items
are completely determined by the essential part
of the matrix analog of the Hamiltonian
of longitudinal motion on the $\nu$-th subband
of size quantization.

\subsection {Essential parts of the effective Hamiltonians
for longitudinal motions in nanotubes for different regimes.}

The goal of this subsection is to realize the ideas from the
previous subsection concerning the problem of quantum waves
in nanotubes with spin taken into account.

As already noted,
the ``rapidity" of the longitudinal mode
is determined by the parameter~$h$.
For $h\ll 1$, the corresponding states
(in particular, the levels of longitudinal motion)
are described by semiclassical asymptotics.
Since the strong electric or magnetic field
increases the longitudinal energy of a particle
while it passes through the tube,
the parameter~$h$ cannot be chosen arbitrarily
and must be consistent
with the strengths of the external fields
contained in the problem.
For mathematical rigor,
we must assume that there is a functional dependence
between the parameters~$\mu$, $h$
and the strengths of the fields.
From the physical viewpoint,
it is natural to speak about the value
of different terms in (\ref{HamAdiabatic})
in dependence on the ``rapidity" of the longitudinal mode
and the strengths of external fields.
For a solution with wavelength $\sim h$ to exist,
it is necessary
that the work of the effective field
$E_{\rm eff}(x)
=-\pa v_{\rm ext}/\pa x-\pa\varepsilon_\perp(x)/\pa x-
\left< \pa_x{\bf R},\pa {\bf A}_0/\pa t\right>$
along the tube do not exceed, in the order of magnitude,
the characteristic kinetic energy of the longitudinal motion:
$\varepsilon_\parallel\sim\mu^2/h^2$.
This implies the following constraints
on the external fields
and the ``oscillation" in the dimensions of the section.
We assume that the potential of the external electric field
is equal to zero at one of the tube ends;
thus, $v_{\rm ext}({\bf R}(x),t)$
will be equal to the work of the field along the tube.
This implies that the external potential
in dimensionless variables has the order of $\mu^2/h^2$.
We introduce the functions $v^1_{\rm ext}(x,t)$ and $\lambda^\nu(x)$
determined by the relations
$v_{\rm ext}({\bf R}(x),t)=(\mu^2/h^2)v^1_{\rm ext}(x,t)$,
$\varepsilon_\perp^\nu(x)=
\varepsilon_\perp^\nu(0)+(\mu^2/h^2)\lambda^\nu(x)$ and we assume that
$v^1_{\rm ext}(x,t)$, $\lambda^\nu(x)$ take values that do not
exceed unity with respect to the parameter.

In what follows, for simplicity, we assume that
the function $v^1_{\rm ext}$, in fact, smoothly depends on $t'$,
and we write $v^1_{\rm ext}(x,t')$.
The constraint on the effective field
implies a constraint on the character
of the time-dependence of the magnetic field.
We assume that ${\bf H}={\bf H}_0+(\mu/h){\bf H}_1(t')$.

We replace the ``adiabatic" momentum operator
$\hat p=-i\mu\pa/\pa x$ by the ``dynamic" operator
$\hat p^h=-ih\pa/\pa x$,
obtain $\hat p=(\mu/h)\hat p^h$,
and then
divide the adiabatic Hamiltonian by~$\mu^2/h^2$
to ``compensate" the redefined momentum.
This leads to the redefined classical momentum
$p=(\mu/h)p^h$ and, by~(\ref{HamAdiabatic}),
to the following formula
for the symbol ${\cal L}^\nu(p^h,x,t',\mu,h)$
of the {\it essential part of the quantum effective Hamiltonian
$\hat {\cal L}^\nu$ of the longitudinal motion}:
\begin{gather} \nonumber{\cal L}^\nu
=
\left[\frac{(p^h)^2}{2}
+v^1_{\rm ext}({\bf R}(x),t')+\phi(x,t')
+\lambda^\nu(x)\right]
E_r\otimes E_s+
\\
\nonumber
+\frac{h^2}{\mu}\left[-E_r\otimes\frac{1}{2}
\langle\boldsymbol{\sigma},{\bf H}\rangle
-\frac{1}{2}\langle \pa_x{\bf R},{\bf H}\rangle\Lambda
\otimes E_s+L_{sy}\right]
+h\Bigl[(\pa_x\Phi)\Lambda
-\langle{\bf Y}_\perp,{\bf H}\rangle \Bigr]\otimes E_s p^h+
\\
+\mu\left\langle {\bf Y}, \nabla v_{\rm ext}^1
+ \frac{\pa {\bf A}_0^1}{\pa t'}+
k(p^h)^2{\bf n}\right\rangle
-\frac{h^2k(x)^2}{8}\ E_r\otimes E_s,
\label{HamSemiClass}
\end{gather}
where ${\bf A}_0^1=(1/2)[{\bf R},{\bf H}_1(t')],
\phi(x,t')=(1/2)\int_{x^*}^x
\left< \pa_x{\bf R}(x'),
[{\bf R}(x'),{\pa {\bf H}_1(t')}/{\pa t'}]\right>dx'$.
Omitting the primes, we obtain the desired reduced equation on
the subregion of dimensional quantization in the form:
\begin{equation}
\label{SemiRedEq}
ih\frac{\pa \psi^\nu}{\pa t}=\hat {\cal L}^\nu \psi^\nu, \qquad
\hat {\cal L}^\nu={\cal L}^\nu(\ph,\x,t,\mu,h).
\end{equation}
Along with formulas~(\ref{newrepr}) and~(\ref{chi0}),
this equation, for different relations between~$\mu$ and~$h$,
determines the leading term of the asymptotic solutions
of the original equation~(\ref{InEq}).
Sometimes, several terms in~(\ref{HamSemiClass}) can be omitted,
and several terms from the ``adiabatic" correction
(sometimes, matrix terms)
can be transferred into the leading part
of the ``semiclassical" effective Hamiltonian
determining, in particular,
the dynamics of the classical motion of a particle
in a thin tube or,
if the terminology introduced in~\cite{MaslovNonstandard,MaslovCharactsPseododiff}
is used, determining
the {\it nonstandard characteristics\/}
of the original equation~(\ref{InEq}).
We describe this ``transfer"
and the corresponding classical systems
in the next subsection.

{\bf Equations of classical mechanics
in nanotubes with spin and the term multiplicity
taken into account.}
Below we perform different expansions
of the function~(\ref{HamSemiClass}).
For $h\ll 1$, we set
${\cal L}^\nu={\cal L}^\nu_0+h{\cal L}^\nu_1+\ldots$,
where ${\cal L}^\nu_0$ denotes the terms larger than~$h$
and $h{\cal L}^\nu_1$ denotes the terms $\sim h$.
According to \cite{MaslovNonstandard,MaslovCharactsPseododiff},
these terms allow one to reconstruct the leading term of the semiclassical
asymptotics completely for $h\ll 1$,
which outside the focal points has the form
of WKB-solutions~$\psi^\nu\approx\exp(iS^\nu(x,t)/h)A^\nu(x,t)$.
Their phases $S^\nu(x,t)$ can be determined by integrating
the one-dimensional Hamiltonian system
\begin{equation}\label{HS}
\dot p^h={\pa H^h_{\rm eff}}/{\pa x},\
\dot x=-{\pa H^h_{\rm eff}}/{\pa p^h}
\end{equation}
with the classical Hamiltonian $H^h_{\rm eff}(p^h,x)$,
which is an eigenvalue of the matrix symbol
${\cal L}^\nu_0$.
If the matrix ${\cal L}^\nu_0$ has distinct eigenvalues,
then the {\it semiclassical splitting of the adiabatic term\/}
occurs, i.e.,
several distinct classical Hamiltonians $H^h_{\rm eff}$
can correspond to the same adiabatic term $H_{\rm eff}$.
The vector part of the asymptotics $\psi^\nu$
can be found from the linear ``polarization" equation
which contains~${\cal L}^\nu_1$.

We shall consider the following situations
corresponding to different relations
between the parameters~$\mu$ and~$h$.

\noindent{\bf a) Short-wave regime:} $h=\mu$.
Then
\begin{gather}\nonumber
{\cal L}^\nu_0=H^h_{\rm eff}E_r\otimes E_s, \qquad
H^h_{\rm eff}=\frac{(p^h)^2}{2}+v^h_{\rm eff},\quad
v^h_{\rm eff}=v_{\rm ext}^1({\bf R}(x),t)+\phi(x,t)
+\lambda^\nu(x),\\\label{sw}
{\cal L}^\nu_1=\left\langle {\bf Y}, \nabla v_{\rm ext}
+ \frac{\pa {\bf A}_0^1}{\pa t}
+ k(p^h)^2{\bf n}\right\rangle\otimes E_s
+ \Bigl[(\pa_x \Phi)\Lambda-\langle {\bf Y}_\perp,{\bf H}\rangle \Bigr]
\otimes E_s p^h+\\\nonumber
+\left[-E_r\otimes\frac{1}{2}\langle\boldsymbol{\sigma},{\bf H}\rangle
-\frac{1}{2}\langle \pa_x{\bf R},{\bf H}\rangle\Lambda
\otimes E_s+L_{sy}\right].
\end{gather}
The Hamiltonian system in this case
is equivalent to the Newton system
$\ddot x=-\pa v^h_{\rm eff}/\pa x$.
By the estimates in items~{\bf 2.8, 2.10},
the leading term of the semiclassical asymptotics,
which is determined by these classical equations,
must give a good approximation in problems
with external fields $|{\bf H}|\lesssim 1$\,T,
$E_{\rm ext}=|\nabla v_{\rm ext}|\lesssim 10^{-3}$\,V/nm.

\noindent{\bf b) Medium-wave regime:} $h=\mu^{1/2}$.
In this case, we have
\begin{gather}
\label{mv}
{\cal L}^\nu_0=\left[\frac{(p^h)^2}{2}
+v_{\rm ext}^1({\bf R}(x),t)+\phi(x,t)+\lambda^\nu(x)\right]
E_r\otimes E_s+W,
\\\nonumber W=
\left[-E_r\otimes\frac{1}{2}\langle\boldsymbol{\sigma},{\bf H}\rangle
-\frac{1}{2}\langle \pa_x{\bf R},{\bf H}\rangle\Lambda
\otimes E_s+L_{sy}\right],\quad
{\cal L}^\nu_1=\Bigl((\pa_x \Phi)\Lambda-
\langle {\bf Y}_\perp,{\bf H}\rangle \Bigr)\otimes E_s p^h.
\end{gather}
Thus, here the symbol ${\cal L}^\nu_0$ is a matrix symbol.
Here the classical effective Hamiltonians
are the eigenvalues of the matrix ${\cal L}^\nu_0$.
Obviously, they can be represented as the sum of
the function ${(p^h)^2}/{2}
+v_{\rm ext}^1({\bf R}(x),t)+\phi(x,t)+\lambda^\nu(x)$
and the eigenvalues of the matrix~$W$
which, in general, depend on~$p^h$.
If they are distinct for all $(p^h,x)$,
then the semiclassical splitting of the adiabatic term
occurs.
Obviously,
these classical Hamiltonians depend on the spin terms;
hence the spin of a particle affects its phase trajectory
as follows:
the energy of longitudinal motion in the tube,
in contrast to the total energy
and the energy of transverse motion,
is comparable with the spin energy.

The eigenvalues of the matrix~$W$ depend on the form
of the transverse section of the tube and on the external fields.
As the simplest example,
we consider the case of
the cylindrically symmetric crystal potential
($y_1=\varrho \cos \phi, y_2=\varrho \sin \phi$)
$v_{\rm int}=v_{\rm int}(\varrho)$, and
$w^\nu=\exp(\pm i \nu\phi)u^\nu(\varrho)$.
In this case,
${\bf Y}=0$,
the eigenvalues of the matrix~$\Lambda$ are $\pm \nu$,
$M^1=M^2=0$, \
$M^0$ is the diagonal matrix with elements
$\pm\kappa \nu$, \
$\kappa=2\pi\int_0^\infty
({\pa v_{\rm int}(\varrho)}/{\pa \varrho})(u^\nu(\varrho))^2\,
d\varrho$,
and the adiabatic term splits into four semiclassical ones
determined
by the classical effective Hamiltonians
$(H_{\rm eff}^h)^{\pm}_{\uparrow\downarrow}=
{(p^h)^2}/{2}+(v^h_{\rm eff})^{\pm}_{\uparrow\downarrow}$,
\begin{gather}
(v^h_{\rm eff})^{\pm}_{\uparrow\downarrow}=
v_{\rm ext}^1({\bf R}(x),t)+\phi(x,t)+\lambda^\nu(x)
\mp\frac{1}{2}\langle \pa_x{\bf R},{\bf H}\rangle \nu
+\sigma_{\uparrow\downarrow}
\left|\frac{1}{2}{\bf H}\pm\nu \mu^{-1}\alpha \kappa
\pa_x{\bf R}\right|,
\label{mwv}
\end{gather}
where $\sigma_{\uparrow\downarrow}=\pm1$.
In this case, we must assume that
$\left|({1}/{2}){\bf H}\pm\mu^{-1}\alpha
\kappa \nu \pa_x{\bf R}\right|\neq 0$.
Otherwise,
the effect of intersection of terms
or of a change in the multiplicity of the characteristics
occurs,
where the standard semiclassical approximation
does not work
(see, e.g.,~\cite{LandLif}).
In this example,
the Hamiltonian system is also equivalent
to the Newton system with the effective potential
$(v^h_{\rm eff})^{\pm}_{\uparrow\downarrow}$.
The semiclassical approximation must work
for external fields
$|{\bf H}|\lesssim 1$\,T,
$E_{\rm ext}=|\nabla v_{\rm ext}|\lesssim 10^{-5}$V/nm.
We also note that Eq.~(\ref{mv})
can be treated as an equation with operator-valued symbol
(its operator symbol is a matrix),
and we can again apply the ``operator" reduction
to this equation.

\noindent{\bf c) Long-wave regime:} $h=1$.
In contrast to the above regimes,
for the ``long-wave" approximation to exist,
it is necessary to impose additional constraints
on the value of the magnetic field.
Formally, this follows from the existence
of a term $\sim h^2/\mu$
in the Hamiltonian~\eqref{HamSemiClass}.
The physical cause consists in the following.
Even without the spin effects taken into account,
the longitudinal and transverse modes are related
due to the interaction between
the magnetic field and the transverse orbital momentum,
which is described by the term
$-(2\mu)^{-1}\langle \pa_x{\bf R},{\bf H}\rangle\Lambda\otimes E_s$.
In this case, the transverse energy
significantly exceeds the longitudinal energy;
hence the dynamics of longitudinal motion
becomes very sensitive to small variations
in the transverse energy.
In turn,
the transverse energy changes because of the variation
in the magnetic field flux through the tube section
due to a change in the angle between the plane
of the tube transverse section
and the vector of the magnetic field ${\bf H}$.
For the magnetic fields $\sim 1$\,T chosen above,
this interaction accelerates the particle to energies
that are not consistent with the parameter~$h$.
To avoid the acceleration,
we must take into account that
$\varepsilon_\perp^\nu(x)
=\varepsilon_\perp^\nu(0)+\mu^2\lambda^\nu(x)$,
and ${\bf H}_0=0,\ {\bf H}=\mu{\bf H}_1$.
This means that we consider magnetic fields
$\lesssim 10^{-2}$\,T.
We also assume that the dimensionless constant
of spin-orbit interaction is $\alpha\sim\mu^2$.
This implies that the leading part of the Hamiltonian
contains only the terms
$L^\nu_2(0,x)|_{{\bf H}=0,\ v_{\rm ext}=0,\ \alpha=0}$.
Subsequent calculations of this term
(see item~A3 in the Appendix)
lead to the last term in~(\ref{HamSemiClass}).
As a result, we obtain the limit reduced equation of the form
\begin{gather}
\Biggl[\left(-\frac{1}{2}\frac{\pa^2}{\pa x^2}+
v_{\rm ext}^1(x)+\phi(x,t)\right)E_r\otimes E_s+
(\pa_x \Phi)\Lambda\otimes E_s
\left(-i\frac{\pa}{\pa x}\right)+W\Biggr]\psi^\nu=0
\nonumber\\
W=\left(\lambda^\nu(x)-\frac{k^2}{8}\right)E_r\otimes E_s
-\frac{1}{2}\langle \pa_x{\bf R},{\bf H}_1\rangle\Lambda\otimes E_s
-E_r\otimes\frac{1}{2}\langle\boldsymbol{\sigma},{\bf H}_1\rangle
+\mu^{-1}L_{sy}.
\label{lw}
\end{gather}
Here the semiclassical approximation cannot be used
and any information about the solutions of the reduced equation
(more precisely, about the systems of equations)
can be obtained either by  general qualitative methods
or by numerical integration
\cite{JensenKoppe,daCosta1,daCosta2,EM1,EM2,SchusterJaffe,Exner,LinJaffe,Geyler}.
As already noted for the Helmholtz operator
in plane one-mode waveguides,
such an equation was first obtained in~\cite{MaslovWaveGuide},
where, in particular,
it was proved that one can organize
a single bound state in the waveguide
by choosing an appropriate curvature of the waveguide.
An equation similar to~\eqref{lw}
and several consequences of it
were obtained in~\cite{EM1,EM2}.
Eq.~(\ref{lw}) works
in the case where the correction $L^\nu_1$ is small.
This imposes several constraints on the value
of the constant-in-time component
of the magnetic field $H_0\sim\mu$
and on the constant of the spin-orbit interaction
$\alpha\sim\mu^2$.

\noindent{\bf d) Ultrashort-wave regime:}
$\mu^{3/2}\ll h\ll \mu$.
In the case of ultrashort-wave approximation,
the external magnetic and electric fields
can be stronger than those considered above.
Namely,
$v_{\rm ext}\sim \mu^2/h^2$,
$|{\bf H}(t)|\sim \mu/h$, $|{\bf H}|\lesssim 10$\,T.
Introducing the notation
$v_{\rm ext}=(\mu^2/h^2)v^1_{\rm ext}$,
${\bf H}(t)=(\mu/h){\bf H}_1(t)$
and omitting the primes, we obtain
\begin{gather}
{\cal L}^\nu_0=H_{\rm eff}^h \ E_r\otimes E_s
+ (h^2\mu^{-2})\varepsilon_\perp^\nu(x)
+ \mu\left(-\langle {\bf Y}_\perp,{\bf H}\rangle p^h
+ \left\langle{\bf Y}, \nabla v_{\rm ext}
+ \frac{\pa {\bf A}_0}{\pa t}
+ k(p^h)^2{\bf n}\right\rangle\right),
\nonumber\\
{\cal L}^\nu_1=\Bigl(p^h (\pa_x \Phi)
-1/2\langle \pa_x{\bf R},{\bf H}_1 \rangle\Bigr)\Lambda
-1/2\langle \boldsymbol{\sigma},{\bf H}_1\rangle
+\mu^{-1}\alpha\Bigl(M^1\otimes\langle\boldsymbol{\sigma},
{\bf n}_1\rangle p^h
+M^2\otimes\langle\boldsymbol{\sigma},{\bf n}_2\rangle p^h\Bigr),
\end{gather}
where $H_{\rm eff}^h=(p^h)^2/2+v_{\rm ext}(x)+\phi$.
Although ${\cal L}^\nu_0$ is a matrix symbol in this case,
its leading part is  $H_{\rm eff}^h \ E_r\otimes E_s$.
Nevertheless, the matrix correction $\sim\mu>h$
can lead to the splitting of the adiabatic term
and, in particular, to additional terms in the phase
of the wave function $\sim\mu/h,\ \mu^2/h, \ldots$,
which are, in general, different for each of the states
contained inside the adiabatic term.

\subsection{Examples of asymptotic solutions for some problems
in nanotubes}
As we mentioned above, here we have no opportunity to describe
in detail asymptotic solutions of the effective equation of
adiabatic motion in nanotubes and particularly to discuss
concrete physical conclusions.
This discussion requires a special publication
(see \cite{TMF-BDT2}) and probably not a single one.
Here we only want very briefly to outline the
structure of semiclassical asymptotics
for some natural problems.

{\bf Wave trains propagation.}
It is natural to formulate the problem of the wave train propagation for the reduced equation \eqref{SemiRedEq}. We consider the general case when ${\cal L}^\nu_0$ and ${\cal L}^\nu_1$ are matrices. Let $H^h_{\rm eff}$ be the certain eigenvalue of the matrix ${\cal L}^\nu_0$. If $H^h_{\rm eff}$ is degenerate, to restrict on the subspace corresponding to it, we can apply the reduction of \S 3 (by parameter $h$ instead of $\mu$) once again. So we have to restrict the operator ${\cal L}^\nu_1$ to the subspace corresponding to the eigenvalue $H^h_{\rm eff}$. After the reduction our problem has the following form: the leading part of the matrix Hamiltonian is proportional to unity $r'\times r'$-matrix with coefficient $H^h_{\rm eff}$, where $r'$ is the multiplicity of degeneracy of eigenvalue $H^h_{\rm eff}$, and ${\cal L}^\nu_1$ is a $r'\times r'$-matrix. Thus we always can reduce the problem of the construction of semiclassical solutions to the problem with the leading term proportional to unity matrix. We construct asymptotic solution assuming that the initial problem is reduced to this form. In order to not overload the notation we will write ${\cal L}^\nu_1$ for the matrix restricted to the proper subspace corresponding to eigenvalue $H^h_{\rm eff}$.

Let
$\psi^\nu(x,0)=\psi_0(x)=\exp\left(\frac{iS_0(x)}{h}\right)\varphi_0(x)$,
where the phase $S_0(x)$ is a smooth function and $\varphi_0(x)$ is $2r$-dimensional smooth vector function of $x\in\mathbb{R}$ with compact support ${\cal M}$. Then in the semiclassical approximation the solution to \eqref{SemiRedEq} is determined by solution to the Hamiltonian system \eqref{HS} with Hamiltonian $H^h_{\rm eff}=\frac{1}{2}p^h+v^h_{\rm eff}$, where $v^h_{\rm eff}$ depending on regimes of longitudinal motion are defined by formulas \eqref{sw},\eqref{mwv}
with initial conditions $ p^h(0)=\pa S_0/\pa x(x_0),\,x(0)=x_0$. Denote its solution $p^h=\pi(t,x_0),\,x=\xi(t,x_0)$.
Suppose that $J(x_0,t)=\frac{d\xi}{d x_0}\neq 0$ for $t\in[0,t^*]$ and $x_0\in {\cal M}$. Then the equation $x=\xi(x_0,t)$ is uniquely solvable, $x_0=X_0(x,t)\in {\cal M}$, and, for $t\in[0,t^*]$, the asymptotic solution to the Cauchy problem for system \eqref{SemiRedEq} with initial condition $\psi|_{t=0}=\psi^0(x,h)$ has the form:
\begin{gather}
\psi(x,t)=\left[\exp\left(\frac{iS(x_0,t)}{h}\right)
\left(\frac{\varphi(x_0,t)}{\sqrt{J(x_0,t)}}+O(h)\right)\right]_{x_0=X_0(x,t)}, \\
S(x_0,t)=S_0(x_0)+\int_0^t\left(\pi_i\frac{\pa H^h_{\rm eff}(\pi,\xi,t)}{\pa \pi_i}-
H^h_{\rm eff}(\pi,\xi,t)\right)dt. 
\end{gather}
Function $\varphi(x_0,t)$ satisfies the following equation:
\begin{gather}
\frac{d\varphi}{dt}+i{\cal L}_1\varphi=0
\label{phi-evolution}
\end{gather}
with initial condition $\varphi(x_0,0)=\varphi_0(x_0)$.

At $t>t^*$ the asymptotics of the solution is specified with the use of the Maslov canonical operator $K_{\Lambda_t}$ (see \cite{Vainberg,MaslovFedoryuk}) on the curves $\Lambda^1_t=\{x=\xi(x_0,t),p=\pi(x_0,t)\}$: $\psi={\bf K}_{\Lambda_t}\psi_0$.

{\bf Remark.} In the case of a finite effective potential the effect of splitting of incoming wave train into two space components can appear. They are partially reflected and partially transmitted, containing harmonics $E<\max{v^h_{\rm eff}}$ and and $E>\max{v^h_{\rm eff}}$, respectively.

{\bf Plane wave scattering.}
Consider the nanotube having the following structure. For $x<x_-$ and $x>x_+$ ($x_\pm={\rm const}$), this tube has rectilinear axis, constant torsion angles $\Phi^-$ and $\Phi^+$, and the contraction coefficients $D_{-}$ and $D_{+}$ at the ends. Suppose that ${\bf H}=0$, $v_{\rm ext}$ and $v_{\rm int}$ do not depend on $t$, $v_{\rm ext}({\bf R }(x))=
\{v^{-}_{\rm ext}$ for $x<x_{-},\ v^{+}_{\rm ext}$ for $x>x_{+}\}$, $v_{\rm int}(x,y)=
\{v^{-}_{\rm int}(y)$ for $x<x_{-},\ v^{+}_{\rm int}(y)$ for $x>x_{+}\}$. Then for $x<x_{-}$ and $E>v_{\rm ext}^{-}+\varepsilon_{-}^\nu$, the system has a solution of the form $\exp((-iEt+ip_{-}x)/h)g^{\nu k}_{-}(y),\ k=1,\ldots,2r$, which represents the plane wave propagating along the tube axis with vector (spinor) amplitude $g^{\nu k}_{-}(y)$.
Vector $g^{\nu k}_{-}(y)$ and $\varepsilon_{-}^\nu$ are the eigenfunction and eigenvalue of the following problem, correspondingly:
\begin{gather}
\left(-\frac{1}{2}\frac{\pa^2 g}{\pa y^2}+
v_{\rm int}^{-}(y)+
\alpha\langle\boldsymbol{\sigma},\hat{\bf M}\rangle
\right)g^{\nu k}_{-}(y)=
\varepsilon_{-}^\nu g^{\nu k}_{-}(y), \qquad
{\bf n}_0=\pa_x{\bf R}, \\
\hat{\bf M}=\pa_x{\bf R}
\Bigl((\pa_1 v_{\rm int})\pa_2-
(\pa_2 v_{\rm int})\pa_1\Bigr)+
{\bf n}_1(\pa_2 v_{\rm int}) p -
{\bf n}_2(\pa_1 v_{\rm int}) p.
\end{gather}

{\bf Remark.} Note that eigenfunction $g^{\nu k}$ is not the product $w^\nu_j(y)\otimes\eta$ of  ``pure state" inside the term $w^\nu_j(y)$ and ``pure spin state" $\eta_k$ ($\eta_k$ doesn't depend on $y$). It means that we can't separate spin states and states inside one term. Since the parameter $\alpha\ll 1$, to construct $g^{\nu k}$ one can use perturbation theory. To construct solutions in the case when $v_{\rm int}=y_1^2+y_2^2$ (parabolic confinement) one has to use second order of perturbation theory.

As exact solutions ``at infinity" are not products of ``pure states," we need to expand them with respect to the basis $w^\nu_j(y)\otimes\eta^k$. We obtain:
\begin{gather}
g^{\nu k}_{-}(y)=\|w^\nu_1(y),\ldots,w^\nu_r(y)\|\otimes E_s\varphi^{-}, \qquad
\varphi^{-}=\{\varphi^{-}_1,\ldots,\varphi^{-}_{2r}\}^T.
\end{gather}
The evolution of ``initial scattering data" for reduced equation is determined by the following transport equation:
\begin{gather}
\frac{d\varphi}{dt}+i{\cal L}^\nu_1\varphi=0, \qquad
\varphi\to \varphi^{-}, \quad t\to-\infty.
\end{gather}

The final scattering data for reduced equation is $\varphi^+=\lim_{t\to\infty}\varphi(t)$. So we obtain the final scattering data for original problem in the form:
\begin{gather}
g^{\nu k'}_{+}(y)=\|w^\nu_1(y),\ldots,w^\nu_r(y)\|\otimes E_s\varphi^{+}.
\end{gather}

In the semiclassical approximation this problem has
$2r$-dimensional family of asymptotic solutions of the form $\psi(x,E)\simeq {\bf K}_{\Lambda^1(E)}\{\varphi(t)\}$, where ${\bf K}_{\Lambda^1(E)}$ is the Maslov canonical operator on the nonclosed curve $\Lambda^1(E)=\left\{H^h_{\rm eff}=E:\ p^h=p^h(t),\ x=x(t)\right\}$, $t$ is the proper time (parameter on $\Lambda^1(E)$): $dx/dt=\pm\sqrt{2(E-v^h_{\rm eff}(x))}$. Let $E>v^h_{\rm eff}(x)$ for each $x$. Then with an accuracy exponential with respect to $h\to 0$, we have a passage of the incident wave above the barrier; as $x\to\pm\infty$
\begin{gather}
\psi(x,E,h)\to\frac{1}{\sqrt{p_\pm}}
\exp\left(\frac{i}{h}p_{\pm}x\right)\varphi^{\pm}, \qquad
p_\pm=\sqrt{2(E-v_{\rm ext}^{-}-\varepsilon^\nu_{-})}.
\end{gather}

If $E<\max{v^h_{\rm eff}(x)}$ at some points of the tube axis, then the incident wave is reflected off the barrier with an accuracy exponential with respect to $h$. In the domain $x<x_f(E)$, where $x_f$ is the rotation point at the energy level $E=v^h_{\rm eff}(x_f(E))$, $\psi(x,E,h)$ is the sum $\psi_{-}+e^{-i\pi/2}\psi_+(x,E,h)$ of the incident and reflected waves $\psi_\pm(x,E,h)$ with $x<x_f(E)$ (at $x>x_f(E)$ $\psi(x,E,h)=O(h^\infty)$). As $x\to-\infty$, we have
\begin{gather}
\psi_{\pm}\to\frac{1}{\sqrt{p}}\exp\left(\pm\frac{i}{h}px\right)\varphi^{\pm}.
\end{gather}
The form of the nanotube after barrier modulo exponentially small corrections doesn't influence the solution. Therefore the part of the tube behind the barrier can be removed. One can see from the formulas \eqref{edep} that the barrier appears not only because of the external potential but also because of the narrowing of the tube ($D(x)\to 0$). Thus the constructed asymptotics model the situation when the end of the tube narrows conically-like, this result in the appearance of a barrier, that is a turning point in the system \eqref{SemiRedEq}. It results in sharp increasing of the wavefunction in the neighborhood of the conical end of the tube, this is probably related to the effect of the luminosity of the tube end.

{\bf Bound states. Asymptotic eigenvalues.}
Suppose that the potential $v_{\rm ext}$ and the magnetic field ${\bf H}$ do not depend on $t$
and the effective potential $v^h_{\rm eff}({\bf R}(x))$ has a stable minimum point $x_0$. In its neighborhood $v^h_{\rm eff}({\bf R}(x))$ has the form of a potential well, which generates a family of $T(E)$-periodic trajectories $\xi(t,E),\,\pi(t,E)$ of system \eqref{HS} parametrized by the energy $E=\frac{(p^h)^2}{2}+v^h_{\rm eff}(x)$.
Substituting them into \eqref{phi-evolution} we obtain the system of the form $\dot\varphi=-G(t)\varphi$ with unitary matrix $T$-periodic with respect to $t$.
We can form a basis in the space of their solutions from
vector-functions of the form $z^j(t,E)\exp(i\beta^j(E)t)$, where $j=1,2,\ldots 2r$,
$z(t,E)$ are $T$-periodic in $t$, and the Floquet exponents $\beta^j(E)$ are real for all $E$. We choose them in such a way that $|\beta^j|$ are minimal.
Let $E^{\nu n}$ be the levels determined by the Bohr-Sommerfeld quantization condition
\begin{gather}
\frac{1}{2\pi}\oint\pi d\xi=\frac{1}{\pi}\int\limits_{x_{\rm min}}^{x_{\rm min}}
\sqrt{2(E-v^h_{\rm eff})}dx=h\left(n+\frac{1}{2}\right), \qquad n=0,1,2,\ldots
\end{gather}


Then the numbers $E^{\nu n}_j=E^{\nu n}+h\beta^j(E^{\nu n})$, where $j = 1,2,\ldots,2r$,  give spectral series (sets of bound states) of the operator $\hat{\cal L}$; the $O(h^2)$-neighborhood of $E^{\nu n}_j(h)$ necessarily contains a point of its spectrum (continuous or discrete). Namely, if the spectrum of the original problem on the interval $E_0-\varepsilon,E_0-\varepsilon)$ is discrete, then the numbers $E^{\nu n}_j$ give the asymptotics of some of its eigenvalues. If the spectrum on this interval is continuous then these numbers apparently approximate the exponentially small bands of the spectrum (c.f. \cite{BrunDobrPank}).

The eigenfunctions corresponding to $E^{\nu n}_j(h)$ are determined with the use of the Maslov canonical operator.

\subsection{Remark on rigorous justification of constructed asymptotic solutions.}

It is natural to discuss the important question about the strict justification
of asymptotic solutions, which can be constructed by using
formal procedures suggested above.
In this paper, we almost do not touch upon this problem (see \S 5.1)
whose solution in general situation is not trivial and
requires an additional investigation.
To study  this problem one can use at least two possible ways.

1) One has to prove that the asymptotic solutions
are close to the exact solution of the initial problem
under some conditions on the coefficients of the initial
equation and on the function classes
to which the solutions of the reduced equation belong.
This way is based on the technique of obtaining a series
of estimates with respect to the parameters $\mu$
and $h$ from different diapasons.

2) The second way of justification is to estimate the accuracy
of functions obtained,
which approximate the exact solution of the initial equation.
This way seems to be more useful, at least from the pragmatic
viewpoint, since the obtained explicit asymptotic formulas
for the solutions of real physical problems can be used.

\subsection{Several effects in nanotubes generated
by their geometry and external fields}
Finally let us described very briefly several effects which one can obtain from the analysis of the asymptotic solutions constructed above. (Some of them have been already discussed in \S 5.1).

{\bf The possibility to model effective one-dimensional
potentials.}
First let us   mention  elementary, but curious, properties
of the above equations,
which are caused by the possibilities
of the nanotechnology:
{\it changing the geometry of a tube placed
in a homogeneous electric field, one can model
different one-dimensional effective potentials}.

First, we consider a tube of constant section
whose axis is a plane curve on the plane $(r_1,r_2)$.
Let the electric field of strength $E_{\rm ext}$
be directed along the $Ox_2$ axis.
Then the effective potential has the form
$\varphi =v_{\rm ext}\bigl({\bf R}(x)\bigr)=E_{\rm ext}R_2(x)$.
If the tube is curved not too much
with respect to the axis~$r_1$,
then we have $x\approx r_1$.
Thus, choosing the tube axis as an appropriate curve
$r_2=r_2(r_1)$,
we can model the potential, the double  potential well,
etc.

As an example of a nonplanar tube,
we consider the spiral
${{\bf R}(x)=(\rho_1\cos (x/\sqrt{\rho_1^2+\rho_2^2}),
\rho_1\sin (x/\sqrt{\rho_1^2+\rho_2^2}),
\rho_2 x/\sqrt{\rho_1^2+\rho_2^2})}$
($\rho_1=\const,\ \rho_2=\const$ are parameters)
in the field $E_{\rm ext}(0,\sin\alpha,\cos\alpha)$.
The effective potential contains
the {\it oscillating and linearly increasing\/}
components:
$\varphi(x)=-\sin\alpha E_{\rm ext}\rho_1\sin(x/\sqrt{\rho_1^2+\rho_2^2})
-x\,\cos\alpha E_{\rm ext}\rho_2/\sqrt{\rho_1^2+\rho_2^2}$.
If $\alpha=\pi/2$, i.e.,
the field is directed perpendicular to the tube axis,
then we obtain a periodic potential
and the equations obtained above
coincide in the first approximation
with the Mathieu equation.
If $\alpha=0$, i.e., the field is directed along the tube,
we obtain the Airy equation.
A more complicated example is the case in which
the tube axis is a winding of a torus:
in this case, in particular,
we can obtain almost periodic potentials.
Similar results can be obtained by changing
the thickness of the tube along the tube axis.

{\bf Dependence of the effective one-dimensional potential
on the tube thickness.}
It is easy to show that for the above potentials
modelling the ``soft" and ``rigid" walls,
the effective one-dimensional potential
of the longitudinal motion depends on the tube thickness
$d_0(x)$ as $1/d_0(x)^2$.
This dependence is a purely quantum effect and
is caused by the effect of the size quantization.
Thus, narrowing the tube,
we obtain a barrier in the one-dimensional motion,
while expanding the tube,
we obtain a potential well or a ``trap."

{\bf Semiclassical splitting of the adiabatic term.}
The ``adiabatic" terms can split
in the semiclassical approximation
due to spin and the magnetic field.
In particular, if the splitting is caused by spin effects,
then spin affects the ``classical" trajectory.
For the case in which the adiabatic term is nondegenerate,
we have $\Lambda=0$. If degeneration exists,
then the momentum matrix~$\Lambda$ is nonzero.
This results in the appearance of an effective ``dipole"
that interacts with the projection
of the magnetic field on the tube axis,
and thus an additional phase of the wave function
(the Berry phase) appears.

{\bf An increase in the electron density
near the endpoint of the nanotube
caused by reflection.} Since the longitudinal energy decreases
with decreasing cross-section of the nanotube,
in a nanotube with a ``closed" endpoint,
the wave packet reflects from the closed endpoint.
In this case,  near the endpoint of the nanotube,
there is a sharp increase in the electron density,
which, apparently, is related to the effect
of luminous emittance observed in nanotubes.

{\bf The Berry phase of the wave function.}
The term
$[\Phi_x\Lambda-\langle Y_2 {\bf n}_1 - Y_1 {\bf n}_2 ,{\bf H}\rangle]\otimes E_s p^h$
in the Hamiltonian (\ref{HamSemiClass})
proportional to the momentum~$p^h$
can be excluded from this Hamiltonian by a change
of the wave function.
This change has the form
\begin{equation}
\psi^j=\exp\left(\int_{x^*}^x\lambda_j(x)dx\right){\psi'}^j,
\end{equation}
where $\lambda_j(x)$ is an eigenvalue of the matrix
$\Phi_x\Lambda-\langle Y_2 {\bf n}_1 - Y_1 {\bf n}_2 ,{\bf H}\rangle$.
In fact, the existence of this term results
in the appearance of an additional phase of the wave function.
This phase must be consistent with the boundary conditions
at the endpoints of the tube,
which gives a correction to the quantization condition.
For example,
we consider the Born--Karman boundary conditions
that are equivalent to the case of a closed tube
(the nanoannulus).
While constructing the eigenfunctions in this case,
it is necessary to require that
the total phase of the wave function be
$2\pi$-periodic, with the above adiabatic correction
taken into account.
It is well known that the adiabatic phase
responsible for the correction to the quantization condition
is the so-called Berry phase.

If the magnetic field is zero,
then the Berry phase is reduced to
$\exp\left(\int_{x^*}^x\Lambda_j(x)d\Phi\right)$,
where $\Lambda_j(x)$ is an eigenvalue
of the momentum matrix~$\Lambda$.

{\bf Ultrashort modes and the state density.}
The correction to~$\mu\chi_1$ becomes comparable with~$1$
in the case $h=\mu^{3/2}$.
This means that the adiabatic asymptotics of the form~(\ref{newrepr})
ceases to hold.
However, for $h<\mu$ and for the same total energy,
there exists a state of the form~(\ref{newrepr})
that belongs to the next subregion of the transverse quantization.
From the physical standpoint,
the fact that there is no asymptotic solution of the form~(\ref{newrepr})
in the case of ultrashort waves in a curved tube means that
such fast modes are destroyed
in passing from a straight tube to a curved tube;
they ``fall" to the next subregions
because of the interaction with the tube ``walls."

The ``instability" of ultrashort modes with parameter $h<\mu^{3/2}$
in a curved tube can lead to a change in the density of states
and, in contrast to the straight nanotube,
effectively increase the Fermi level (cf. \cite{Poklonski}).

{\bf Spin beats.}
The operator
$L_{sy}$ in the semiclassical Hamiltonian,
which corresponds to the spin-orbit interaction,
leads to splitting of the adiabatic term
into $2r$ semiclassical terms
of the longitudinal motion.
To each of the semiclassical terms,
there corresponds, in general, its own phase
$S^\nu_j(x,t),\ j=1,\ldots 2r$.
If the variation in the phase
due to the spin-orbit interaction
is small, then the phase can be expanded in a series
in the constant of the spin-orbit interaction~$\alpha$.
The zero term in the expansion
corresponds to the classical motion of the ``spinless" particle,
and the correction results in the appearance of an envelope
for the fast-oscillating exponential.
Apparently, the appearance of such an envelope
is related to the observed effect of the electron density beating
along the tube \cite{spin}.

\section{Concluding remarks}
Let us briefly summarize the results of this paper,
sometimes repeating the above arguments.
The adiabatic approximation is one of the main tools for
analyzing the solutions of linear stationary and
nonstationary problems in modern mathematical and
theoretical physics.
Different versions of adiabatic approximation
are used in problems of molecular physics,
solid-state physics, plasma physics,
theory of internal and surface waves in fluids,
averaging theory, quantum gravitation theory, etc.

The adiabatic approximation is used in situations
in which
the study of some classes
of physically interesting wave process
described by a ``large" system with $N$ degrees of freedom
can be reduced, on some time intervals,
to the study of a simplified ``effective" system
with $n<N$ degrees of freedom.
In this case,
on the one hand,
a sufficiently wide range of states and solutions
with some general characteristic properties
(but not a set of individual concrete states or solutions)
is considered,
and, on the other hand,
it is not assumed
that all the solutions of the original ``large" system
can be described in the same way.
From the physical viewpoint,
this possibility is usually ensured
by the fact that the problem has different
spatial or spatio-temporal scales,
which, from the mathematical viewpoint,
means that
there is a small parameter characterizing
the different scales of the problem.
As a rule,
the different scales are manifested
in different dependences of the coefficients
of the original equation or boundary conditions
on various variables (or groups of variables),
i.e., on $(N-n)$ ``fast" and~$n$ ``slow" variables.
Thus, it is natural to divide the study
of such distinguished wave processes
into two stages:
1) to derive ``effective" reduced systems,
2) to find their concrete solutions
and then to reconstruct the total solution
describing the entire process.
Of course, these considerations
appear in different fields of physics and mechanics;
the problem is to realize them in concrete formulas.
Moreover, in each field, its own terminology is used.
For example, such reduced systems correspond to terms
in molecular physics,
to modes in hydrodynamic problems,
to subbands of dimensional quantization in nanophysics,
etc.
For various reasons,
it is convenient for us to use the terminology
accepted in physics of low-dimensional systems.

In turn, roughly speaking,
the idea to derive simplified systems has two stages:
1) first,
after the $n$ above-mentioned degrees of freedom are ``frozen"
(it is assumed that
the differentiation operators with respect to the slow variables
commute with the slow variables),
the operator determining the effective equation
can be obtained as an eigenvalue of
an auxiliary spectral problem with $N-n$ degrees of freedom;
realizing this stage,
we otbain effective Hamiltonians (or dispersion relations),
well-known in molecular physics;
2) the operator
(the ``quantization" or the Peierls substitution)
determining the effective reduced equation
with~$n$ degrees of freedom
is reconstructed simply by taking into account the fact
that
the differentiation operator with respect to slow variables
does not actually commute with the slow variables.
We note that the distinguishing of the ``frozen" (slow) variables
can be natural and obvious as, for example,
in problems of molecular physics or
in mechanical problems with different spatial scales,
but can be more ``latent" as in electron waves
in crystals and in the averaging theory.
In this case, an additional degree of freedom
appears in the regularization of the problem.

In some problems, it is sufficient
to perform the ``naive" quantization
of the effective Hamiltonian
(of the dispersion relation),
but, in many problems,
a more accurate analysis is required
and certain difficulties arise
(see, e.g., \cite{LifPit},~\S56).
We note that, as a rule, it is impossible to write
the effective reduced equation exactly;
the problem is to construct
a minimally reasonable number of terms
in the expansion of the operator
in the effective reduced system,
which allows a correct construction of the equation
describing a wide range of wavelengths.
Indeed, the most popular approach
in the adiabatic approximation,
originating from Born and Oppenheimer works,
is based on the assumption that
the desired solution depends smoothly
on all the variables.
Thus, for example,
only the lower energy levels are usually
``grasped" in spectral problems.
At the same time,
in many physical situations,
for example,
in describing the valence electrons in crystals,
the higher energy levels are most interesting,
but, strictly speaking,
they cannot be considered under this approach.
In this case, one can use the semiclassical theory
proposed by Maslov, which, however,
is based on the assumption that
there is a sufficiently rigid relation
between the excitation level and the parameter
characterizing the different scales of the problem.
From the mathematical viewpoint,
this means that, in the problem under study,
along with the parameter
characterizing the different scales,
which is naturally called an adiabatic parameter,
there is another (``semiclassical") parameter
characterizing the excitation level
of the state under study;
in this case, the form of the approximate
(asymptotic) solution depends, as a rule,
on the relations between these parameters.
This can easily be verified comparing
the averaging method,
the Born--Oppenheimer method,
and the Maslov method,
which give  solutions
with ``slow," ``medium," and ``fast" variations
(with respect to the slow variables).

In this paper, we propose an approach
based on the above considerations.
This approach allows one, first,
to describe all the states listed above,
in the range from slowly varying states
to fast or, sometimes, even ``superfast" varying states,
and, second, to classify them appropriately.
In particular,
this approach explains why the states obtained
by the Born--Oppenheimer and Maslov methods
can be treated as the states
on the same subregions (terms)
corresponding to different  ``longitudinal states."
So the effective reduced equation thus obtained
describes not only the states
at the ``bottom" of the subregion,
but also the states corresponding to
the higher energy levels.
We note that, without this reduction,
choosing the subregion for higher energy levels
can be a rather complicated problem by itself.

The approach proposed here
is based on Maslov's observation that the problems
in which the adiabatic approximation is used
can be interpreted as problems with operator-valued symbol,
and these problems are well known in mathematical physics.
This means that the operator
determining the original ``large" system
is a function of two groups of operators
with ``small" and ``large" commutators
(or ``large" anticommutators,
as in the electron--phonon interaction problem).
The use of the concept proposed in this paper,
which is based on equations with operator-valued symbol,
allows one to deal with different linear adiabatic problems
from the unified viewpoint.

The realization of our approach
is based on representing the solution in the form~(3.5)
and obtaining the effective reduced equation (3.7),
which is a generalization of the Peierls substitution.
These formulas, together with the algorithm described in~\S3,
give one of the main results of the present paper.
In these constructions, the key role is played
by the techniques of noncommuting operators
based on elementary notions from the Maslov operator method.
The constructed algorithm allows us,
with any prescribed accuracy,
to calculate the operator determining
the effective reduced equation (3.7)
and the intertwining operator reconstructing
the solution of the original problem
from the solution of Eq.~(3.7).
In particular,
the algorithm thus constructed allows one
to obtain accurate estimates for
the minimal number of terms necessary to construct
the leading part of the asymptotic solution.
It should be noted that, in the reduction procedure,
the possible degeneration of the term
(the effective Hamiltonian) must be taken into account.
In~\S\S4--5,
these formulas are illustrated by several examples
from different fields of physics and mathematics.
Some heuristic arguments leading to formulas (3.5) and (3.7)
are given in~\S2.

In~\S3, we also show that the ``operator separation"
of variables can be treated as
a ``quantum" (or wave) analog of the procedure
of excluding (holonomic) constraints in classical mechanics
(\cite{SchusterJaffe,Arnold} and others).
Indeed, the imposed quantum constraints
can be treated as the restrictions
arising due to the confinement potential
in the ambient configuration space.
In this case, it is natural to assume that
the ``condition of dimensional quantization" is satisfied,
i.e.,
the wavelength in the directions normal to the manifold
corresponding to the degrees of freedom of the effective system
(i.e., the ``limit" manifold)
is compatible with the ``width of the film"
surrounding the limit manifold.
Here the most important is the fact justified in this paper
that, excluding the constraints,
one can, in general, obtain
different effective Hamiltonians
depending on the energy of the ``longitudinal" motion.
In the case of fast oscillating longitudinal states,
this leads to different classical Hamiltonians
determining the motion on the ``limit manifold."

The study of solutions of reduced equations
on some distinguished subregions
is the second part of our approach.
Depending on the relations between the parameters,
this study can generally be performed by different methods.
In~\S5,
taking into account the fact that
the problem contains two parameters,
an adiabatic and a semiclassical,
we classify the solutions,
and this classification shows that
the excited states are constructed differently
than the lower states.
In the construction of excited states,
the momentum in the intertwining operator cannot be neglected,
i.e., the intertwining operator is {\it indeed\/}
an {\it operator} and cannot be replaced by a function,
as it is usually done in different versions
of the Born--Oppenheimer method
(in particular, in solid-state physics).
The existence of the momentum operator
in the intertwining operator
shows that,
from the viewpoint of the Born--Oppenheimer method
for excited states,
a ``distortion" of the term occurs.
For fast varying solutions,
the main methods in this case
are the semiclassical approximation and the WKB-method;
if there are turning points and caustics,
then the WKB--Maslov method is used.
It is well known that,
in the realization of this method,
one must pass to classical Hamiltonian systems.
One of the elementary, but important, consequences
is the fact that
the classical systems can be different
for different excitation levels
and
the ``small" terms in the original equation
can significantly affect the semiclassical characteristics
for some values of the longitudinal energies.
In particular,
in the case of a degenerate adiabatic effective Hamiltonian,
the degeneration can be removed
in the semiclassical approximation;
in this case, the adiabatic term ``splits"
into several semiclassical terms (Hamiltonians).
We consider an example (nanotubes) and
show that the interaction of spin
with the confinement potential
can change the classical trajectories
of the longitudinal motion.

As was already noted,
the possibility to obtain numerical solutions,
graphs, etc. at this stage
significantly depends on the specific character
of a concrete problem
and requires separate publications.
In the present paper,
we briefly describe this procedure
for problems related to the modern field of nanophysics
and restrict ourselves to rather general formulas.
In~\S5, we derive some simplest conclusions
for models arising in nanophysics.
In particular,
we show that, placing nanotubes of various geometry
in a constant electric field,
we can model various one-dimensional potentials,
for example, ``double well" type potentials,
periodic potentials, etc.;
the degenerate adiabatic term
(for example, in the case of a tube of circular section)
can split into several semiclassical terms
(effective Hamiltonians), etc.
The problems concerning detailed studies
of how the spin affects the classical trajectories,
the electron density pulsation due to spin, etc.
are beyond the framework of this paper.

We believe that the arguments and formulas
given in this paper can be helpful
in studying the problems
arising in solid-state physics,
hydrodynamics (waves on water),
the theory of shells, plates, and rods,
and in nanophysics.
It seems possible that this method can be used
in weakly nonlinear situations.

\bigskip

In conclusion,
we make a remark concerning the list of references.
As was already noted,
the number of works dealing with
the adiabatic approximation and its applications
is extremely large;
our list of references does not absolutely pretend
to be complete;
here we present only several papers
that are to some extent close to our approach.


\bigskip
\subsection*{Acknowledgments}
The authors are grateful to S.~Albeverio, J.~Br\"uning, V.A.~Geyler, K.V.~Pankrashkin  and~I.V.~Tyutin for useful
discussions and valuable remarks.
The work is~supported in part by grants INTAS (00-257) and DFG-RAS (DFG 436 RUS 113/785, DFG 436 RUS 113/572/0-2).

\end{document}